\makeatletter
\def\input@path{{latex/}}
\makeatother
\documentclass[cernpreprint,atlasdraft=true,texlive=2016,UKenglish,texmf,texmf]{atlasdoc}
 
\usepackage[subfigure=true]{atlaspackage}
 
\usepackage{atlasbiblatex}
 
\usepackage{atlasphysics}
\usepackage{multirow}
\usepackage{siunitx}
\usepackage{arydshln}
\usepackage{enumitem}

\graphicspath{{logos/}{figures/}}
 
\usepackage{ANA-HDBS-2018-13-PAPER-defs}

\AtlasTitle{Search for a heavy Higgs boson decaying into a $Z$ boson
and another heavy Higgs boson in the $\ell\ell bb$ and $\ell \ell WW$ final states in
$pp$ collisions at $\sqrt{s}=13$~$\text{Te\kern -0.1em V}$ with the ATLAS detector}
 
\AtlasAbstract{A search for a heavy neutral Higgs boson, $A$, decaying into a $Z$ boson
and another heavy Higgs boson, $H$, is performed using
a data sample corresponding to an integrated luminosity of
139~fb$^{-1}$ from proton--proton collisions at
$\sqrt{s} = 13$~$\text{Te\kern -0.1em V}$ recorded by the ATLAS detector at the LHC.
The search considers the $Z$ boson decaying into electrons or muons
and the $H$ boson into a pair of $b$-quarks or $W$ bosons.
The mass range considered  is 230--800~$\text{Ge\kern -0.1em V}$ for the $A$ boson
and  130--700~$\text{Ge\kern -0.1em V}$ for the $H$ boson.
The data are in good agreement with the background predicted by
the Standard Model, and therefore 95\% confidence-level upper limits
for $\sigma\times B(A\to ZH)\times B(H\to bb \; \textrm{or} \; H \to WW)$
are set. The upper limits are
in the range 0.0062--0.380~pb for the $H\to bb$ channel and
in the range 0.023--8.9~pb for the $H\to WW$ channel.
An interpretation of the results in the context of two-Higgs-doublet
models is also given.}
 
\author{The ATLAS Collaboration}
 
\AtlasRefCode{HDBS-2018-13}
\PreprintIdNumber{CERN-EP-2020-191}
\AtlasJournal{EPJC}
\AtlasJournalRef{Eur. Phys. J. C. 81 (2021) 396}
\AtlasDOI{10.1140/epjc/s10052-021-09117-5}

\hypersetup{pdftitle={ATLAS document},pdfauthor={The ATLAS Collaboration}}
 
\begin{document}
 
\maketitle

\section{Introduction}
\label{sec:intro}
 
% The next lines are included from the .//introduction.tex input file
After the discovery of a Higgs boson at the Large Hadron Collider
(LHC)~\cite{HIGG-2012-27,CMS-HIG-12-028},
detailed measurements of its
properties~\cite{HIGG-2018-57,Aad:2020ivc,HIGG-2018-14,HIGG-2013-17,Sirunyan:2018sgc,Sirunyan:2018koj,Sirunyan:2020sum,Khachatryan:2016vau}
have shown
excellent compatibility with the Standard Model (SM) Higgs boson
\cite{PhysRevLett.13.321,HIGGS1964132,PhysRevLett.13.508,PhysRevLett.13.585,PhysRev.145.1156,PhysRev.155.1554}.
These results indicate that the scalar sector of the theory of
the electroweak interaction contains at least a doublet of complex
scalar fields. In addition, they constrain the possibilities for
additional spin-0 field content in the theory and disfavour
parts of the parameter space in models with extended Higgs sectors.
These results, however, still allow several extensions of the Higgs sector,
such as the two-Higgs-doublet model (2HDM)~\cite{Lee:1973iz,Branco:2011iw}, in
which large parts of the parameter space are compatible with the existence of a
Higgs boson like the one in the SM.
In the 2HDM, a second complex doublet of the Higgs fields is added to the
single SM Higgs doublet. The model has a weak decoupling limit
\cite{PhysRevD.67.075019} in which one of its predicted Higgs bosons
has couplings to fermions and vector bosons
that are the same as those of the SM Higgs boson at lowest order.
In addition, a Higgs sector structure with two complex doublets of
fields appears
in several new physics scenarios, including
supersymmetry~\cite{Djouadi:2005gj}, dark-matter models~\cite{Abdallah:2015ter},
axion models~\cite{Kim:2008hd}, electroweak baryogenesis~\cite{Cohen:1993nk}  and
neutrino mass models~\cite{King:2003jb}.
 
The addition of a second Higgs doublet leads to five Higgs bosons
after electroweak symmetry breaking. The phenomenology of such a model
is very rich and depends on many parameters, such as the ratio of the
vacuum expectation values of the two Higgs doublets ($\tan\beta$)
and the Yukawa couplings of the scalar sector~\cite{Branco:2011iw}.
When CP conservation is assumed, the model contains two CP-even Higgs bosons,
$h$ and  $H$ with $m_{H} > m_{h}$,
one that is CP-odd, $A$, and two charged scalars, $H^{\pm}$.
There have been many searches for a CP-even Higgs boson at the LHC,
in channels that include
$H\to WW/ZZ$~\cite{Aaboud:2017itg,Aaboud:2017fgj,HDBS-2018-10-p,HDBS-2018-31,CMS-HIG-17-033,CMS-HIG-17-012} and $H\to hh$ \cite{HDBS-2018-58,CMS-HIG-17-030}, as well as
dedicated searches for the heavy CP-odd Higgs boson, as in
the $A\to Zh$ channel \cite{EXOT-2016-10,CMS-HIG-18-023}.
Some 2HDM searches are agnostic with respect to whether the heavy Higgs bosons
are CP-even or CP-odd, for example searches in the
$A/H\to \tau\tau/bb$ \footnote{To simplify the notation, antiparticles are not explicitly labelled in this paper.}
channels~\cite{HDBS-2018-46,CMS-HIG-16-018,CMS-HIG-17-020}.
In the interpretation of this last category of channels it is usually
assumed that both heavy Higgs bosons are degenerate in mass, a hypothesis
that is motivated in certain supersymmetric models~\cite{Djouadi:2005gj}.
Finally, there have been searches for signatures that
explicitly assume different masses for the heavy Higgs bosons,
for example searches in the $A\to ZH\to \ell\ell bb$/$\ell\ell\tau\tau$
channels~\cite{CMS-HIG-15-001,EXOT-2016-34,CMS-HIG-18-012}.
 
The case in which the heavy Higgs bosons have different masses,
in addition to being in an allowed part of the parameter
space, is further motivated  by electroweak baryogenesis
scenarios in the context
of the 2HDM~\cite{Dorsch:2014qja,Turok:1990zg,Fromme:2006cm,Basler:2016obg}.
For 2HDM electroweak baryogenesis to
occur, the requirement $m_A > m_H$ is favoured \cite{Dorsch:2014qja} for
a strong first-order phase transition to take place in the early universe.
The $A$ boson mass  is also constrained to be less than
approximately $ 800$~\GeV, whereas the lighter CP-even Higgs boson, $h$,
is required to have properties similar to those of a SM Higgs boson
and is assumed to be the Higgs boson with a mass of $125$~\GeV\ that
was discovered at the LHC~\cite{Dorsch:2014qja}.
Under such conditions and for large parts of the 2HDM parameter space,
the CP-odd Higgs boson, $A$, decays
into $ZH$~\cite{Coleppa:2014hxa,Dorsch:2014qja}.
At the LHC, the production of the $A$ boson in the relevant 2HDM parameter space proceeds
mainly through gluon--gluon fusion and in association with $b$-quarks
($b$-associated production).

\begin{figure}
\centering
\subfigure[]{\includegraphics[width=0.49\textwidth]{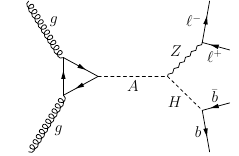}\label{fig:llbb:ggA}}
\subfigure[]{\includegraphics[width=0.49\textwidth]{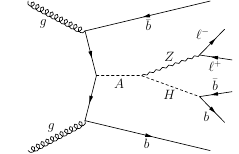}\label{fig:llbb:bbA}}
\subfigure[]{\includegraphics[width=0.49\textwidth]{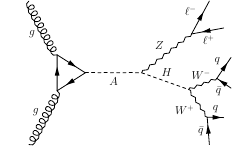}\label{fig:llWW:ggA}}
\caption{Example lowest-order Feynman diagrams for (a) gluon--gluon fusion production of $A$ bosons
decaying into $ZH\to \ell\ell bb$, (b) $b$-associated production of $A$ bosons decaying into $ZH\to \ell\ell bb$,
and (c) gluon--gluon fusion production of $A$ boson decaying into $ZH\to \ell\ell WW$.}
\label{fig:feyn}
\end{figure}

This search for $A\to ZH$ decays uses proton--proton collision
data at $\sqrt{s} = 13$~\TeV\
corresponding to an integrated luminosity of 139~fb$^{-1}$
recorded by the ATLAS detector at the LHC.
The search considers $Z\to \ell\ell$, where $\ell=e,\ \mu$,
to take advantage of the clean leptonic  final state. The $H$ boson
is studied in the $H\to bb$ and $H\to WW$ decay channels.
The $H\to bb$ channel takes advantage of the high branching ratio in large
parts of the 2HDM parameter space, especially in the weak decoupling
limit,
where the $H$ boson decays into weak vector bosons are suppressed.
The $H\to WW$ decay channel is considered in the case where both $W$ bosons
decay hadronically. This heavy Higgs boson decay is dominant in parts of the 2HDM parameter space
close to, but not exactly at, the weak decoupling limit~\cite{Dorsch:2014qja}
and it provides a new way to look for $\ell\ell WW$ resonances
in a final state that has been less explored by other LHC searches.
Both final states considered allow full reconstruction
of the $A$ boson's decay kinematics.
This search considers both the gluon--gluon fusion (see Figure~\ref{fig:llbb:ggA})
and $b$-associated production mechanisms (see Figure~\ref{fig:llbb:bbA})
for the $A\to ZH\to \ell\ell bb$ channel.
The $b$-associated production mode of the $A\to ZH\to \ell\ell WW$ channel
is theoretically allowed, but leads to more complicated jet combinatorics
and would necessitate changing the event reconstruction strategy.
For this reason, only the gluon--gluon fusion production mode (see Figure~\ref{fig:llWW:ggA})
is considered here.
 
This article is organised as follows.
Section~\ref{sec:atlas} introduces the ATLAS detector.
A description of the collision and simulated data samples used in this
article is given in Section~\ref{sec:samples}. The
algorithms used to reconstruct the objects used in this search
are described in Section~\ref{sec:objects}. The event selection
and background estimates
for the two channels considered and the modelling of the signal
are discussed in Sections~\ref{sec:analysis} and \ref{sec:signal},
respectively. Section~\ref{sec:systematics} is devoted to the
description of the systematic uncertainties. The results are discussed
in Section~\ref{sec:results} and the conclusions are given in
Section~\ref{sec:conclusion}.
 
\section{ATLAS detector}
\label{sec:atlas}
 
The ATLAS experiment~\cite{PERF-2007-01} at the LHC is a general-purpose particle detector
with cylindrical geometry and forward--backward symmetry.
It includes an inner-detector tracker surrounded by a
2~T superconducting solenoid, electromagnetic and hadronic
calorimeters, and a muon spectrometer with a toroidal magnetic field.
The inner detector consists of a high-granularity silicon pixel detector,
including the insertable B-layer~\cite{ATLAS-TDR-2010-19-with-add,Abbott:2018ikt},
a silicon microstrip tracker,
and a straw-tube tracker. It provides precision tracking of charged particles
with pseudorapidity $|\eta| < 2.5$.\footnote{ ATLAS uses a right-handed coordinate
system with its origin at the nominal interaction point (IP)
in the centre of the detector and the $z$-axis along the beam pipe.
The $x$-axis points from the IP to the centre of the LHC ring, and the
$y$-axis points upward. Cylindrical coordinates ($r$,$\phi$) are used
in the transverse plane, $\phi$ being the azimuthal angle around
the beam pipe. The pseudorapidity is defined
in terms of the polar angle, $\theta$, as $\eta = -\ln \tan(\theta/2)$.
Transverse momenta are computed from the three-momenta, $\vec{p}$,
as $\pT = |\vec{p}| \sin\theta$.}
The calorimeter system covers
the pseudorapidity range $|\eta| < 4.9$. It is composed of sampling
calorimeters with either lead/liquid-argon, steel/scintillator-tiles,
copper/liquid-argon or tungsten/liquid-argon as the absorber/sensitive material.
The muon spectrometer provides muon identification and momentum
measurement for $|\eta| < 2.7$. A two-level trigger system~\cite{TRIG-2016-01} is employed
to select events to be recorded at an average rate of about 1~kHz for offline analysis.
% End of text imported from the .//introduction.tex input file
 
\section{Data and simulated event samples}
\label{sec:samples}
% The next lines are included from the .//samples.tex input file
 
The data used in this search were collected between
2015 and 2018 from $\sqrt{s}=13~\TeV$ proton--proton collisions and
correspond to an integrated luminosity of 139~fb$^{-1}$~\cite{DAPR-2010-01,DAPR-2011-01,DAPR-2013-01,LUCID2}, which
includes only data-taking periods where all relevant detector
subsystems were operational~\cite{DAPR-2018-01}. The data sample was collected using
a set of single-muon~\cite{Aad:2020uyd} and single-electron
triggers~\cite{TRIG-2018-05}.
The single-muon triggers had \pt\ thresholds in the range of 20--26~\GeV\ for
isolated muons and 50~\GeV\ for muons without any isolation requirement.
The single-electron triggers employed a range of \pt\ thresholds in the range 24--300~\GeV\
and a combination of quality and isolation requirements depending on the
data-taking period and the \pt\ threshold.

Simulated signal events with $A$ bosons produced by gluon--gluon fusion
were generated at leading order (LO) with {\MGMCatNLO}~2.3.3 \cite{Alwall:2011uj,Alwall:2014hca}, using
{\PYTHIA}~8.210~\cite{Sjostrand:2014zea} with a set of tuned parameters called
the A14 tune~\cite{ATL-PHYS-PUB-2014-021} for parton showering.
The decays of $H\rightarrow bb$ and $WW$ were considered.
Additionally, in the $A\to ZH\to \ell\ell bb$ channel, $A$ bosons produced in association with $b$-quarks were
generated at next-to-leading-order (NLO)
with {\MGMCatNLO}~2.1.2 \cite{Alwall:2014hca,Frixione:2002ik,Wiesemann:2014ioa} following
Ref.~\cite{deFlorian:2016spz} together with
{\PYTHIA}~8.212 and the A14 tune for parton showering.
The gluon--gluon fusion production used \nnpdftwo~\cite{Ball:2012cx}
as the parton distribution functions~(PDFs),
while the $b$-associated production used \texttt{CT10nlo\_nf4}~\cite{Lai:2010vv}.
The signal samples were generated for $A$ bosons with masses in the range
of $230$--$800$~\GeV\ ($300$--$800$~\GeV)
and widths up to 20\% of the $A$ mass, and for narrow-width $H$ bosons with
masses in the range of $130$--$700$~\GeV\ ($200$--$700$~\GeV) for the $\ell \ell bb$ ($\ell \ell WW$) channel.

Background events from the production of $W$ and $Z$ bosons in association with jets were
simulated with {\SHERPA}~v2.2.1~\cite{Bothmann:2019yzt}
using NLO matrix elements (ME) for up to two partons, and LO matrix elements for up to four partons calculated with the Comix~\cite{Gleisberg:2008fv}
and OpenLoops~\cite{Cascioli:2011va,Denner:2016kdg} libraries. They
were matched with the {\SHERPA} parton shower~\cite{Schumann:2007mg} using the MEPS@NLO
prescription~\cite{Hoeche:2011fd,Hoeche:2012yf,Catani:2001cc,Hoeche:2009rj}
using the set of tuned parameters developed by the {\SHERPA} authors.
The \nnpdfnnlo set of PDFs~\cite{Ball:2014uwa} was used and the samples
were normalised to a next-to-next-to-leading-order (NNLO)
prediction~\cite{Anastasiou:2003ds}.
Production of $WW$, $ZZ$ and $WZ$ pairs was simulated using
the same generator and parameters as for the $W$ and $Z$ boson samples.
 
The production of \ttbar\ 
events was modelled using the
\POWHEGBOXV{v2}~\cite{Frixione:2007nw,Nason:2004rx,Frixione:2007vw,Alioli:2010xd}
generator at NLO with the \nnpdfnlo~\cite{Ball:2014uwa} PDF set
and the \hdamp\ parameter\footnote{The
\hdamp\ parameter is a resummation damping factor and one of the
parameters that controls the matching of \powheg matrix elements to
the parton shower and thus effectively regulates the
high-\pt\ radiation against which the \ttbar\ system recoils.} set
to 1.5~\mtop~\cite{ATL-PHYS-PUB-2016-020}.  The events were interfaced
to {\PYTHIA}~8.230 to model the parton shower,
hadronisation, and underlying event, with parameters set according
to the A14 tune and using the \nnpdftwo set of PDFs.
The decays of bottom and charm hadrons were performed by \evtgen~v1.6.0~\cite{EvtGen}.
The associated production of a single top quark and $W$ boson ($tW$)
and single top production in the $s$-channel were
modelled using the
\POWHEGBOXV{v2}~\cite{Re:2010bp,Nason:2004rx,Frixione:2007vw,Alioli:2010xd,Alioli:2009je}
generator at NLO in QCD using the five-flavour scheme and the
\nnpdfnlo
set of PDFs.
The diagram removal scheme~\cite{Frixione:2008yi} was used to
remove interference and overlap with \ttbar\ production in the
case of $tW$ production.
The production of \ttV\ events was modelled using the
\mgamc~v2.3.3
generator at NLO with the
\nnpdfnlo
PDF set.
The events were interfaced to {\PYTHIA}~8.210
using the A14 tune
and the
\nnpdftwo
PDF set. The decays of bottom and charm
hadrons were simulated using the \evtgen~v1.2.0 program.

Finally, SM Higgs boson
production in association with a vector boson was simulated using
\powheg~\cite{Nason:2009ai,Alioli:2010xd,Nason:2004rx,Frixione:2007vw} and interfaced with {\PYTHIA}~8.186~\cite{Sjostrand:2007gs} for parton shower and non-perturbative effects. The \powheg prediction is accurate to NLO for the $Vh$ boson plus one
jet production. The loop-induced $gg\to Zh$ process was generated separately at LO. The PDF4LHC15 PDF
set~\cite{Butterworth:2015oua} and the AZNLO tune~\cite{STDM-2012-23} of {\PYTHIA}~8.186 were used. The simulation prediction was normalised to cross sections calculated at NNLO in QCD with NLO electroweak corrections for $q\bar{q}/qg \to Vh$  and at NLO and next-to-leading-logarithm accuracy in QCD for $gg \to
Zh$~\cite{Ciccolini:2003jy,Brein:2003wg,Brein:2011vx,Altenkamp:2012sx,Denner:2014cla,Brein:2012ne,Harlander:2014wda}.

The effect of multiple interactions in the same and neighbouring bunch
crossings (pile-up) was modelled by overlaying the original hard-scattering event with simulated inelastic
proton--proton events generated with {\PYTHIA}~8.186
using the \nnpdftwo set of PDFs and the A3 tune~\cite{ATL-PHYS-PUB-2016-017}.
The simulated events were weighted to reproduce the
distribution of the average number of interactions per bunch crossing
($\left<\mu \right>$) observed in the data. The $\left<\mu \right>$
value in the simulation was rescaled by a factor of $1.03 \pm 0.07$ to improve
agreement between data and simulation in the visible inelastic
proton--proton cross section~\cite{STDM-2015-05}.
All generated background samples were passed through the
{\GEANT}4-based~\cite{Agostinelli:2002hh}
detector simulation~\cite{SOFT-2010-01} of the ATLAS detector.
The ATLFAST-II simulation~\cite{SOFT-2010-01}
was used for the signal samples to allow for the generation of many different
$A$ and $H$ boson masses.
The simulated events were reconstructed in the same way as the data.
 
% End of text imported from the .//samples.tex input file
 
\section{Object reconstruction}
\label{sec:objects}
% The next lines are included from the .//objects.tex input file
Selected events are required to contain at least one vertex having at least two associated tracks with $\pT > 500$~\MeV, and the primary vertex is chosen to be the vertex reconstructed with the largest $\Sigma p_{\mathrm{T}}^2$ of its associated tracks.
 
Electrons are reconstructed from energy clusters in
the electromagnetic calorimeter that are matched
to tracks in the inner detector~\cite{PERF-2018-01}.
Electrons are required to have $|\eta| < 2.47$ and $\pt > 7~\GeV$.
The associated track must have
$|d_0|/\sigma_{d_0}<5$ and $|z_0|\sin \theta<0.5$~mm, where $d_0$
($z_0$) is the transverse (longitudinal) impact parameter relative to
the primary vertex and $\sigma_{d_0}$ is the error in $d_0$.
To distinguish electrons from jets, isolation and quality requirements
are applied.
The quality requirements refer to
both the inner detector track and the calorimeter shower shape.
The isolation requirements are defined
using tracking and calorimeter measurements.
Electrons used in this search satisfy the `Loose' quality and isolation requirements.

Muons are reconstructed by matching tracks reconstructed
in the inner detector to tracks or track
segments in the muon spectrometer~\cite{PERF-2015-10}.
Muons used for this search must have $|\eta| < 2.5$, $\pt > 7~\GeV$, $|d_0|/\sigma_{d_0}<3$,
and $|z_0|\sin \theta<0.5$~mm. They are also
required to satisfy `Loose' isolation requirements, similar to those
used for electrons, as well as `Loose' quality criteria for tracks in the inner detector and muon
spectrometer~\cite{ATLAS-CONF-2020-030}.

Jets are reconstructed from topological clusters in the calorimeter system~\cite{PERF-2014-07},
using the anti-$k_t$ algorithm~\cite{Cacciari:2008gp,Cacciari:2011ma} with radius
parameter $R = 0.4$.
Candidate jets are required to have
$\pt > 20~\GeV$ ($\pt > 30~\GeV$) for $|\eta| < 2.5$
$(2.5 < |\eta| < 4.5)$~\cite{Aad:2020flx}. Low-\pt\ jets from pile-up are rejected
by a multivariate algorithm that uses properties of
the reconstructed tracks in the event for jets with $\pt < 60$~\GeV
and $|\eta| < 2.4$~\cite{PERF-2014-03}.

Jets containing $b$-hadrons
are identified using a multivariate tagging algorithm ($b$-tagging)~\cite{ATL-PHYS-PUB-2017-013, FTAG-2018-01}, which makes use of track impact parameters and reconstructed secondary vertices.
The $b$-tagging algorithm output is used to define
a criterion to select jets originating from $b$-quark hadronisation for jets
with $|\eta| < 2.5$. The jets that
are selected in this way are referred to as $b$-jets in the following.
The criterion in use has an average efficiency of 70\% for jets from $b$-quarks
in simulated \ttbar{} events, with rejection factors of
8.9, 36 and 300 for jets initiated by $c$-quarks, hadronically decaying
$\tau$-leptons and light-flavour quarks or gluons, respectively~\cite{FTAG-2018-01}.

Electrons, muons and jets are reconstructed and identified independently. When those objects are spatially close, these algorithms can lead to ambiguous identifications. An overlap removal
procedure~\cite{Aaboud:2017xsd} is therefore applied to remove ambiguities.
 
The missing transverse momentum, whose magnitude is denoted by \met, is computed as the negative vectorial sum of the transverse momenta of calibrated leptons and jets, plus an additional soft term constructed from all tracks that originate from the primary vertex but are not associated with any identified lepton or jet~\cite{PERF-2016-07,ATLAS-CONF-2018-023}.

% End of text imported from the .//objects.tex input file

\section{Event selection and background estimation}
\label{sec:analysis}
% The next lines are included from the .//analysis.tex input file
The final states for the $A\to ZH\to \ell\ell \ bb$/$WW$ decays feature a pair of oppositely
charged, same-flavour leptons and either two $b$-jets or four mostly
light-flavour jets from the $W$ bosons decays. Three resonances
can be formed by combining the selected objects:
(i) the $Z$ boson ($\ell\ell$),
(ii) the $H$ boson ($bb$ or $WW \rightarrow 4j$), and
(iii) the $A$ boson ($ZH$ system).
 
Events are required to contain exactly two muons or two electrons. The two muons must have opposite electric charges. This requirement is not applied to electrons, since they have a non-negligible charge misidentification rate due to conversions of bremsstrahlung photons.
The highest-\pt\ lepton must satisfy $\pt>27$~\GeV\
in the $\ell\ell bb$ final state, to ensure full efficiency of the single-lepton triggers.
This requirement is raised to $\pt>30$~\GeV\ for the $\ell\ell WW$ final state.
The invariant mass of the lepton pair, $m_{\ell\ell}$, must be in
the range of 80--100~\GeV\ to be compatible with the mass of the $Z$ boson.
 
Further event selection criteria are channel-specific, and are described separately in the following sections.
 
% The next lines are included from the .//llbb.tex input file
\subsection{The $\ell\ell bb$ final state}
\label{sec:llbb}
 
The events that are used for the  $A\to ZH\to \ell\ell bb $  search
are required to have at least two $b$-jets, with
at least one of them having $\pt > 45~\GeV$.
The two highest-\pt\ $b$-jets of the event form the $H\to bb$ system
candidate. The $A$ boson candidate is formed by these two $b$-jets
and, in addition, the two leptons that are matched to the
$Z$ boson.

\begin{figure}[htbp]
\centering
\subfigure[]{\includegraphics[width=0.45\textwidth]{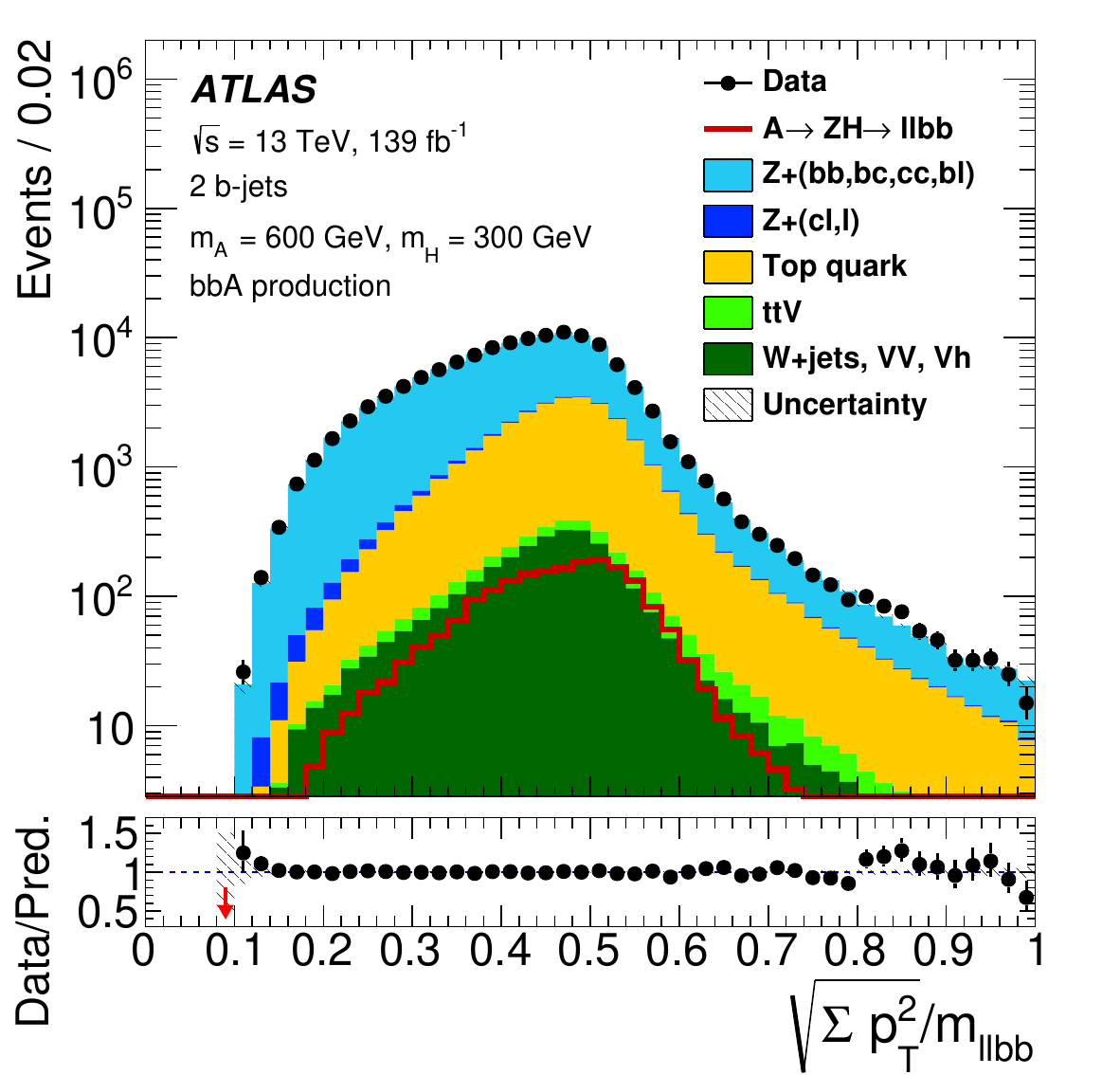}}
\subfigure[]{\includegraphics[width=0.45\textwidth]{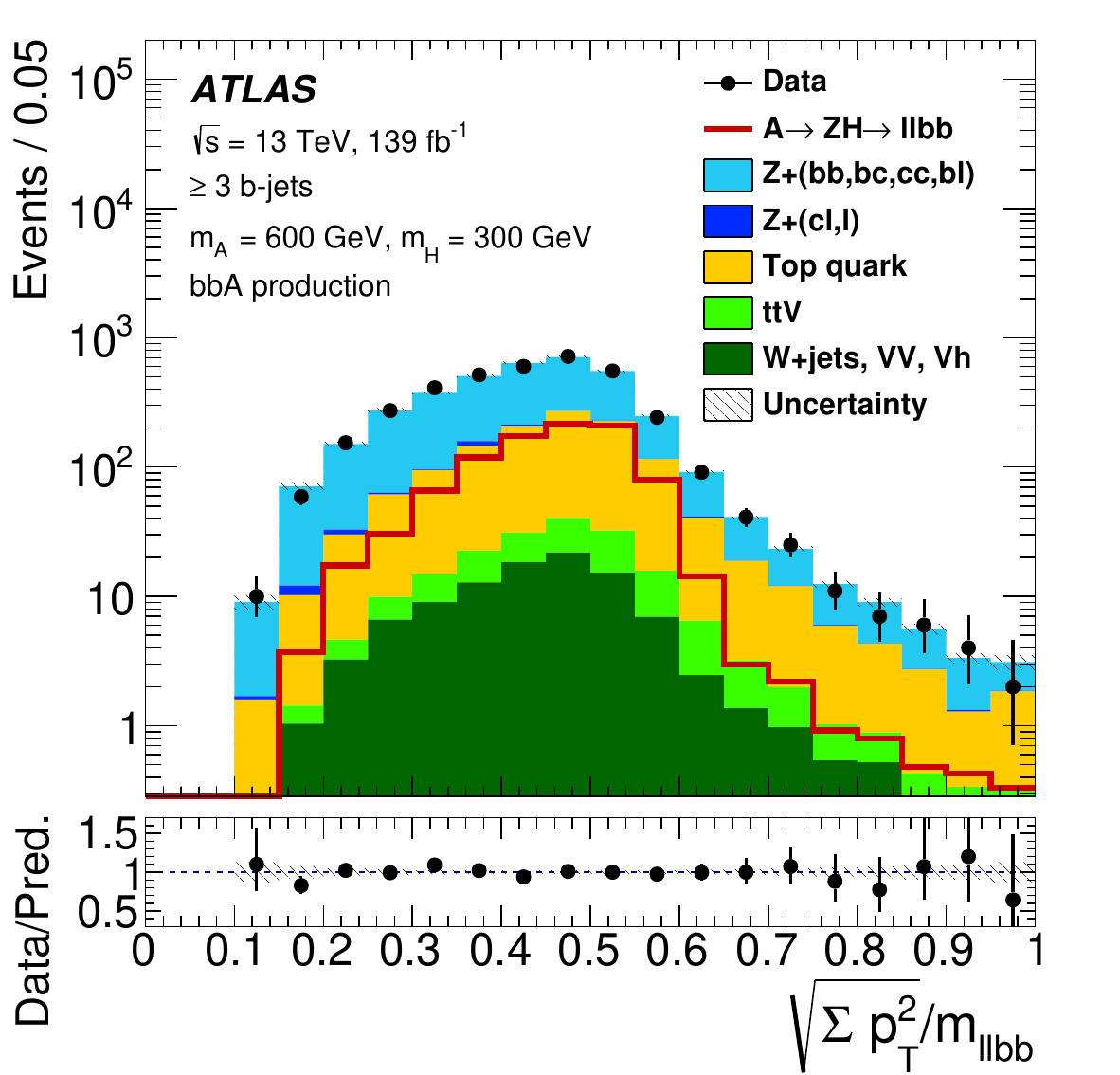}}
\caption{
The $\sqrt{\Sigma \pT^2}/m_{\ell\ell bb}$ distributions
shown before the requirement on this variable is applied
for events with (a) exactly two $b$-jets and (b) three or more $b$-jets.
Corrections from a fit to the data are applied to the simulation,
as described in Sections~\ref{sec:llbb} and \ref{sec:fit}.
The signal distribution for ($m_A,m_H) = (600, 300)$~\GeV\ is also shown, and is
normalised such that the production cross section times the
branching ratios $B(A\to ZH)$ and $B(H\to bb)$ corresponds to 1~pb.
The signal shown includes only $A$ bosons produced in association with $b$-quarks.
The lower panel shows the ratio of the data to the background prediction
(black filled circles) and the relative uncertainty,
which includes both statistical and systematic components,
in the background prediction
(hatched area). The notations $ttV$, $VV$ and $Vh$ refer to top-pair production in association
with a vector boson, diboson production and SM Higgs boson production in association with
a vector boson, respectively. The production of a $Z$ boson in association with jets
is split based on jet flavour. The notation Z+(bb,bc,cc,bl) refers to the case
where the jets originate from heavy flavour, which includes
at least one jet originating from a $b$-quark or two jets originating from $c$-quarks,
whereas the notation Z+(cl,l) includes all the remaining cases.
}
\label{fig:llbb-jvar}
\end{figure}

The requirement of a same-flavour lepton pair along with several
$b$-jets implies that the signal region is contaminated by
$Z$ boson production in association with jets
and backgrounds including top quarks, like \ttbar\ production.
The presence of neutrinos in semileptonic top-pair production
provides a handle to reduce this background by requiring
$\met /\sqrt{H_{\textrm T}} < 3.5~\GeV^{1/2}$,
where $H_{\textrm T}$ is the scalar sum of
the \pt\ of all jets and leptons in the event.
The $Z$+jets background is reduced by requiring
$\sqrt{\Sigma \pT^2}/m_{\ell\ell bb} > 0.4$, where
$m_{\ell\ell bb}$ is the four-body invariant mass of the two-lepton,
two-$b$-jet system assigned to the $A$ boson candidate
and the summation is performed over the $\pt^2$ of these objects.
This discriminating variable is chosen because
the signal distributions are similar across the two production mechanisms
and the $m_A$ and $m_H$ range used in the search.
The requirement is optimised separately for each signal hypothesis and,
subsequently, an average value is chosen.
The distribution of the $\sqrt{\Sigma \pT^2}/m_{\ell\ell bb}$ variable
is shown in Figure~\ref{fig:llbb-jvar} separately
for the cases where exactly two $b$-jets and three or more $b$-jets
are present in the event. The distribution is shown before
the $\sqrt{\Sigma \pT^2}/m_{\ell\ell bb} > 0.4$ requirement is applied.

The two signal production mechanisms, gluon--gluon fusion and
$b$-associated production, differ mainly in the number of heavy-flavour jets
that are produced in association with the $A$ boson. This motivates a
categorisation based on the number of $b$-jets present in the event.
In particular, two categories are defined: the
$n_b=2$ category, which contains events with exactly two $b$-jets, and the
$n_b \geq 3$ category, which contains events with three  or more $b$-jets.
For gluon--gluon fusion production, more than 95\% of the events
passing the above selection fall into the $n_b=2$ category.
For $b$-associated production, only 25--35\% of the selected events fall into
the $n_b \geq 3$ category, and the others enter
the  $n_b=2$ category. This is because of the relatively
soft \pt\ spectrum  of the associated $b$-jets and the
geometric acceptance of the tracker.
 
Finally, the invariant mass $m_{bb}$ of the $b$-jets
that are assigned to the $H$ boson
must be compatible with the assumed $H$ boson mass.
This is ensured by requiring $m_{bb}$ to be within
optimised boundaries that depend on the assumed $m_H$:
$ 0.85\cdot m_H - 20~\GeV < m_{bb} < m_H + 20~\GeV  $ for the
$n_b=2$ category, and
$ 0.85\cdot m_H - 25~\GeV < m_{bb} < m_H + 50~\GeV  $ for the
$n_b \geq 3$ category.
The wider window for $n_b\geq 3$ is motivated by a slightly poorer
resolution due to potential $b$-jet misassignments.
The $b$-jets that are matched to the $H$ boson are the highest-\pt\
$b$-jets in the event and, hence, in the case of $b$-associated
production, where more $b$-jets are present, may not be the ones
that actually come from the $H\to bb$ decay. In $b$-associated production,
the fraction of $A$ bosons for which the correct
$b$-jets are chosen is in the range 50--90\% for the
$n_b\geq 3$ category and is at least 65\% for the  $n_b=2$ category.
 
The signal efficiency in the $n_b=2$ category after the $m_{bb}$ window
requirement is 5.1--11\% (2.5--6.6\%) for gluon--gluon fusion
($b$-associated) production, depending
on the $m_A$ and $m_H$ values. Similarly, the efficiency in
the $n_b \geq 3$ category after the $m_{bb}$ window requirement
is 1.3--3.2\% for $b$-associated production.
The quoted numbers refer to the efficiencies for $A$ bosons decaying into $ZH$, with
$Z\to ee/\mu\mu/\tau\tau$ and $H\to bb$, to pass the event selection for
each of the categories.
The inclusion of $Z\to\tau\tau$ in this definition lowers the quoted signal efficiency
because these decays have a very small efficiency to pass in this selection (which aims at  $Z\to ee/\mu\mu$).
The signal region selection is summarised in Table~\ref{tab:selections}.
 
\begin{table}[h!]
\caption{Summary of the event selection for signal and control regions in the  $A\to ZH\to \ell\ell bb$ channel.}
\vspace{0.2cm}
\centering
\small
\setlength\tabcolsep{5pt}
\begin{tabular}{l|lll}
\hline
\multicolumn{4}{c}{Single-electron or single-muon trigger} \\
\multicolumn{4}{c}{Exactly 2 leptons ($e$ or $\mu$) ($\pt > 7$~\GeV)
with the leading one having $\pt > 27$~\GeV} \\
\multicolumn{4}{c}{Opposite electric charge for $\mu\mu$ pairs;~~ $80\ \GeV < m_{\ell\ell,\ e\mu}< 100\ \GeV$, $\ell = e,\ \mu$} \\
\multicolumn{4}{c}{At least 2 $b$-jets ($\pt > 20\, \GeV$) with one of them having $\pt > 45\, \GeV$} \\
\multicolumn{4}{c}{$\met /\sqrt{H_{\textrm T}} < 3.5~\GeV^{1/2}$, $\sqrt{\Sigma \pT^2}/m_{\ell\ell bb} > 0.4$} \\
\hline \\[-3mm]
&    $n_b =2$ category   &\hspace*{0.5cm}&  $n_b \ge 3$ category \\
\hline \\[-3mm]
& Exactly 2 $b$-tagged jets    && At least 3 $b$-tagged jets \\
\hdashline
\\[-3mm]
Signal & $ee$ or $\mu\mu$ pair && $ee$ or $\mu\mu$ pair \\
region   & $ 0.85\cdot m_H\! -\! 20~\GeV\! < m_{bb}\! < m_H\! +\! 20~\GeV $ && $ 0.85\cdot m_H\! -\! 25~\GeV\! < m_{bb}\! < m_H\! +\! 50~\GeV$ \\
\hdashline
\\[-3mm]
$Z+$jets & $ee$ or $\mu\mu$ pair && $ee$ or $\mu\mu$ pair \\
control region   & $m_{bb} < 0.85\cdot m_H - 20~\GeV$  && $ m_{bb} < 0.85\cdot m_H - 25~\GeV $ \\
& or $ m_{bb} > m_H + 20~\GeV$  &&  or $ m_{bb} > m_H + 50~\GeV$ \\
\hdashline
\\[-3mm]
Top & $e\mu$  pair && $e\mu$  pair \\
control region  & $ 0.85\cdot m_H\! -\! 20~\GeV\! < m_{bb}\! < m_H\! +\! 20~\GeV$ && $ 0.85\cdot m_H\! -\! 25~\GeV\! < m_{bb}\! < m_H\! +\! 50~\GeV$ \\
\\[-3mm]
\hline
\end{tabular}
\label{tab:selections}
\end{table}
 
The $m_{\ell\ell bb}$ distribution after the $m_{bb}$ requirement
is the final discriminating variable, which is fitted to obtain
the result of the search in this channel.
To improve the $\mllbb$  resolution, the
$bb$ system's four-momentum components are scaled to match the
assumed $H$ boson mass and the $\ell\ell$ system's four-momentum
components are scaled to match the $Z$ boson mass. This procedure, performed
after the event selection,
improves the $m_{\ell\ell bb}$ resolution by a factor
of two without significantly distorting the background distributions,
resulting in an $A$ boson mass resolution
that is at best about 1\% and up to 4\% for gluon--gluon fusion, up to 10\%
for $b$-associated production in the $n_b=2$ category and up to 16\%
for $b$-associated production in the $n_b\geq 3$ category, depending
on the $m_A$ and $m_H$ values.

Despite the dedicated selection criteria against $Z$+jets and top-quark production, these background processes dominate the signal region: the $Z$+jets contribution is $\sim$60--70\% depending on the $n_b$ category, while the top-quark contribution is $\sim$30--35\%. In the $n_b \geq 3$ category, other processes ($t\bar{t}V$, dibosons, $Vh$) contribute up to $\sim$5\% of the total background, while their contribution to the $n_b = 2$ category is less than 1\%.
The accurate determination of $Z$+jets and top-quark contributions is paramount for the sensitivity of this search. Their estimation employs a combination of data-driven corrections to simulated events.
 
\begin{figure}[htbp]
\centering
\subfigure[]{\includegraphics[width=0.45\textwidth]{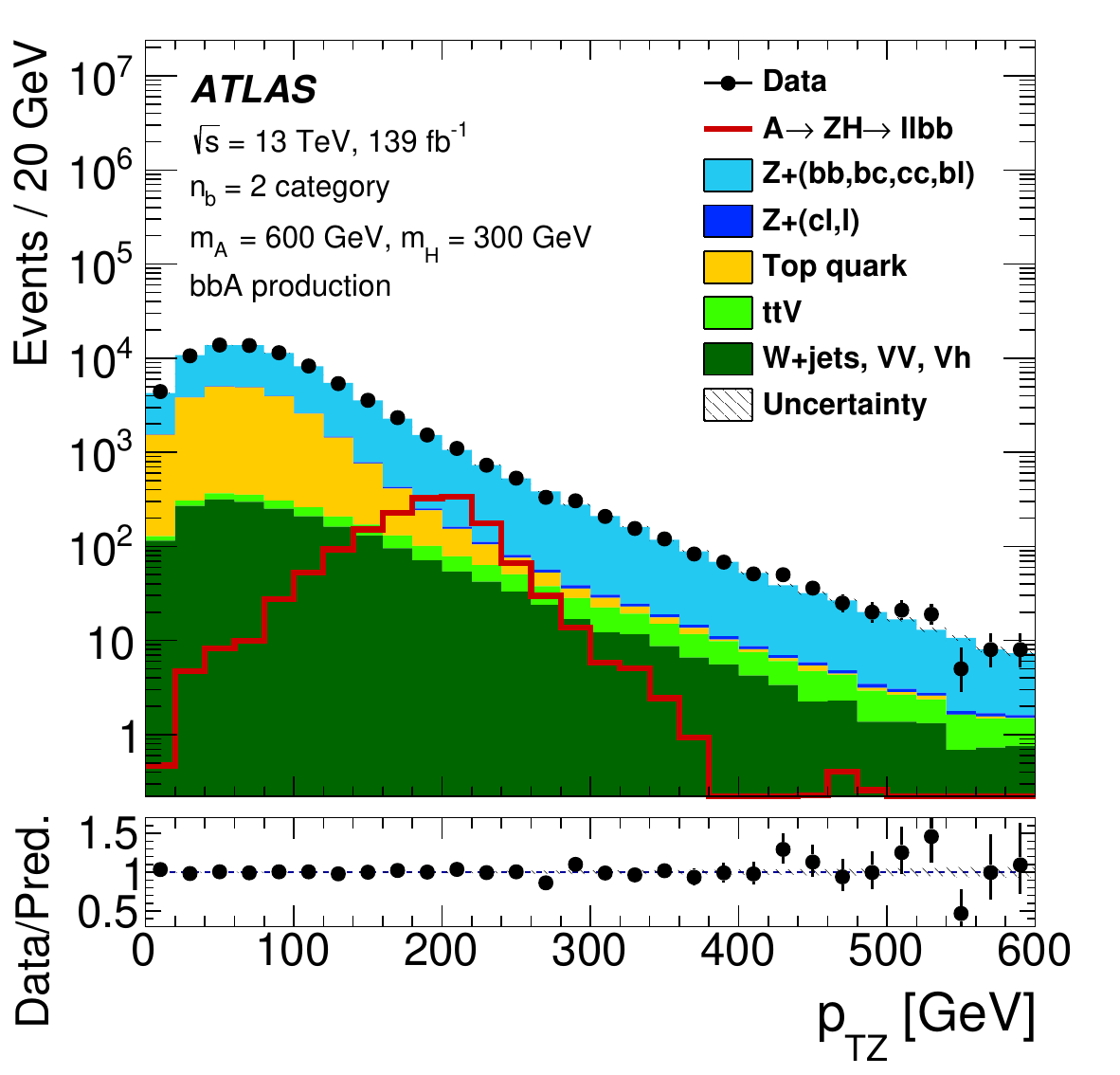}}
\subfigure[]{\includegraphics[width=0.45\textwidth]{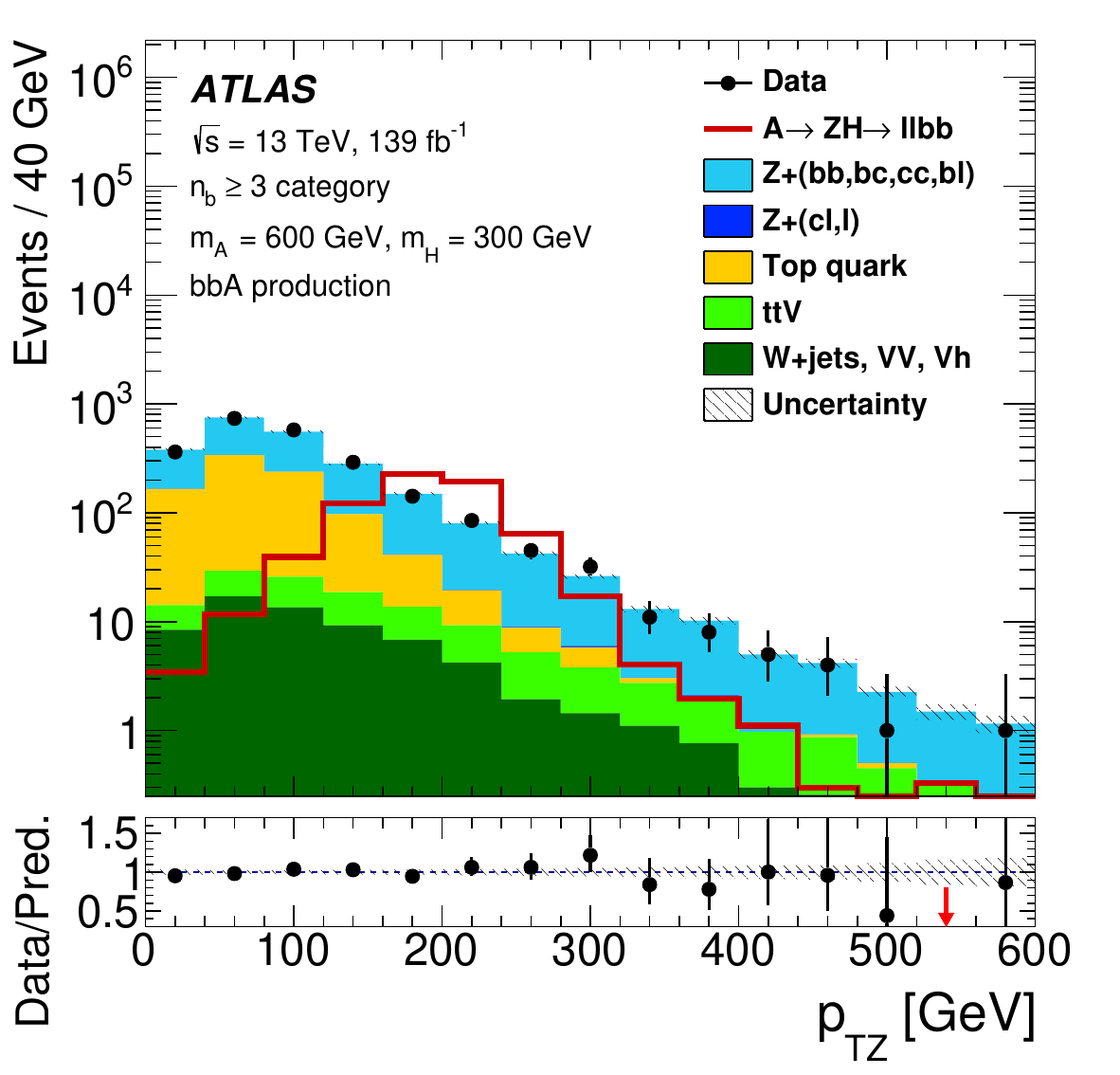}}
\caption{
The \ptz\ distributions for (a) the $n_b=2$ and (b) the $n_b \geq 3$ category.
The events are required to satisfy all the signal region criteria with the
exception of the $m_{bb}$ window requirement.
The same conventions as in Figure~\ref{fig:llbb-jvar} are used.
}
\label{fig:llbb-zpt}
\end{figure}

The most abundant background in this
channel is from $Z$+jets production.
The normalisation of this process is constrained
by a control region defined by inverting the $m_{bb}$
window criterion for each $H$ boson mass hypothesis
(see also Table~\ref{tab:selections}).
The control regions are distinct for the $n_b = 2$ and $n_b \geq 3$
categories, since the accuracy of the background simulation
depends on the number of $b$-jets present in the event.
The modelling of the $Z$+jets simulated events is examined
extensively in a number of kinematic variables, including the
$\pt$  of the $Z$ boson ($\ptz$), the $m_{bb}$ distribution and
the $\sqrt{\Sigma \pT^2}/m_{\ell\ell bb}$ distribution.
The simulated distributions of these variables are compared
against data in a control region
that requires two jets with exactly one of them being a $b$-jet,
as well as an early selection
stage, before the $m_{bb}$ window and
$\sqrt{\Sigma \pT^2}/m_{\ell\ell bb}$ requirements.
For this early selection stage, it was verified that even those signals that were
already excluded in Ref.~\cite{EXOT-2016-34} would be washed
out by the background and would not bias the results.
These regions are not used in the likelihood fit described in Section~\ref{sec:fit}
and thus they are not included in Table~\ref{tab:selections}.
As a result of these studies,
corrections to the distributions of  $\ptz$, $m_{bb}$ and
$\sqrt{\Sigma \pT^2}/m_{\ell\ell bb}$ in the simulated
$Z$+jets events are applied. The corrections are found
to be uncorrelated and they are applied sequentially.
The most significant effect on the
sensitivity of this search (see also Section~\ref{sec:systematics})
is due to the corrections to the modelling of the $\ptz$ distribution,
which range from $+5\%$ to $-10\%$ for most
of the $Z$+jets events.
As an example, Figure~\ref{fig:llbb-zpt} compares the \ptz\ distributions in data with the
background model after all corrections used in this search for events that satisfy all
the requirements of the signal region with the exception of the
$m_{bb}$ window requirement, separately for $n_b = 2$ and $n_b \geq 3$ categories.
 
Top-quark production is heavily
dominated by \ttbar\ production in which both top quarks decay
semileptonically.
Therefore, it is possible to define a pure top-quark control
region by keeping the same selection as discussed previously,
apart from an opposite-flavour lepton criterion, i.e.\
an $e\mu$ pair is required instead of
an $ee$ or $\mu\mu$ pair (see also Table~\ref{tab:selections}).
This region is used for top-pair production normalisation, and also
to check that kinematic distributions such as the top-quark \pt\ spectrum are
adequately modelled in simulation.
Different control regions are used in the $n_b =2$ and $n_b \geq 3$ categories.
This is because in the $n_b \geq 3$ category
the top-quark background is dominated by top-quark pair production
in association with jets, which is more difficult to model than the inclusive
top-quark pair production that dominates the top-quark
background in the $n_b =2$ category.
Finally, the $m_{bb}$
window requirement is also applied to the top-quark control
region, resulting in a separate control region for
each $m_H$ hypothesis tested in the search.
Good agreement within uncertainties is observed between data and simulation in
the shape of all variables considered.
 
Backgrounds from diboson, single top-quark, and SM Higgs boson production, as well as
\ttbar\ production in association with a vector boson are minor contributions to the
total background composition.
The shapes of their distributions are taken from simulation, whereas they are normalised using
precise inclusive cross sections calculated from theory.
The diboson samples are normalised using NNLO cross sections~\cite{Gehrmann:2014fva,Grazzini:2016swo,Grazzini:2015hta,Cascioli:2014yka}.
Single-top-quark production and top-quark-pair production in
association with vector bosons are normalised to NLO cross sections from
Refs.~\cite{Kidonakis:2010tc,Kidonakis:2011wy,Kidonakis:2010ux} and
Ref.~\cite{Alwall:2014hca}, respectively.
The normalisation of SM Higgs boson production in association with
a vector boson follows the recommendations of Ref.~\cite{deFlorian:2016spz}
using NNLO QCD and NLO electroweak corrections.

% End of text imported from the .//llbb.tex input file
 
% The next lines are included from the .//llWW.tex input file
\subsection{The $\ell\ell WW$ final state}
\label{sec:llWW}

The decay $A\to ZH\to \ell\ell WW $ features a pair of
electrons or muons and four jets from the hadronic $W$ boson
decays. The selected events are required to have at least
four jets with  the highest- and second-highest-\pt\
jets satisfying $\pt > 40~\GeV$ and $\pt > 30~\GeV$, respectively.
In addition, the lowest-\pt\ electrons or muons are required
to have $\pt > 15$~\GeV.
 
The selection of the correct jet pairs in the reconstruction
of the two $W$ boson candidates is important for
improving the signal resolution and suppressing
backgrounds.
For this task, all possible
jet pairs that can be formed by considering up to the
five highest-\pt\ jets in the event are taken into account.
A set of requirements on
kinematic variables, such as the angular
distances between the jets within a pair, the
jet transverse momenta and the reconstructed masses of the
$W$, $H$ and $A$ boson candidates, is optimised to
test the various combinations for compatibility with the
signal hypothesis so that the signal efficiency and
background rejection are maximised. This procedure
results in a signal efficiency that ranges
from 50\% to 70\% depending on $m_{A}$ and $m_{H}$, whereas
for background processes the efficiency is about 40\%.
The fraction of events in which the correct jet pairs
are assigned to the $W$ boson candidates after this procedure
is in the range from 50\% to 70\%, depending on the $m_A$ and
$m_H$ values.
 
\begin{figure}[htbp]
\centering
\includegraphics[width=0.45\textwidth]{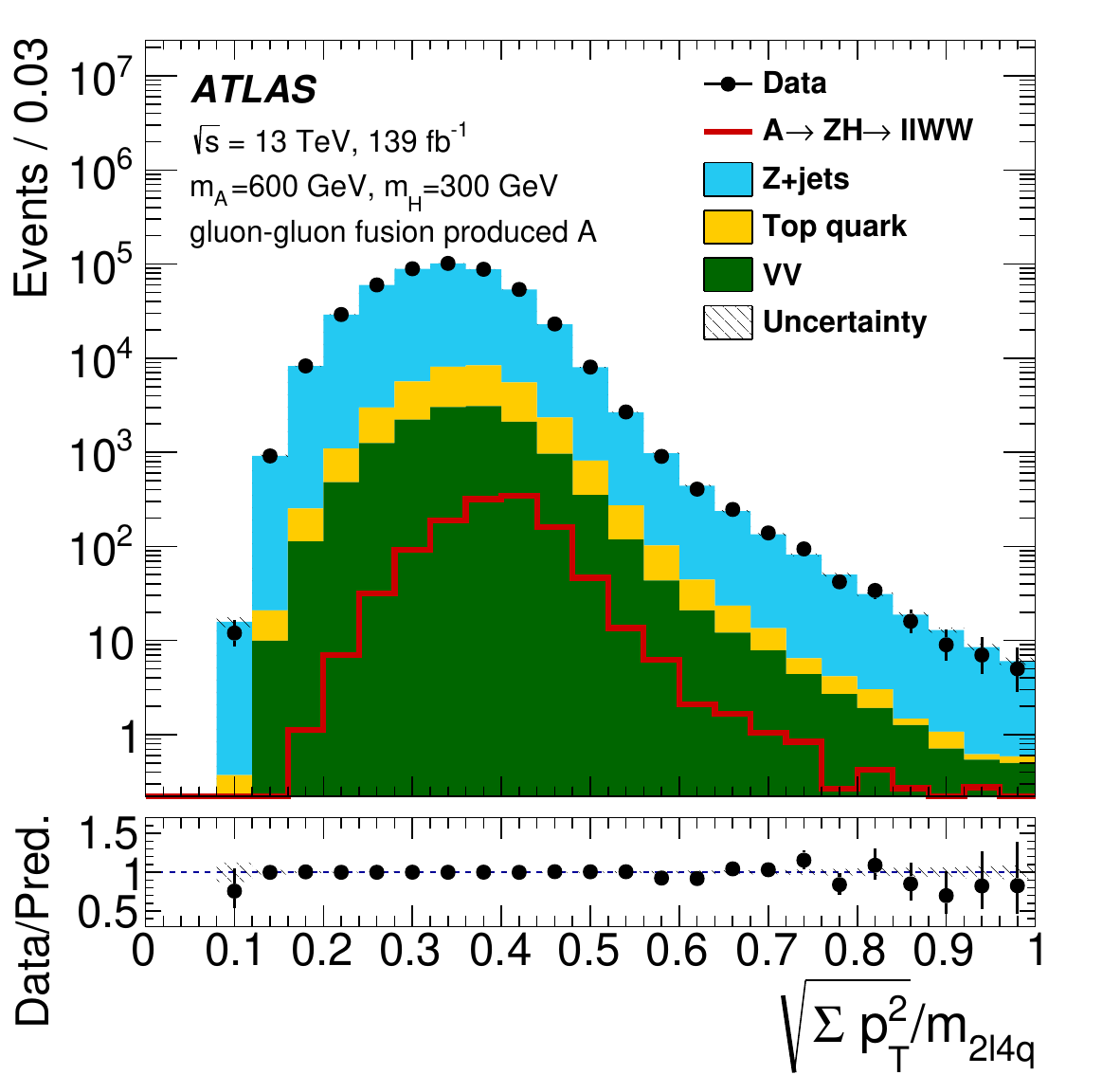}
\caption{
The $\sqrt{\Sigma \pT^2}/\mllqqqq$ distribution
shown before the requirement on this variable is applied.
Corrections from a fit to the data are applied to the simulation,
as described in Sections~\ref{sec:llWW} and \ref{sec:fit}.
The notation VV in the legend corresponds to the production of diboson events.
The signal distribution for ($m_A,m_H) = (600, 300)$~\GeV\ is also shown, and is
normalised such that the production cross section times the
branching ratios $B(A\to ZH)$ and $B(H\to WW)$ corresponds to 1~pb.
The lower panel shows the ratio of the data to the background prediction
(black filled circles) and the relative uncertainty,
which includes both statistical and systematic components,
in the background prediction
(dashed area).
}
\label{fig:Jvar_llWW}
\end{figure}

The main background in this channel is from the production
of a $Z$ boson in association with jets.
A criterion similar to that in the $\ell\ell bb $ channel is employed to
discriminate against it:
$\sqrt{\Sigma \pT^2}/\mllqqqq > 0.3$, where
$\mllqqqq$ is the six-body invariant mass of the two-lepton,
four-jet system assigned to the $A$ boson
and the summation is performed over the $\pt^2$ of these objects.
This discriminating variable is chosen for the same reasons as in the
$\ell\ell bb $ channel and it is optimised following the same considerations.
The distribution of this variable before the requirement is
applied is shown in Figure~\ref{fig:Jvar_llWW}.

Finally, the invariant mass of the four selected jets, $\mqqqq$,
must be compatible with the assumed $H$ boson mass.
This is ensured by requiring $\mqqqq$ to be within
optimised boundaries that depend on $m_H$:
$m_{H} - 53~\GeV < \mqqqq < 0.97 \cdot m_{H} + 54$~\GeV.
After this requirement the signal efficiency for $A$ bosons decaying into $ZH$ with $Z\to ee/\mu\mu/\tau\tau$
and $H\to WW\to qqqq$ is 6.5--11\%, depending on the $m_A$ and $m_H$ values.
The signal region selection is summarised in Table~\ref{tab:selectionsllWW}.

\begin{table}[h!]
\caption{Summary of the event selection for signal and control regions in the  $A\to ZH\to \ell\ell WW$ channel.}
\vspace{0.2cm}
\centering
\small
\setlength\tabcolsep{5pt}
\begin{tabular}{l|l}
\hline
\multicolumn{2}{c}{Single-electron or single-muon trigger} \\
\multicolumn{2}{c}{Exactly 2 leptons ($e$ or $\mu$) ($\pt > 15$~\GeV)
with the leading one having $\pt > 30$~\GeV} \\
\multicolumn{2}{c}{Opposite electric charge for $\mu\mu$ pairs;~~ $80\ \GeV < m_{\ell\ell,\ e\mu} < 100\ \GeV$, $\ell = e,\ \mu$} \\
\multicolumn{2}{c}{At least 4 jets ($\pt > 20\, \GeV$) with leading and second leading jets having $\pt > 40, \;30\, \GeV$} \\
\multicolumn{2}{c}{Jets chosen with a dedicated discriminant}\\
\multicolumn{2}{c}{$\sqrt{\Sigma \pT^2}/\mllqqqq > 0.3$} \\
\hline \\[-3mm]
Signal & $ee$ or $\mu\mu$ pair \\
region   & $m_{H} - 53~\GeV < \mqqqq < 0.97 \cdot m_{H} + 54$~\GeV \\
\hdashline
\\[-3mm]
$Z+$jets & $ee$ or $\mu\mu$ pair \\
control region   & $\mqqqq < m_{H} - 53~\GeV$ \\
& or $ \mqqqq > 0.97 \cdot m_{H} + 54$~\GeV   \\
\hdashline
\\[-3mm]
Top & $e\mu$  pair  \\
control region  & $m_{H} - 53~\GeV < \mqqqq < 0.97 \cdot m_{H} + 54$~\GeV  \\
\\[-3mm]
\hline
\end{tabular}
\label{tab:selectionsllWW}
\end{table}

The $\mllqqqq$ distribution after the $\mqqqq$ requirement
is the final discriminating variable, which is fitted to obtain
the results of the search in this channel.
To improve the $\mllqqqq$ resolution, the
four-jet system's four-momentum components are scaled to match the
assumed $H$ boson mass and the $\ell\ell$ system's four-momentum
components are scaled to match the $Z$ boson mass. The final $A$ boson mass resolution
is in the range from 1\% to 17\% of $m_A$, depending on the
$m_A$ and $m_H$ values.

The dominant backgrounds after the event selection are from
$Z$+jets ($\sim$90\% of total background), top-quark ($\sim$5\%), and diboson ($\sim$5\%) production.
Smaller backgrounds ($W$+jets, $t\bar{t}h$, $t\bar{t}V$, and $Vh$) contribute less than $1\%$ to the
total background and are not included in the background composition.
 
\begin{figure}[htbp]
\centering
\subfigure[]{\includegraphics[width=0.45\textwidth]{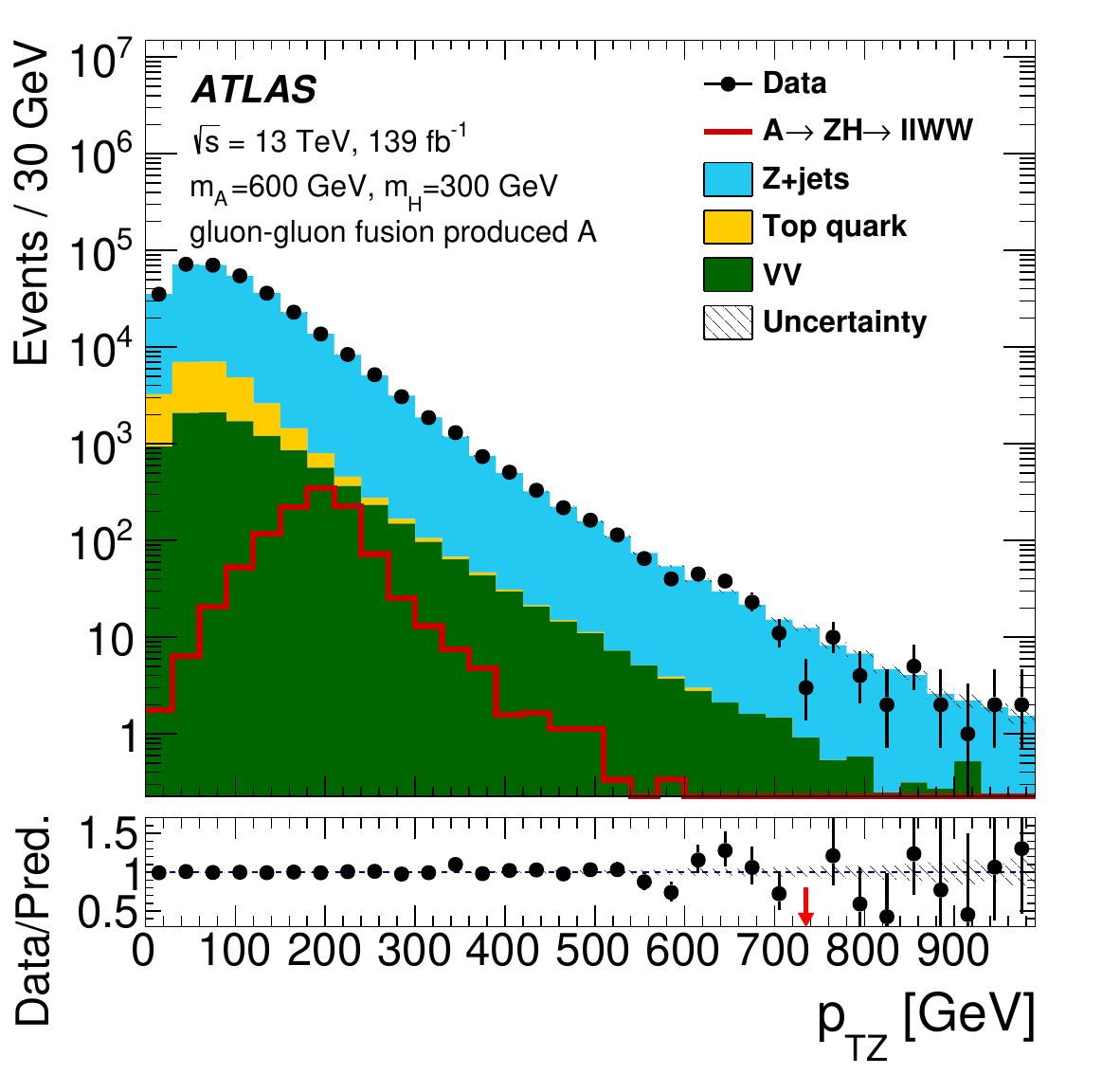}}
\subfigure[]{\includegraphics[width=0.45\textwidth]{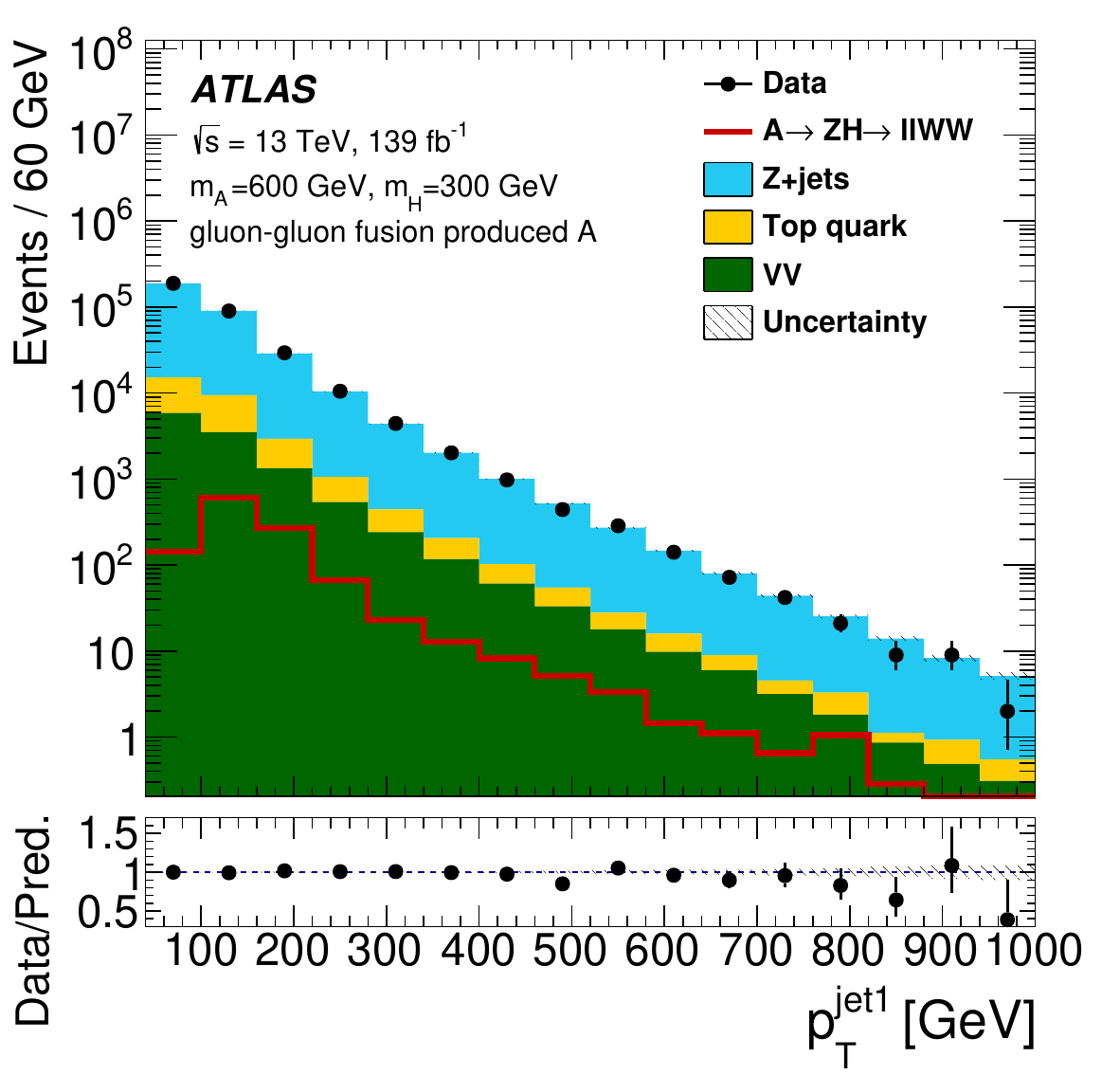}}
\caption{
The distributions of (a) the \pt\ of the $Z$ boson candidates
and (b) the leading jet's \pt\ in the $\ell \ell WW$ channel.
The events are required to satisfy all the signal region criteria with the
exception of the $\mqqqq$ window requirement.
Data-driven corrections are applied, as described in the text.
The same conventions as in Figure~\ref{fig:Jvar_llWW} are used.
}
\label{fig:CR_llWW}
\end{figure}

The shape of the $Z$+jets background is taken from simulation combined with
data-driven corrections, and the normalisation is constrained by
the control region outside the $\mqqqq$ mass window of each signal region  (see Table~\ref{tab:selectionsllWW}),
using a procedure similar to that in the $\ell \ell bb$ channel.
To address shape differences between distributions of kinematic variables
in data and simulated backgrounds,
two corrections are applied to the \pt\ of the $Z$ boson candidates
and to the leading jet's \pt.
Those corrections are derived from a control region orthogonal to the signal region, obtained by
selecting $\sqrt{\Sigma \pT^2}/\mllqqqq < 0.3$. This region is not used subsequently in the
likelihood fit described in Section~\ref{sec:fit} and therefore it is not included in Table~\ref{tab:selectionsllWW}. The corrections are found to be uncorrelated and they are applied
sequentially. The correction to the $\ptz$ distribution in the simulation is as large as
20\% at low $\ptz$ values and it becomes smaller as $\ptz$ increases, whereas the correction to the
leading jet's $\pt$ does not exceed $\pm 10$\%.
The distributions of the \pt\ of the $Z$ boson candidates
and of the leading jet's \pt, after the reweighting, are shown in Figure~\ref{fig:CR_llWW} for events satisfying all
requirements for the signal region with the exception of the $\mqqqq$ window cut.
 
The top-quark background shape is taken from simulated events. The normalisation is constrained using a high-purity control region defined by keeping the same selection as for the signal region, but replacing the electron or muon pairs by opposite-flavour leptons ($e\mu$ pairs), as indicated in Table~\ref{tab:selectionsllWW}. The single-top-quark, $Z$+jets and  diboson production contributions in this control region are estimated from simulation. The diboson background shape and normalisation are taken from the simulated samples, using the same cross-section calculation as in the $\ell \ell bb$ channel.

% End of text imported from the .//llWW.tex input file
% End of text imported from the .//analysis.tex input file
 
\section{Signal modelling}
\label{sec:signal}
% The next lines are included from the .//signal.tex input file
 
This analysis searches for two new particles, with
their mass hypotheses considered in the
two-dimensional space $m_A$--$m_H$, with good mass
resolution of the $A$ and $H$ reconstructed final states.
The investigation of the relevant phase space requires
a large number of signal mass hypotheses to be tested.
In addition, various new physics scenarios which are of interest
for this search, like the 2HDM, include $A$ bosons with natural
widths comparable to, or larger than, the experimental mass resolution
for large parts of the  parameter space in which this search has sensitivity.
The $H$ bosons are considered to always have negligible natural width,
in accordance with the 2HDM scenarios used to interpret this search
(see Section~\ref{sec:fit}).
For these reasons, the $m_{\ell \ell bb}$ and $\mllqqqq$
distributions can be simulated only for some ($m_A$, $m_H$) points
and an interpolation using analytic
functions is employed for the rest,
following a procedure similar to that
used in Ref.~\cite{EXOT-2016-34}.
 
In the cases where the natural
widths of both the $A$ and $H$ bosons are much smaller than the
experimental mass resolution,
the modelling of the mass distributions
uses two types of parametric functions.
First, an {\em ExpGaussExp}~(EGE) function~\cite{Das:2016stf,EXOT-2016-34}
provides a good description of gluon--gluon fusion production of
$A$ bosons in the $n_b = 2$ category of the $\ell \ell bb$ channel.
Second, a double-Gaussian Crystal Ball (DSCB)
function~\cite{EXOT-2016-34,Oreglia:1980cs} gives a good description of
gluon--gluon fusion production in the  $\ell \ell WW$ channel
and $b$-associated production
in both the $n_b = 2$ and $n_b \geq 3$ categories of the
$\ell \ell bb$ channel.

Both the EGE and DSCB functions have a Gaussian core but they differ in the way
the tails are treated. The tails of the EGE function are exponential, described
by two parameters, whereas DSCB has power-law tails described by four extra
parameters. The values of the function parameters are extracted from
unbinned maximum-likelihood fits to the simulated $\mllbb$ and $\mllww$
distributions.
Polynomial functions are used to interpolate the parameters to
mass points that were not simulated. These interpolated parametric functions are used to
model the signal mass shapes for all the signal assumptions considered in this search.
The fit uncertainties of the DSCB and EGE function parameters, as well as the parameters
of the polynomial functions used for the interpolation,
are used to derive a shape uncertainty for each of the interpolated distributions.
 
\begin{figure}[htbp]
\centering
\subfigure[]{\includegraphics[width=0.49\textwidth]{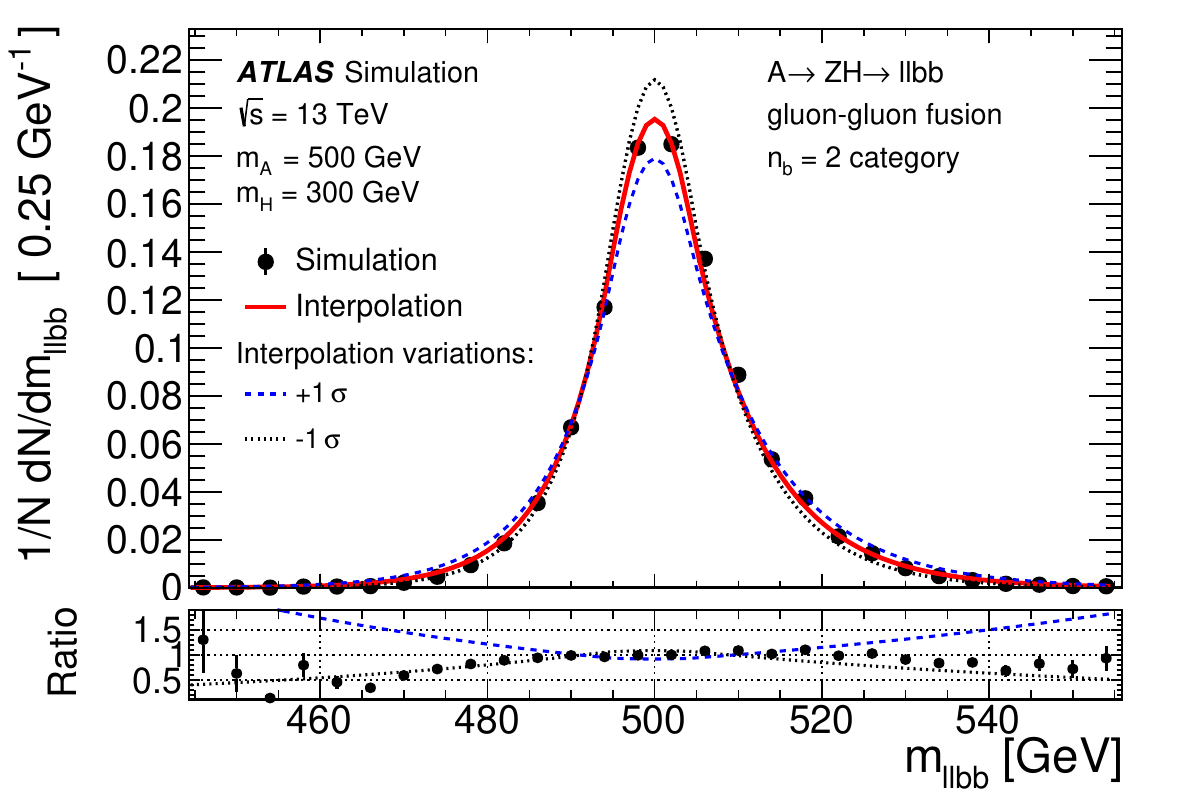}}
\subfigure[]{\includegraphics[width=0.49\textwidth]{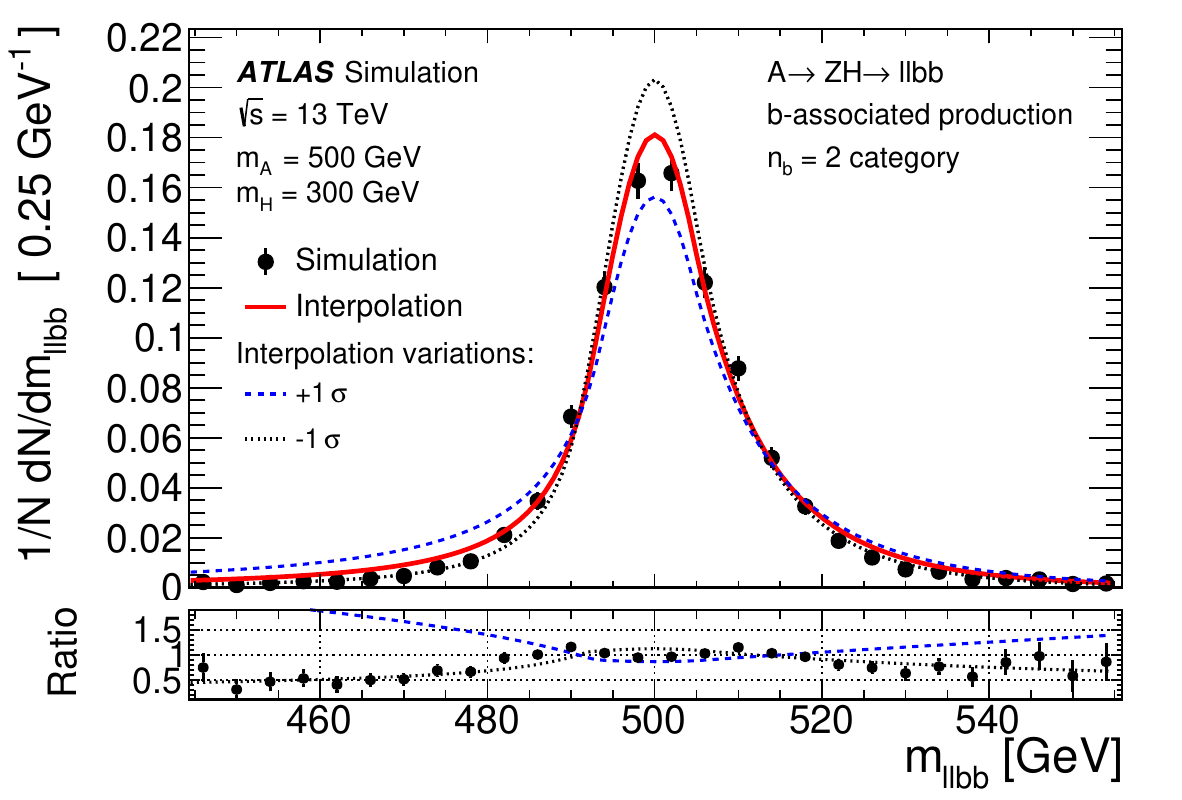}}
\subfigure[]{\includegraphics[width=0.49\textwidth]{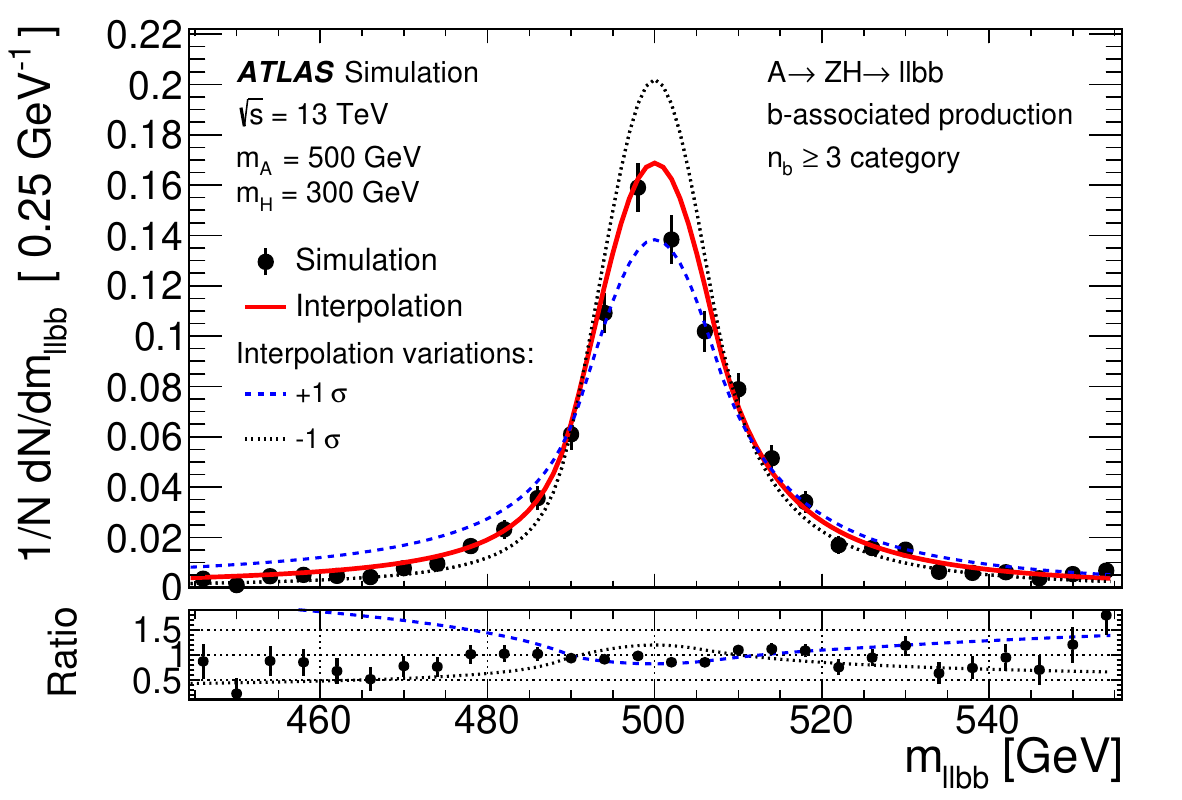}}
\subfigure[]{\includegraphics[width=0.49\textwidth]{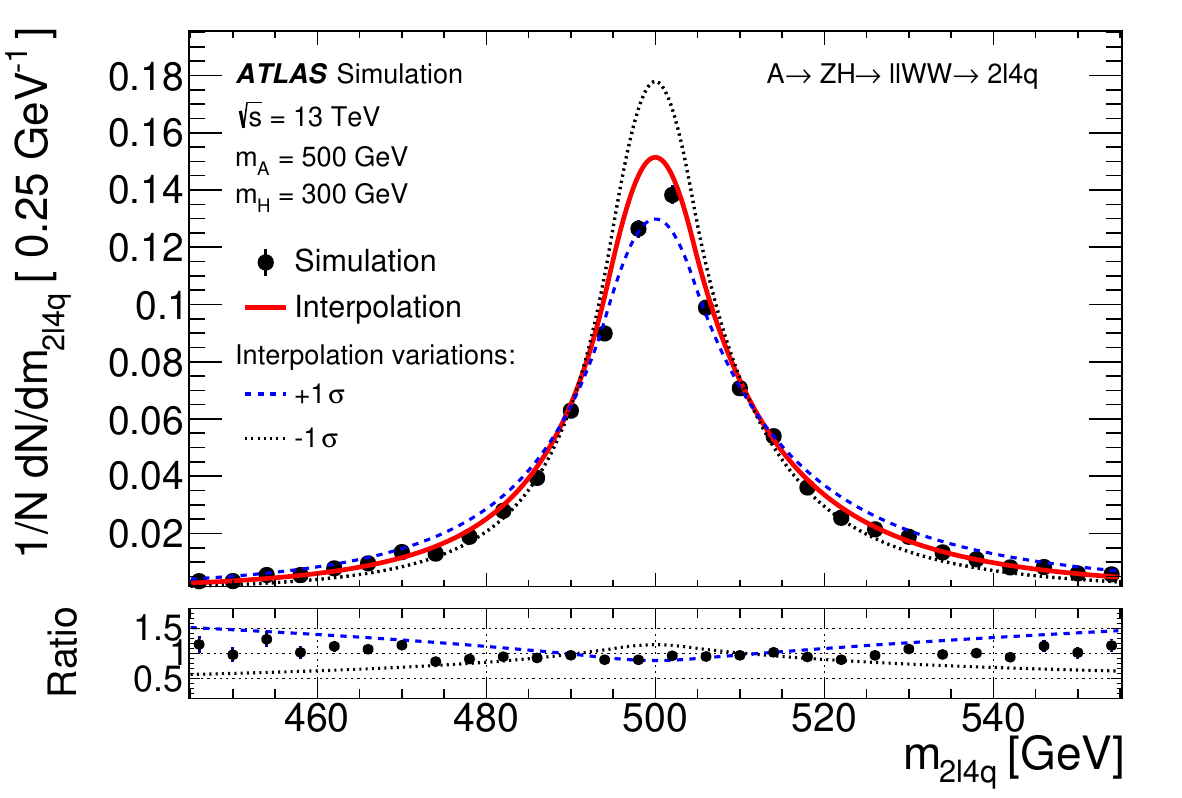}}
\caption{
Signal $\mllbb$ or $\mllqqqq$ distributions
assuming  $m_A=500$~\GeV\ and $m_H=300$~\GeV\ for the following cases:
$\ell \ell bb$ channel:
(a) gluon--gluon fusion in the $n_b = 2$ category,
(b) $b$-associated production in the $n_b = 2$ category, and
(c) $b$-associated production in the $n_b \geq 3$ category;
(d) $\ell \ell WW$ channel.
In the upper panels, the black filled circles correspond to the simulated
distributions, which are compared against the
interpolated parameterised signal distributions shown as solid red curves.
Also in the same panels, the shape variations of the interpolated
parameterised signal distributions are shown in dotted blue ($+1\sigma$)
and black ($-1\sigma$) lines.
In the lower panels, the black filled circles correspond to the ratio of
the simulation to the interpolated parameterised curve. The dotted blue
(black) line corresponds to the ratio of the $+1\sigma$ ($-1\sigma$) shape variation of the
interpolated curve to the interpolated curve. }
\label{fig:signal_model}
\end{figure}

A typical example of the result of the signal parameterisation is shown in
Figure~\ref{fig:signal_model} for  the $(m_A, m_H)=(500, 300)$~\GeV\  mass point.
The figure shows a comparison of the simulated mass
distribution and the interpolated parametric function, as well as the shape
variation that is taken as an estimate of the systematic uncertainty from
the procedure. In general, the cores of the
$\mllbb$ and $\mllww$ distributions
are well-parameterised by the chosen functional forms.
There are some small differences between the function description
and the simulated distribution in the tails of the distributions,
but those have negligible effects on the final results and they are covered by
interpolation uncertainties.

The parameterisation procedure described in the previous paragraph
is modified to allow for cases where the width of the $A$ boson
is comparable to, or larger than, the experimental mass resolution.
This can be modelled by convolving a modified Breit--Wigner
distribution\footnote{The
modification is the multiplication of the Breit--Wigner distribution with a
log-normal distribution to account for the distortion due to the event
selection.} with the EGE or DSCB function.
This procedure is valid as long as the width of the $H$ boson remains
narrow relative to the experimental resolution, which is the case
for the 2HDM scenarios considered in Section~\ref{sec:fit}.
Widths of up to approximately 20\% of the $A$ boson mass are considered,
which is the range relevant to the sensitive parameter
space of the 2HDM scenarios that are of interest for this search.

Finally, the signal efficiencies for the interpolated mass points are
obtained through separate
two-dimensional interpolations on the $(m_A, m_H)$ plane using
thin-plate splines~\cite{Duchon1976}.
 
% End of text imported from the .//signal.tex input file
 
\FloatBarrier
\section{Systematic uncertainties}
\label{sec:systematics}
% The next lines are included from the .//systematics.tex input file
 
Several sources of systematic uncertainty in the signal and background
estimates are considered, including experimental and theoretical
sources.
Experimental uncertainties comprise those in the luminosity measurement~\cite{ATLAS-CONF-2019-021}
(obtained using the LUCID-2 detector \cite{LUCID2}),
trigger, object identification, energy/momentum scale and resolution as well as underlying-event
and pile-up modelling~\cite{Aad:2020flx,PERF-2015-10,ATLAS-CONF-2020-030,STDM-2015-05}.
These uncertainties impact the simulations of signal and background processes.

The signal and background modelling have associated theoretical uncertainties.
For the signal modelling, the uncertainties due to the factorisation and renormalisation
scale choice, the initial- and final-state radiation treatment and the PDF choice are
considered. No additional signal modelling uncertainties related to model-specific
cross-section predictions, such as the 2HDM predictions used in Section~\ref{sec:fit}, are considered.
The renormalisation and factorisation scales are varied up and down separately by
a factor of two, and the largest deviation from the nominal signal is taken
as the estimated uncertainty. The uncertainties due to initial- and final-state radiation
as well as the multiple parton interaction modelling are estimated using a subset of A14
tuning variations~\cite{ATL-PHYS-PUB-2014-021}. PDF uncertainties are computed using
the prescription from  PDF4LHC15~\cite{Butterworth:2015oua}, which include the envelope
of three PDF sets, namely
\texttt{CT14}\xspace, \texttt{MMHT2014}\xspace and \texttt{NNPDF3.0}.
 
Additional systematic uncertainties are assigned to cover the differences
in signal efficiencies and $\mllbb$ and $m_{\ell \ell WW}$
resolution differences
between
the interpolations and the simulations,
as shown by the dotted blue and black lines in the lower panels of Figure~\ref{fig:signal_model}.

For the background modelling, the most important sources of systematic uncertainty
are the modelling of shapes of several kinematic distributions of $Z$+jets events.
In the $\ell \ell bb$ channel, they arise from the shape corrections for the \ptz ,
$\sqrt{\Sigma \pT^2}/m_{\ell\ell bb}$ and $m_{bb}$ variables described in Section~\ref{sec:llbb}.
An uncertainty is estimated  by comparing the corrections and the
agreement between the background prediction and the data for various variables and among  various control regions.
For each of the corrections, the applied uncertainty is half the size of the correction in the $n_b = 2$ category,
and the full size of the correction in the $n_b \geq 3$ category.
In the $\ell \ell WW$ channel, the uncertainties are due
to the shapes of the \ptz\ and leading-jet $\pt$ distributions (Section~\ref{sec:llWW}).
The uncertainty is estimated similarly to that in the $\ell\ell bb$ channel and is half the size of the correction.
For other background processes, modelling uncertainties are obtained by varying the factorisation and
renormalisation scales, and the amount of initial- and final-state radiation.

The effect of these systematic uncertainties on the search is studied using a
signal-strength parameter $\mu$ for hypothesised signal production (see also Section~\ref{sec:fit}).
The uncertainties found to have the
largest impact depend on the choice of $(m_A, m_H)$ signal point.
Table~\ref{tab:systematics} shows the relative uncertainties in the $\mu$ value from the
leading sources of systematic uncertainty for two example mass points of gluon--gluon fusion
and $b$-associated production for the $\ell \ell bb $ channel. The uncertainties
are evaluated using an Asimov dataset \cite{Cowan:2010js} generated with the
signal cross section
set to the expected limits for the particular $(m_A, m_H)$ signal point, considering
a narrow-width $A$ boson.
Table~\ref{tab:systematicsllWW} shows the
same information for the $\ell \ell WW$ channel.
The leading sources of systematic uncertainty are similar for other mass points studied and for larger $A$ boson widths.
 
For the $\ell \ell bb$ channel, the most relevant sources of systematic uncertainty are the background modelling,
the signal interpolation, and the jet energy scale and resolution.
The limited size of the simulated samples has
a higher impact at low masses, since at higher masses other sources are more dominant.
Other systematic uncertainties with non-negligible impact include
those associated with $b$-tagging and theoretical errors.
In the $\ell \ell WW$ channel, the most relevant systematic uncertainties are those related to the jet energy scale and resolution,
as expected in a channel with four jets in the final state.
The limited size of the simulated samples, the background modelling and the signal interpolation also have a non-negligible impact on the signal-strength parameter.
In both channels, the data statistical uncertainties have lower impact at low masses compared to the systematic uncertainties.
In addition, the search sensitivity is affected at high masses by the limited size of the data sample,
an effect which is more pronounced in the $\ell \ell bb$ channel.
 
\begin{table}[h!]
\caption{The effect of the most important sources of uncertainty on the signal-strength parameter at two example mass points of $(m_A, m_H)=(230, 130)$~\GeV{} and $(m_A, m_H)=(700, 200)$~\GeV{} in the $\ell \ell bb$ channel, for both gluon--gluon fusion and $b$-associated production of a narrow-width $A$ boson. The signal cross sections are taken to be the expected median upper limits (see Section~\ref{sec:results}) and they correspond to values that are shown next to the indicated mass points.
JES and JER stand for jet energy scale and jet energy resolution, `Sim. stat.' for simulation statistics,
`Sig.\ interp.' for signal interpolation,
and `Bkg.\ model.' for the background modelling. `Theory'
refers to theoretical uncertainties in the signal samples due to the PDF choice, factorisation and
renormalisation scales, and initial- and final-state radiation.}
\vspace{0.2cm}
\centering
\small
\setlength\tabcolsep{5pt}
\begin{tabular}{lSlSclSlS}
\hline
\multicolumn{9}{c}{$A \to ZH \to \ell \ell bb$} \\
\hline
\multicolumn{4}{c}{Gluon--gluon fusion production}          && \multicolumn{4}{c}{$b$-associated production} \\
\multicolumn{2}{c}{$(230, 130)$~\GeV, 0.31~pb} &\multicolumn{2}{c}{$(700, 200)$~\GeV, 0.017~pb}  && \multicolumn{2}{c}{$(230, 130)$~\GeV, 0.16~pb} & \multicolumn{2}{c}{$(700, 200)$~\GeV, 0.018~pb} \\ \cline{1-4} \cline{6-9}
{Source}   & {$\Delta\mu/\mu$ [\%]} & {Source}  & {$\Delta\mu/\mu$ [\%]} && {Source}  & {$\Delta\mu/\mu$ [\%]}  & {Source}  & {$\Delta\mu/\mu$ [\%]} \\ \hline
Data stat.   &   28  & Data stat.   &  45 && Data stat.   &  33  & Data stat.   & 46  \\
Total syst.  &   36  & Total syst.  &  26 && Total syst.  &  33  & Total syst.  & 25  \\
\hline
Sim.\ stat.   &   19  & Sim.\ stat.   & 7.2 && Sim.\ stat.   &  18  & Sim.\ stat.   & 7.2 \\
Sig.\ interp. &  9.9  & Sig.\ interp. & 8.7 && Sig.\ interp. &  13  & Sig.\ interp  & 13 \\
Bkg.\ model.  &   19  & Bkg.\ model.  &  18 && Bkg.\ model.  &  15  & Bkg.\ model.  & 16 \\
JES/JER      &   20  & JES/JER      &  18 && JES/JER      &  14  & JES/JER      & 16 \\
$b$-tagging  &  7.5  & $b$-tagging  &  12 && $b$-tagging  &  9.5 & $b$-tagging  & 12 \\
Theory       &  7.4  & Theory       & 9.5 && Theory       &  5.0 & Theory       & 7.1 \\
\hline
\end{tabular}
\label{tab:systematics}
\end{table}
 
\begin{table}[h!]
\caption{The effect of the most important sources of uncertainty on the signal-strength parameter at two example mass points of $(m_A, m_H)=(500, 300)$~\GeV{} and $(m_A, m_H)=(700, 200)$~\GeV{} in the $\ell \ell WW$ channel for gluon--gluon fusion production of a narrow-width $A$ boson.
The same notation as in Table~\ref{tab:systematics} is used.
}
\vspace{0.2cm}
\centering
\small
\setlength\tabcolsep{5pt}
\begin{tabular}{lSlSclSlS}
\hline
\multicolumn{9}{c}{$A \to ZH \to \ell \ell WW$} \\
\hline
\multicolumn{4}{c}{Gluon--gluon fusion production}    \\
\multicolumn{2}{c}{$(500, 300)$~\GeV, 0.70~pb} &\multicolumn{2}{c}{$(700, 200)$~\GeV, 0.38~pb} \\ \cline{1-4} \cline{6-9}
{Source}   & {$\Delta\mu/\mu$ [\%]} & {Source}  & {$\Delta\mu/\mu$ [\%]} \\ \hline
Data stat.       & 32    & Data stat.     & 33 \\
Total syst.      & 42    & Total stat.    & 38 \\ \hline
Sim.\ stat.       & 24    & Sim.\ stat.     & 19 \\
Sig.\ interp.     & 14    & Sig.\ interp.   & 12 \\
Bkg.\ model.      & 14    & Bkg.\ model.    & 16 \\
JES/JER          & 30    & JES/JER        & 23 \\
Theory           & 6.5   & Theory         & 7.6 \\
\hline
\end{tabular}
\label{tab:systematicsllWW}
\end{table}

% End of text imported from the .//systematics.tex input file
 
\FloatBarrier
\section{Results}
\label{sec:results}
% The next lines are included from the .//results.tex input file
\label{sec:fit}

\def \minGGAlim {6.2}
\def \maxGGAlim {380}
\def \minGGAlimExp{15}
\def \maxGGAlimExp{240}
\def \minGGAlimPoint{(780, 129)}
\def \maxGGAlimPoint{(250, 150)}

\def \minGGAlimllWW {0.023}
\def \maxGGAlimllWW {8.9}
\def \minGGAlimExpllWW{0.041}
\def \maxGGAlimExpllWW{3.6}
\def \minGGAlimPointllWW{(770, 660)}
\def \maxGGAlimPointllWW{(340, 220)}
 
\def \minBBAlim {6.8}
\def \maxBBAlim {210}
\def \minBBAlimExp{15}
\def \maxBBAlimExp{160}
\def \minBBAlimPoint{(760, 220)}
\def \maxBBAlimPoint{(230, 130)}

The $\mllbb$ and $\mllqqqq$ distributions are expected to exhibit a resonant
structure if signal events are present, while background events result in a smoothly falling spectrum.
Therefore, those are chosen as the final variables to discriminate between signal and background.
The shape differences between the signal and background contributions in the $\mllbb$ and $\mllqqqq$
distributions are exploited through binned maximum-likelihood fits of the
signal-plus-background hypotheses to extract potential signal contributions.
The fits are based on the statistical framework described in Refs.~\cite{Moneta:2010pm,RooFit,Cowan:2010js}.
For a given mass hypothesis of $(m_A, m_H)$, the likelihood is constructed as the product
of Poisson probabilities for event yields in the $\mllbb$ or $\mllqqqq$ bins:
\[L(\left.\mu,\vec{\alpha},\vec{\theta} | m_A,m_H) = \prod_{i={\textrm{bins}}} {\textrm{Poisson}}\left(N_i \bigg |\left(\mu\times S_i(m_A,m_H,\vec{\theta})+B_i(\vec{\alpha},\vec{\theta})\right)\right.\right)\cdot G(\vec{\theta})~, \]
where $N_i$ is the number of observed events, and $S_i(m_A, m_H, \vec{\theta})$
and $B_i(\vec{\alpha},\vec{\theta})$ are the expected number of signal events and
estimated number of background events in bin $i$. The vector $\vec{\alpha}$ represents
free background normalisation scale factors (described later) and the vector $\vec{\theta}$
denotes all non-explicitly listed parameters of the likelihood function such as nuisance
parameters associated with systematic uncertainties.
Systematic uncertainties are incorporated in the likelihood as nuisance parameters with either
Gaussian or log-normal constraint terms, denoted by $G(\vec{\theta})$ in the formula above.
The parameter of interest, $\mu$, is a multiplicative factor applied to the
expected signal rate.
The $\mllbb$ and $\mllqqqq$ bin widths are
chosen according to the expected detector resolution and taking into account the statistical
uncertainty in the number of simulated background events.
The bin centres are adjusted such that at least 65\% of the test signal is contained
in one bin.

For each bin, $S_i$ is calculated from the total integrated luminosity, the
assumed cross section times branching ratio
for the signal and its selection efficiency. The sum of all background contributions in the bin, $B_i$,
is estimated from simulation, which includes the modelling corrections discussed in
Sections~\ref{sec:llbb} and \ref{sec:llWW}.
The number of events in the $t\bar{t}$ and $Z$+jets control regions is included in the
likelihood calculation to constrain their normalisation in the signal regions.
This is achieved by introducing two free normalisation scale factors per channel,
represented by $\vec{\alpha}$ in the likelihood description earlier in this section.
In the $\ell \ell bb$ channel these scale factors apply to the  $t\bar{t}$
contribution and the heavy-flavour component of the $Z$+jets contribution, whereas the
rest of the contributions in the control region are estimated from simulation.
In the $\ell \ell WW$ channel the scale factors apply to
the $t\bar{t}$ contribution and the flavour-inclusive $Z$+jets contribution.
Typical values of the scale factors are close to unity with the exception of
$Z$+jets in the $\ell\ell bb$ channel, which is scaled by a factor of 1.2,
and $t\bar{t}$ in the $\ell\ell bb$ $n_{b} \ge 3$ category, which
is typically scaled by a factor of 1.4.
 
The signals that are fitted in each category are motivated
by signal efficiency considerations and the interpretation of the
search in the context of the 2HDM.
In the $\ell\ell bb$ channel the following fits are performed.
First, $A$ bosons produced by gluon--gluon fusion are
considered in the $n_b=2$ category.
Second, a combined fit for the $b$-associated production mechanism
in both the $n_b=2$ and $n_b\ge 3$ categories is performed.
Finally, there is a combination of the $b$-associated production
fit with the gluon--gluon fusion fit,
which is interpreted  in the context of the 2HDM.
In the $\ell \ell WW$ channel, only $A$ bosons produced by
gluon--gluon fusion are considered and, hence, it is the only fit that
is considered.

\subsection{$A \to ZH \to \ell\ell bb$ results}
\label{sec:res_llbb}

The $\mllbb$ distributions from different $m_{bb}$ mass windows are scanned for potential
excesses beyond the background expectations through signal-plus-background fits.
The scan is performed in steps of
10~\GeV{} for both the $m_A$ range 230--800~\GeV{} and the $m_H$ range 130--700~\GeV, such that $m_A-m_H \geq 100~\GeV$.
The step sizes are chosen to be
compatible with the detector resolution for $\mllbb$ and $m_{bb}$. In total, there
are 58 $m_{bb}$ windows that are probed for the $n_b=2$ and $n_b\ge 3$ categories.
The overall
number of ($m_A$, $m_H$) signal hypotheses that are tested is 1711 per category.
 
\begin{figure}
\centering
\subfigure[]{\includegraphics[width=0.49\textwidth]{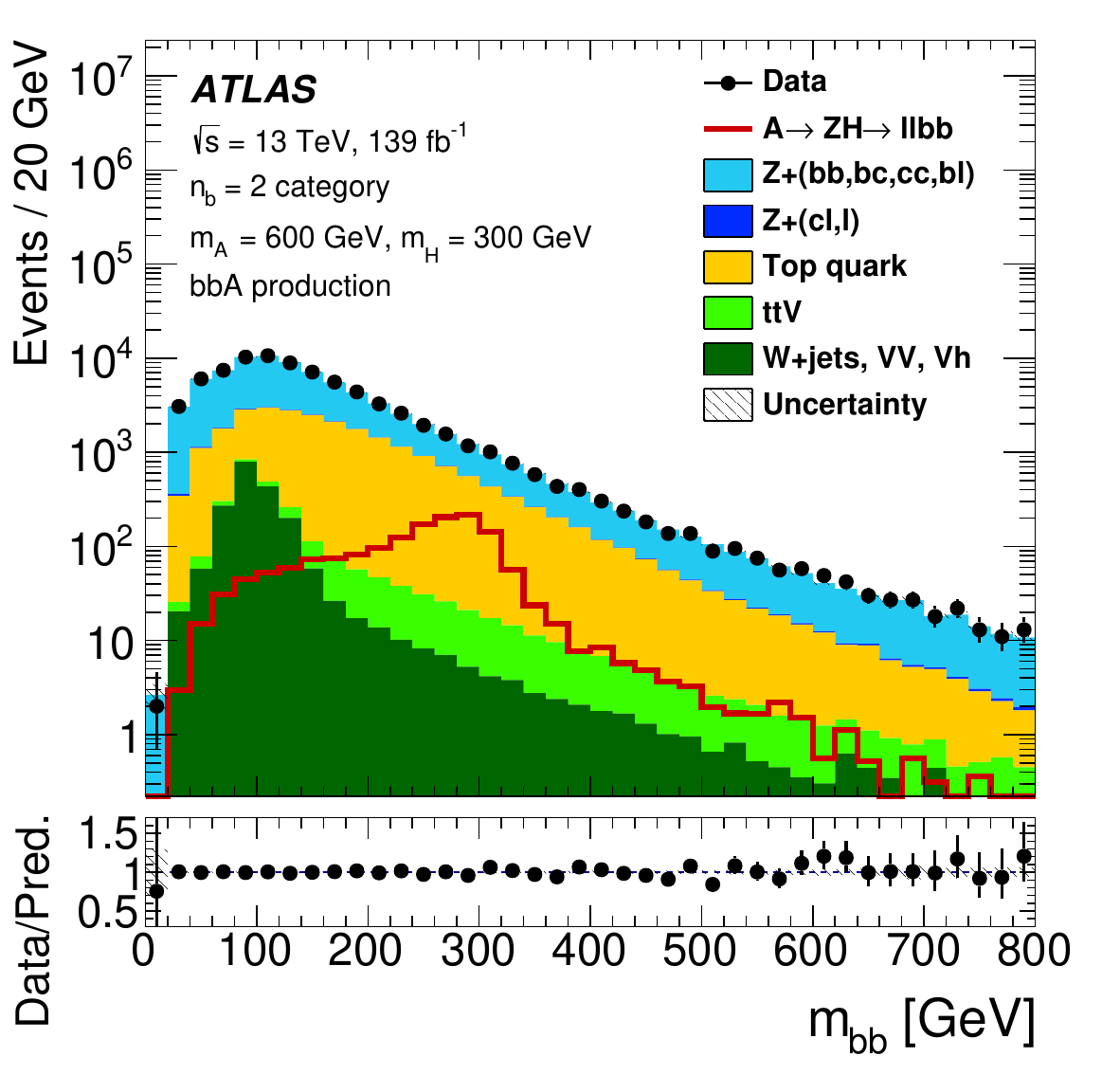}\label{fig:mbb-2b}}
\subfigure[]{\includegraphics[width=0.49\textwidth]{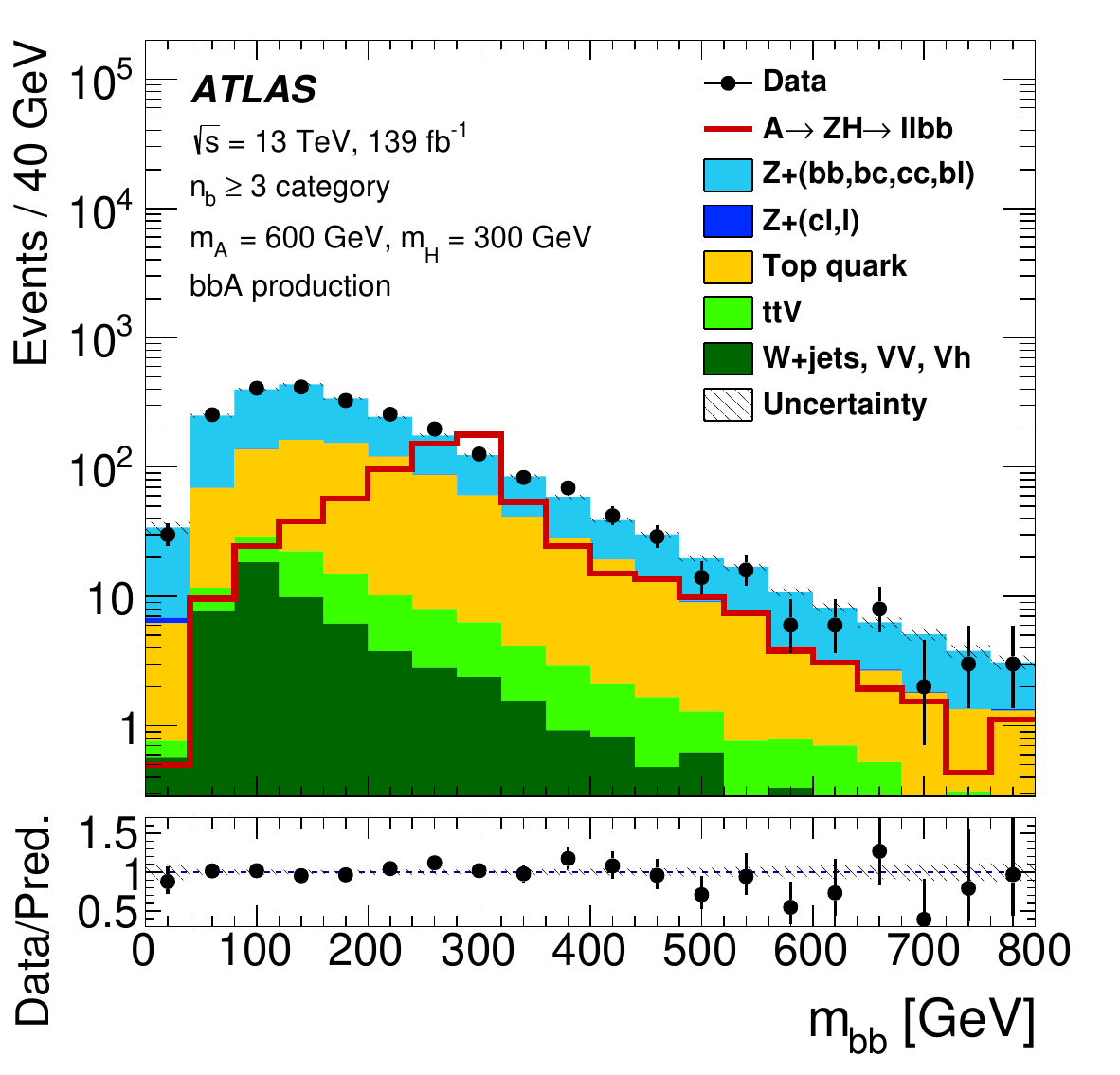}\label{fig:mbb-3b}} \\
\caption{The $m_{bb}$ distribution before any $m_{bb}$ window cuts for the
(a) $n_b=2$ and (b) $n_b \ge 3$ categories. The signal distribution for ($m_A$, $m_H$) = (600, 300)~\GeV\ is also shown, and is
normalised such that the production cross section times the
branching ratios $B(A\to ZH)$ and $B(H\to bb)$ corresponds to 1~pb.
The same conventions as in Figure~\ref{fig:llbb-jvar} are used.
}
\label{fig:mbb}
\end{figure}

\begin{figure}
\centering
\subfigure[]{\includegraphics[width=0.49\textwidth]{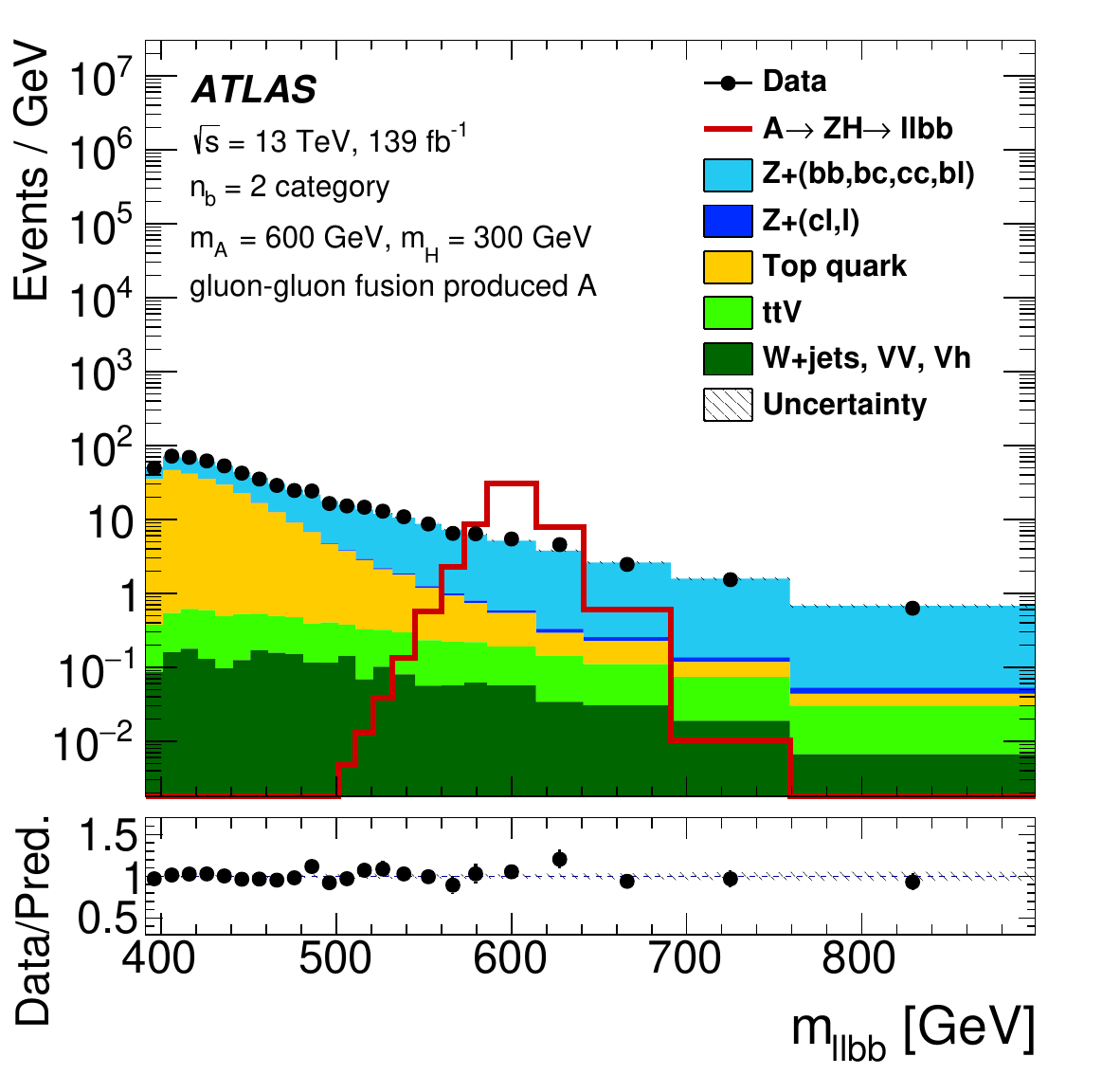}\label{fig:mllbb-1}}
\subfigure[]{\includegraphics[width=0.49\textwidth]{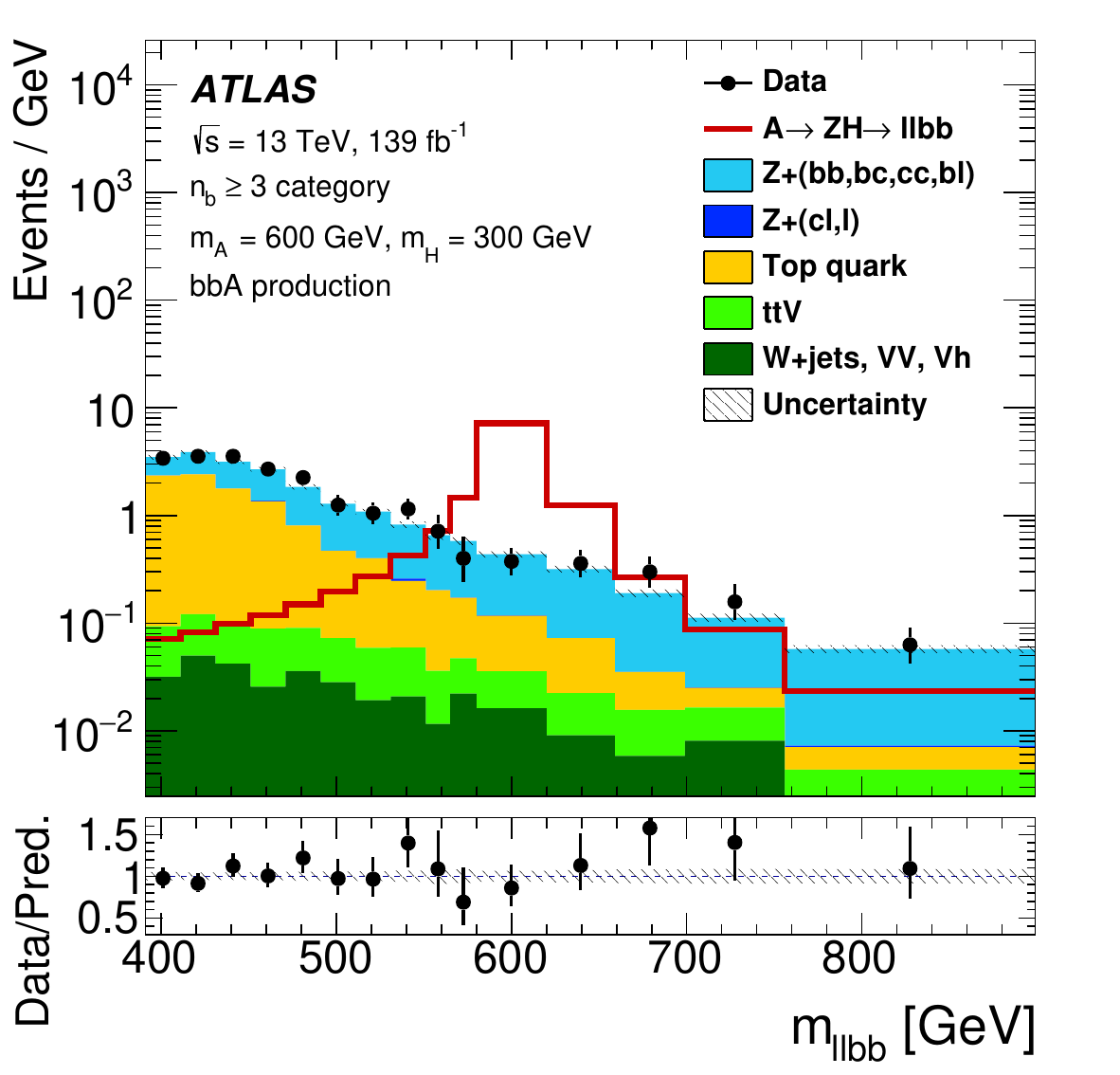}\label{fig:mllbb-2}} \\
\subfigure[]{\includegraphics[width=0.49\textwidth]{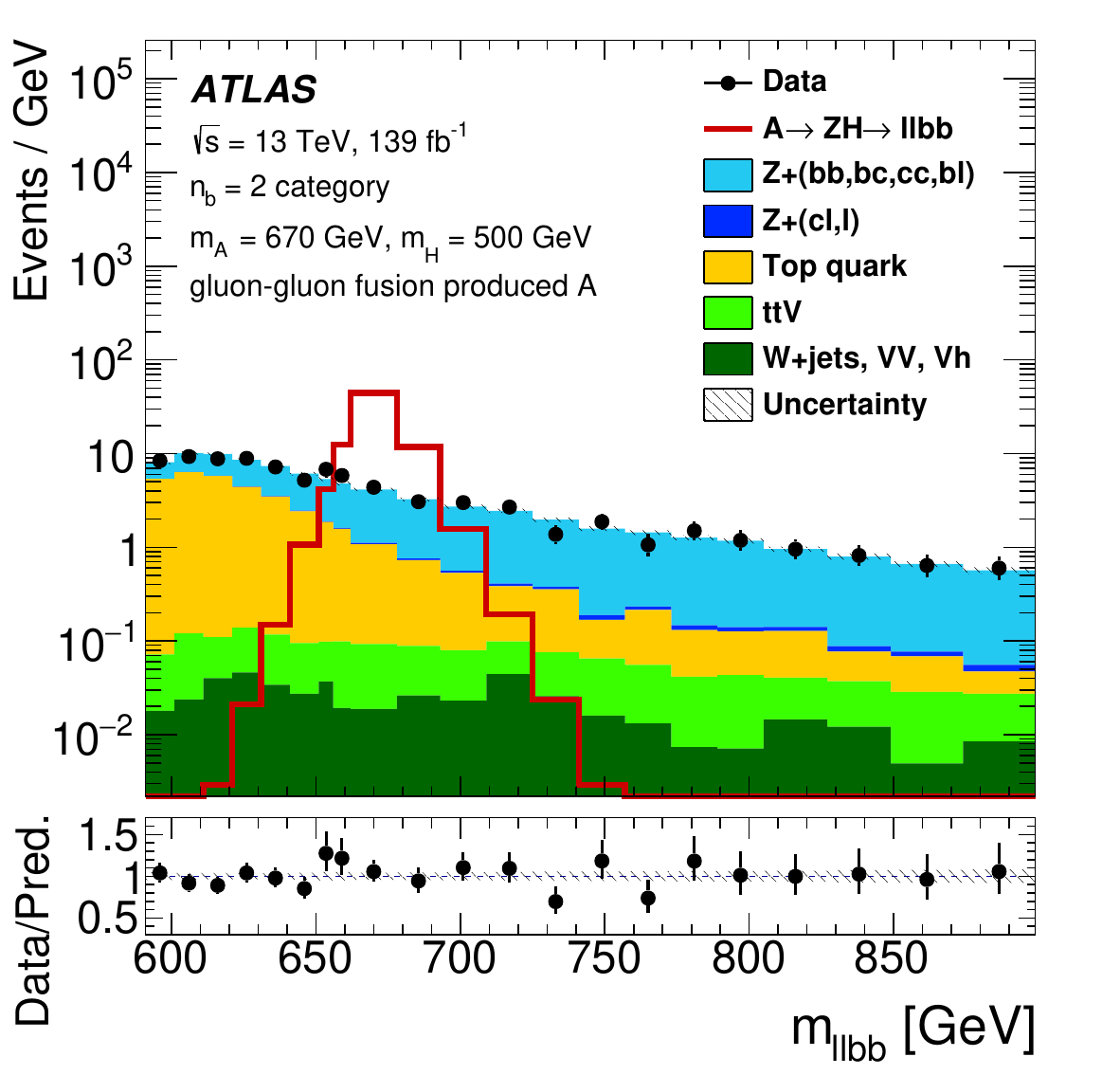}\label{fig:mllbb-3}}
\subfigure[]{\includegraphics[width=0.49\textwidth]{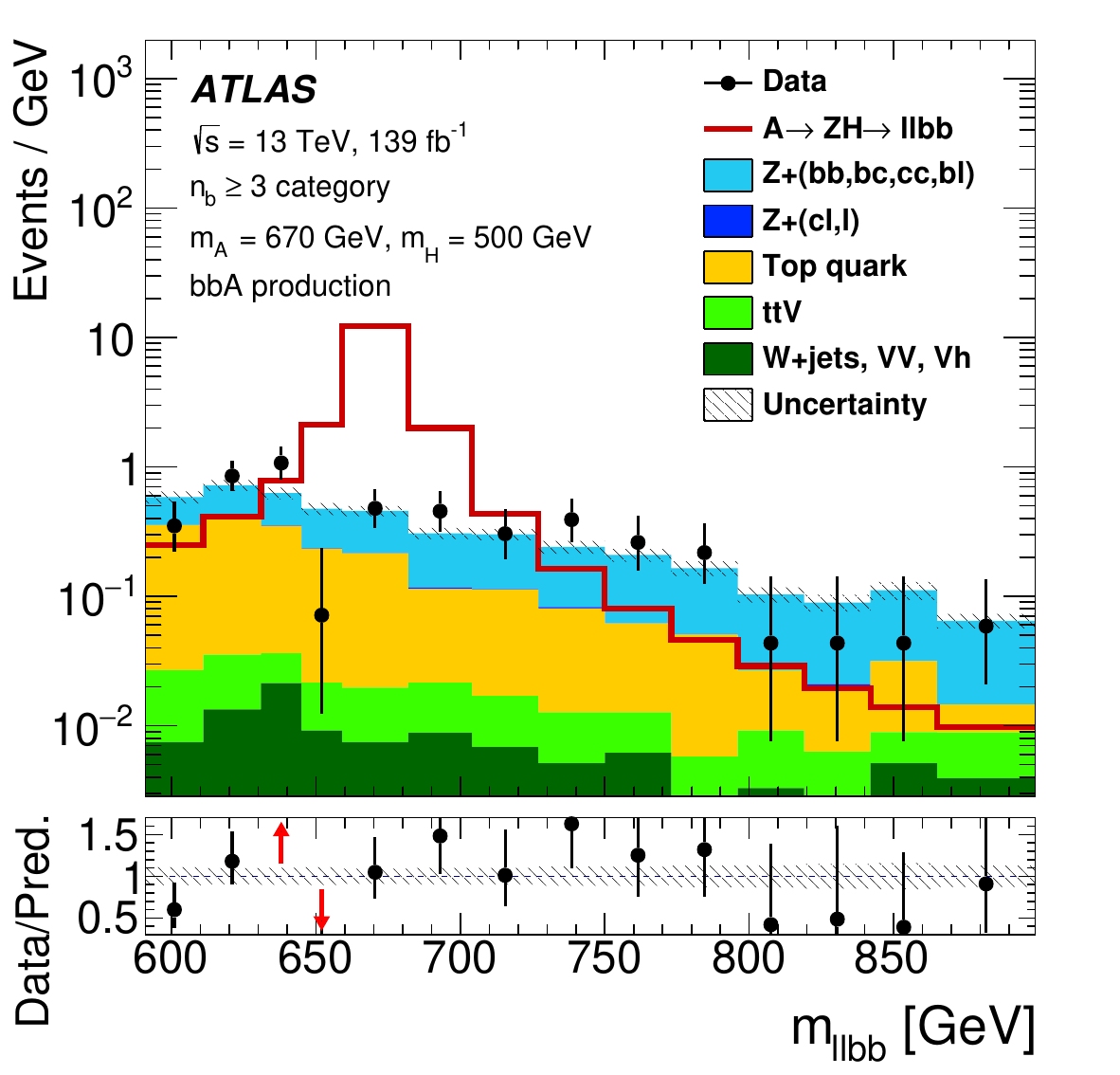}\label{fig:mllbb-4}}
\caption{The $m_{\ell\ell bb}$ mass distribution for the $m_{bb}$ windows
defined for $m_H=300~\GeV$ and $m_H=500~\GeV$ for (a, c)
the $n_b=2$ and (b, d) the $n_b\ge 3$ category, respectively.
Signal distributions with $(m_A, m_H) = (600, 300)$~\GeV\ and $(m_A, m_H) = (670, 500)$~\GeV\
are also shown for
gluon--gluon fusion production in (a, c) and $b$-associated production in (b, d).
The number of entries shown in each bin is the number of events in that bin divided by the width of the bin.
The same conventions as in Figure~\ref{fig:llbb-jvar} are used.
}
\label{fig:mllbb}
\end{figure}

Figure~\ref{fig:mbb} shows the distribution of the $H$ boson candidate mass $m_{bb}$
before the $m_{bb}$ window requirement in
each of the two categories.
Typical examples of $\mllbb$ distributions after the application of the
$m_{bb}$ window requirement are shown in Figures~\ref{fig:mllbb-1}--\ref{fig:mllbb-4}.
In particular, the $m_{bb}$ window defined for  $m_H = 300$~\GeV\ is shown in
Figures~\ref{fig:mllbb-1} and \ref{fig:mllbb-2} for the
$n_b=2$ and $n_b\ge 3$ categories, respectively.
On the same figures, a signal distribution is shown as well, which corresponds to
gluon--gluon fusion production in Figure~\ref{fig:mllbb-1}
and $b$-associated production in  Figure~\ref{fig:mllbb-2}, for
the $(m_A, m_H) = (600, 300)$~\GeV\ signal point.
Similarly, an $m_{bb}$ window defined for $m_H = 500$~\GeV\ is shown
Figures~\ref{fig:mllbb-3} and \ref{fig:mllbb-4} for the
$n_b=2$ and $n_b\ge 3$ categories, respectively.
The signal distribution for the
$(m_A, m_H) = (670, 500)$~\GeV\ signal point is also shown
for gluon--gluon fusion production in Figure~\ref{fig:mllbb-3}
and $b$-associated production in Figure~\ref{fig:mllbb-4}.
 
In all cases, the data are found to be well described by the background model.
The most significant excess for the gluon--gluon fusion production signal
assumption is at the $(m_A, m_H) = (610, 290)$~\GeV{} signal point, for which the
local (global) significance~\cite{Vitells:2011da} is 3.1 (1.3)
standard deviations. For $b$-associated production, the most significant
excess is at the $(m_A, m_H) = (440, 220)$~\GeV{} signal point, for which the
local (global) significance is 3.1 (1.3) standard deviations. The significances are
calculated for each production process separately, ignoring the contribution from the other.

\begin{figure}
\centering
\subfigure[]{\includegraphics[width=0.49\textwidth]{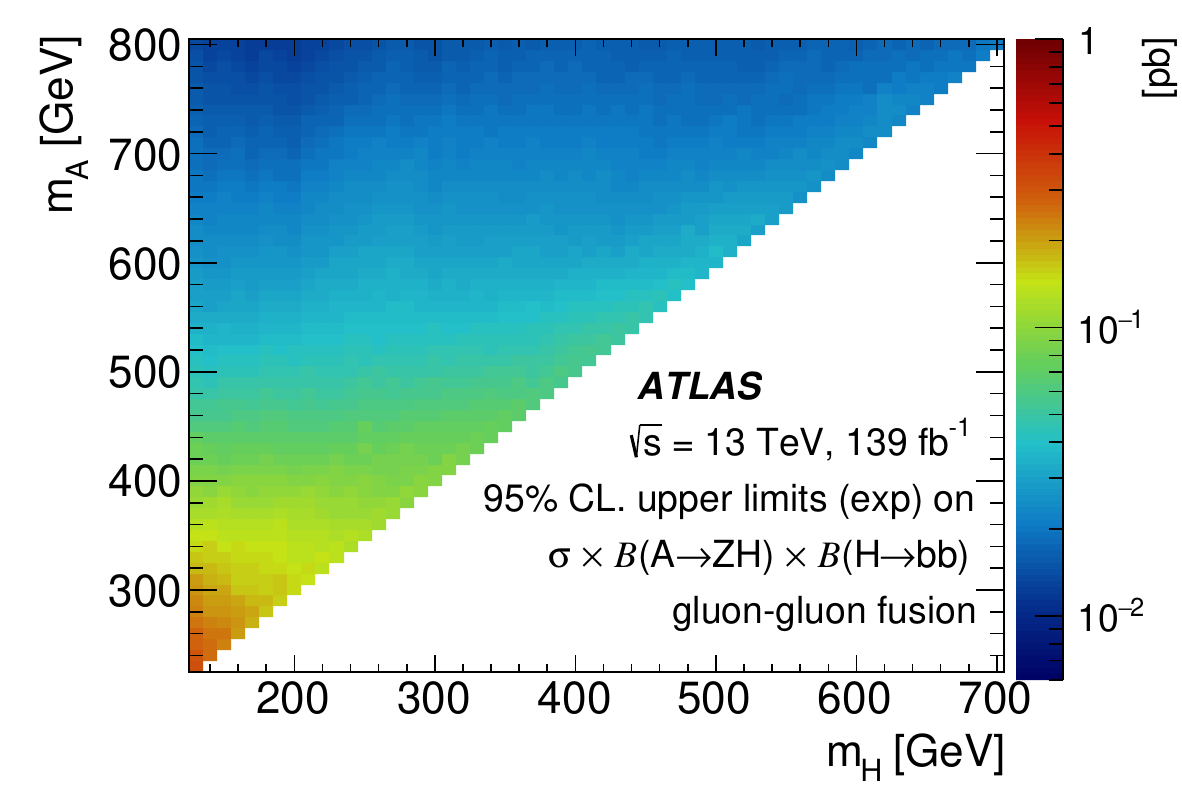}\label{fig:limits:ggA:exp}}
\subfigure[]{\includegraphics[width=0.49\textwidth]{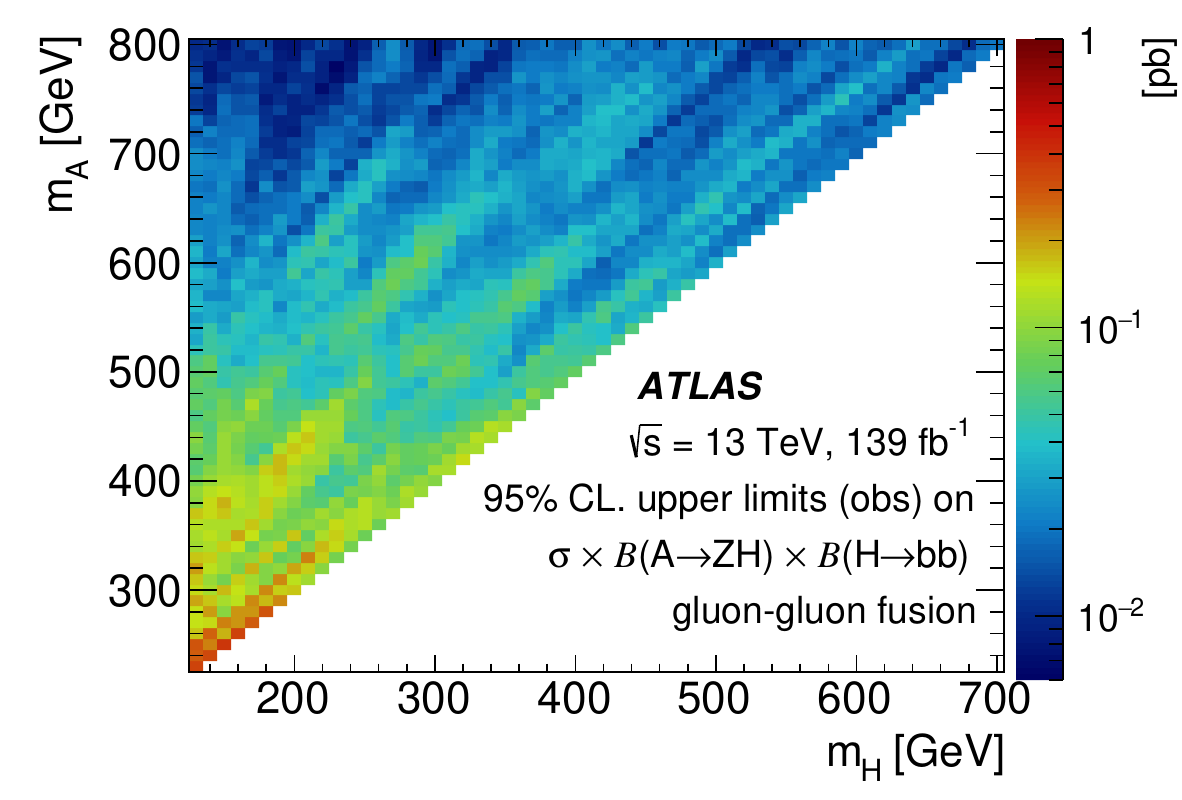}\label{fig:limits:ggA:obs}} \\
\subfigure[]{\includegraphics[width=0.49\textwidth]{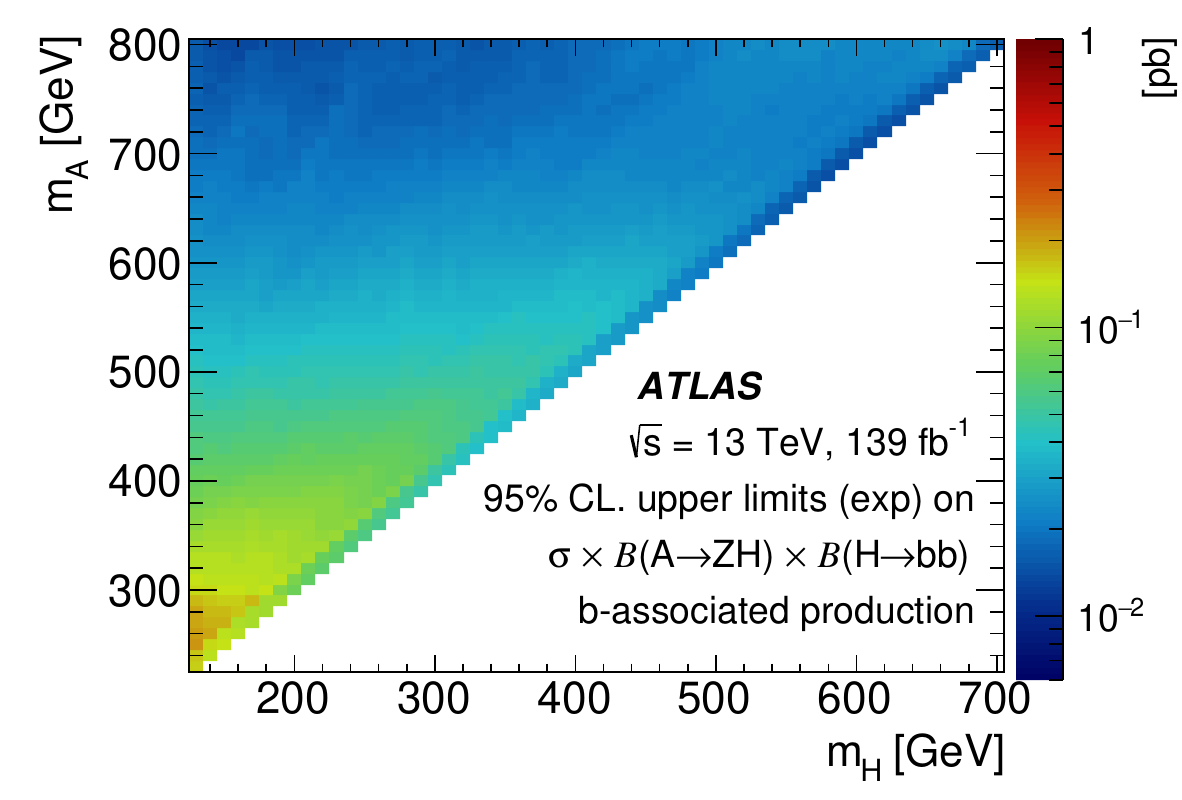}\label{fig:limits:bbA:exp}}
\subfigure[]{\includegraphics[width=0.49\textwidth]{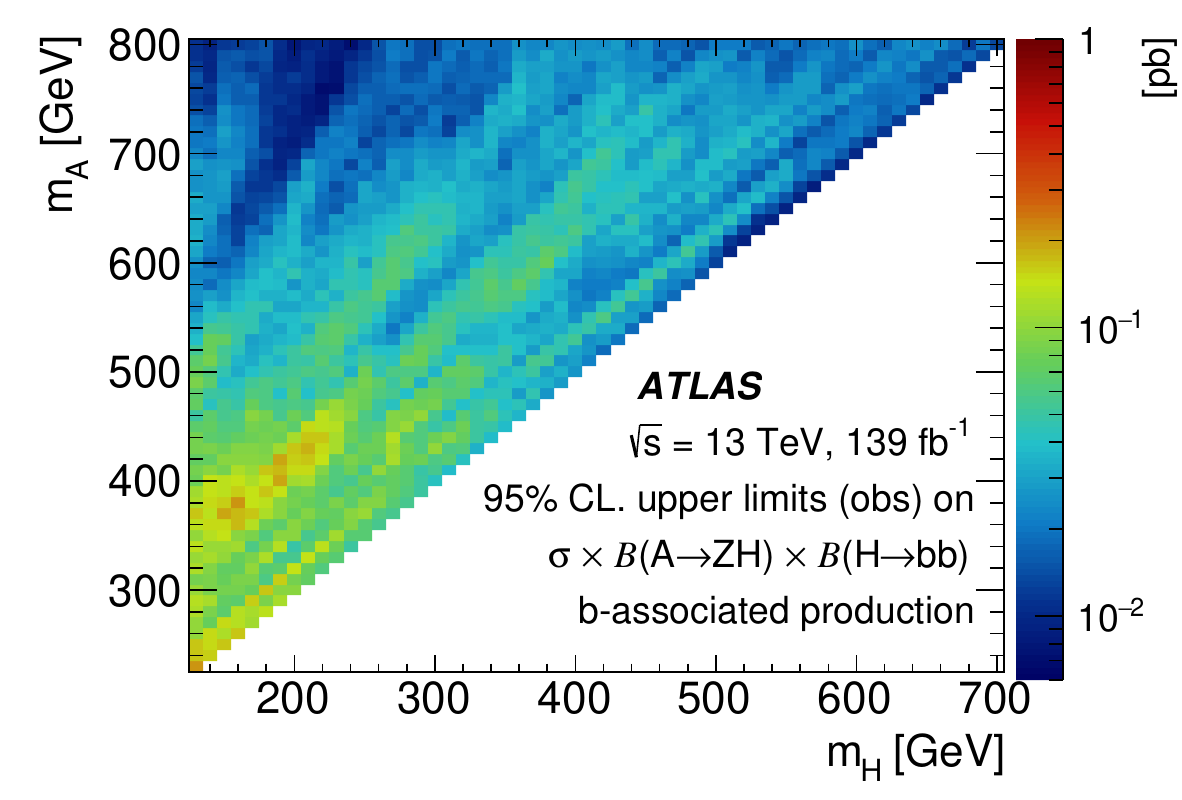}\label{fig:limits:bbA:obs}}
 
\caption{Upper bounds at 95\% CL on the production cross section times the branching ratio $B(A\to ZH)\times B(H\to bb)$ in pb for (a, b) gluon--gluon fusion and (c, d) $b$-associated production. The expected upper limits are shown in (a) and (c) and the observed upper limits are shown in (b) and (d).
}
\label{fig:limits}
\end{figure}

In the absence of any statistically significant excess, the results of the search in this channel are
interpreted as upper limits on the production cross section of an $A$ boson decaying
into $ZH$ followed by the $H\to bb$ decay,
$\sigma\times B(A\to ZH)\times B(H\to bb)$.
The cross-section upper limits consider $A$ bosons
that are produced only by a single mechanism, i.e.\ either gluon--gluon
fusion or $b$-associated production.
Modified frequentist~\cite{Read:2002hq}
95\% confidence level~(CL) upper limits on the production
cross section of this process are obtained using the
asymptotic approximation~\cite{Cowan:2010js} for
the various signal hypotheses that are tested.
In particular, expected and observed upper limits for
gluon--gluon fusion production of narrow-width $A$ bosons
in the $n_b=2$ category are shown in Figures~\ref{fig:limits:ggA:exp}
and \ref{fig:limits:ggA:obs}, respectively.
For $b$-associated production of narrow-width $A$ bosons,
the expected and observed limits for the combination of the
$n_b=2$ and $n_b\ge 3$ categories are shown in
Figures~\ref{fig:limits:bbA:exp}
and \ref{fig:limits:bbA:obs}, respectively.
The upper limits  for gluon--gluon fusion vary from \minGGAlim~fb for
$(m_A, m_H) = \minGGAlimPoint$~\GeV{} to \maxGGAlim~fb for
$(m_A, m_H) = \maxGGAlimPoint$~\GeV. This is to be compared with
the corresponding expected limits of \minGGAlimExp~fb and \maxGGAlimExp~fb
for these two signal hypotheses. For $b$-associated production
the upper limit varies from \minBBAlim~fb for
$(m_A, m_H) = \minBBAlimPoint$~\GeV{} to \maxBBAlim~fb for
$(m_A, m_H) = \maxBBAlimPoint$~\GeV, whereas the corresponding
expected limits are \minBBAlimExp~fb
and \maxBBAlimExp~fb.
 
Upper limits are also calculated for signal assumptions where the
natural width of the $A$ boson is large in comparison with the experimental mass resolution,
which is needed for the interpretation of the search in the context of the 2HDM.
The cross-section upper limit decreases
as the natural width of the $A$ boson increases.
In particular, a gluon--gluon produced
$A$ boson with a natural width of 10\% of its mass has a cross-section
upper limit that is reduced on average by a factor of approximately 3 from the narrow-width case.
This factor becomes approximately 4 when the natural width increases to 20\%.
The $A$ bosons from $b$-associated production have worse experimental
mass resolution and the deterioration of the limit is on average smaller:
the upper limits are reduced by a factor of about 1.9 (2.3) for
a natural width of 10\% (20\%).

The results for $A$ boson natural widths that are comparable to, or larger than,
the experimental mass resolution are used for the interpretation
of the search in the context of the CP-conserving 2HDM. The 2HDM benchmark against which
the search results are compared has three free parameters: $m_A$, $m_H$
and $\tan\beta$. In addition, there are four ways to assign the Yukawa
couplings to fermions, defining type-I, type-II, lepton-specific and flipped 2HDMs.
The remaining parameters are fixed. The mass of the lightest Higgs boson
in the model is fixed to $125~\GeV$ and its couplings are set to be
the same as those of the SM Higgs boson by choosing $\cos(\beta-\alpha)=0$ \cite{PhysRevD.67.075019},
which is known as the 2HDM weak decoupling limit.
The charged Higgs boson is assumed to have the
same mass as the $A$ boson and the potential parameter $m^2_{12}$ \cite{Branco:2011iw} is fixed to
$m^2_{A}\tan\beta/(1+\tan^2\beta)$.
 
The cross sections for $A$ boson production in the 2HDM are
calculated using corrections at up to NNLO in QCD for gluon--gluon fusion and
$b$-associated production in the five-flavour scheme as implemented in
{\texttt SusHi}~\cite{Harlander:2012pb,Harlander:2005rq,Harlander:2003ai,Harlander:2002wh}.
For $b$-associated production a cross section in the four-flavour
scheme is also calculated as described in Refs.~\cite{Dawson:2003kb,Dittmaier:2003ej} and the results are combined with the five-flavour scheme calculation
following Ref.~\cite{Harlander:2011aa}. The Higgs boson widths and branching
ratios are calculated using {\texttt 2HDMC}~\cite{Eriksson:2009ws}. The procedure for the calculation of the cross sections
and branching ratios, as well as for the choice of 2HDM parameters, follows Ref.~\cite{deFlorian:2016spz}.

\begin{figure}
\centering
\subfigure[]{\includegraphics[width=0.49\textwidth]{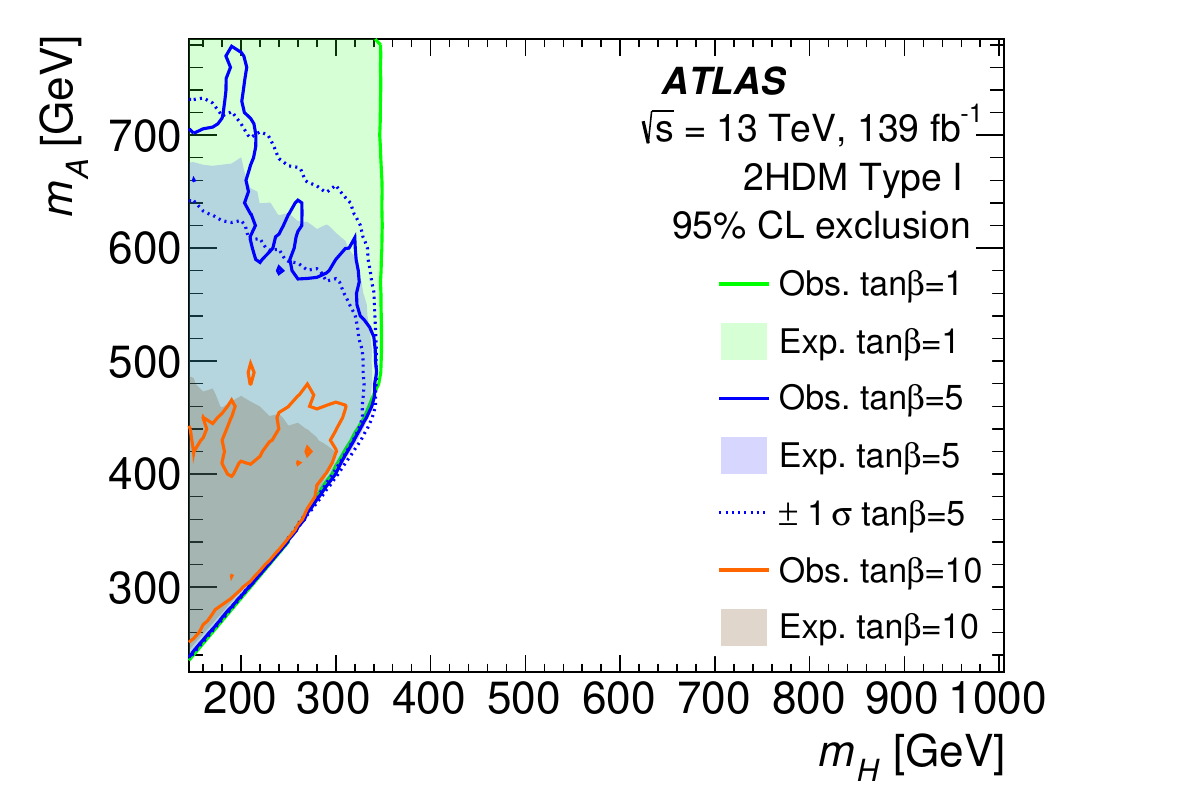}\label{fig:2hdm-type1}}
\subfigure[]{\includegraphics[width=0.49\textwidth]{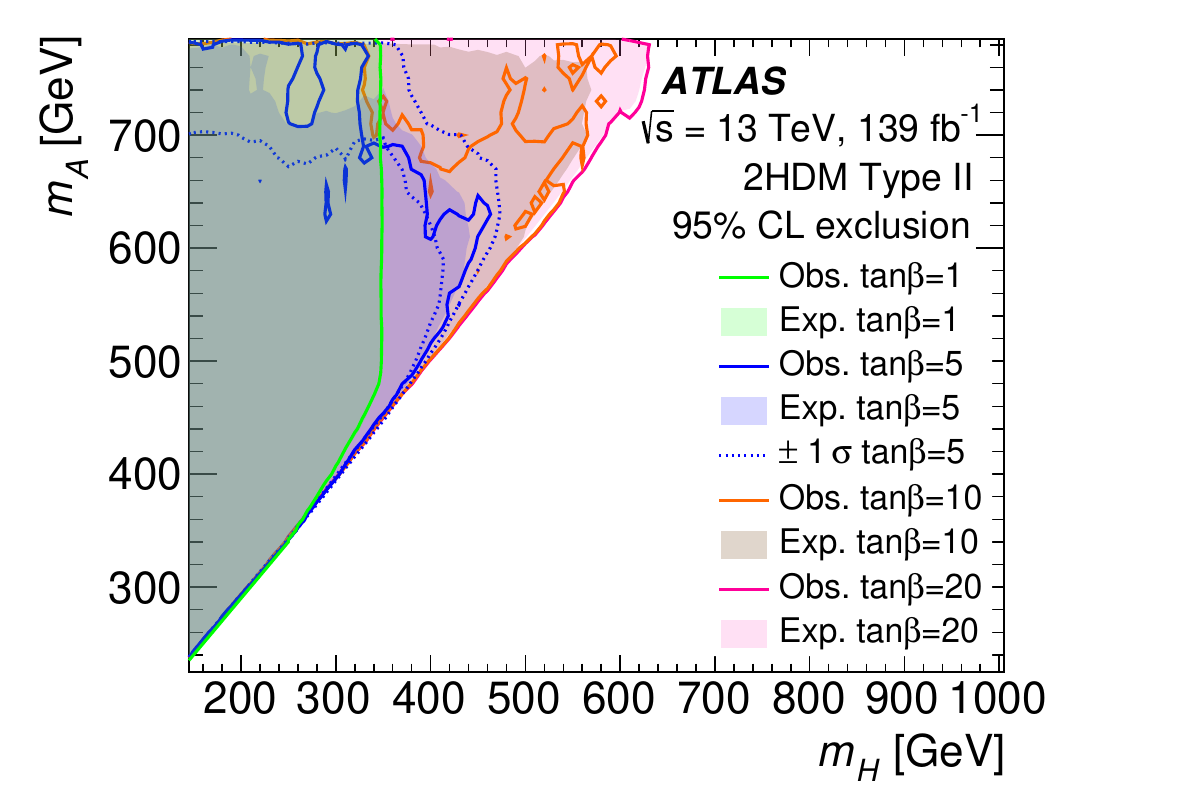}\label{fig:2hdm-type2}} \\
\subfigure[]{\includegraphics[width=0.49\textwidth]{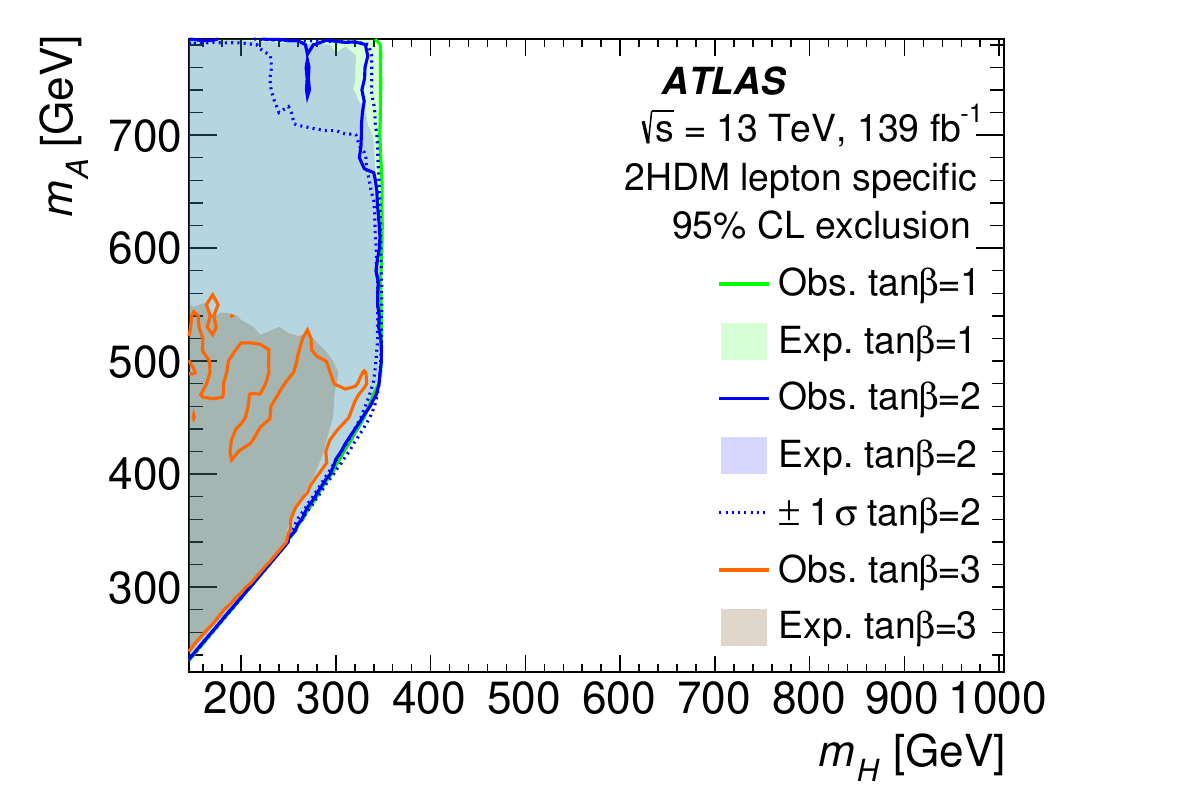}\label{fig:2hdm-type3}}
\subfigure[]{\includegraphics[width=0.49\textwidth]{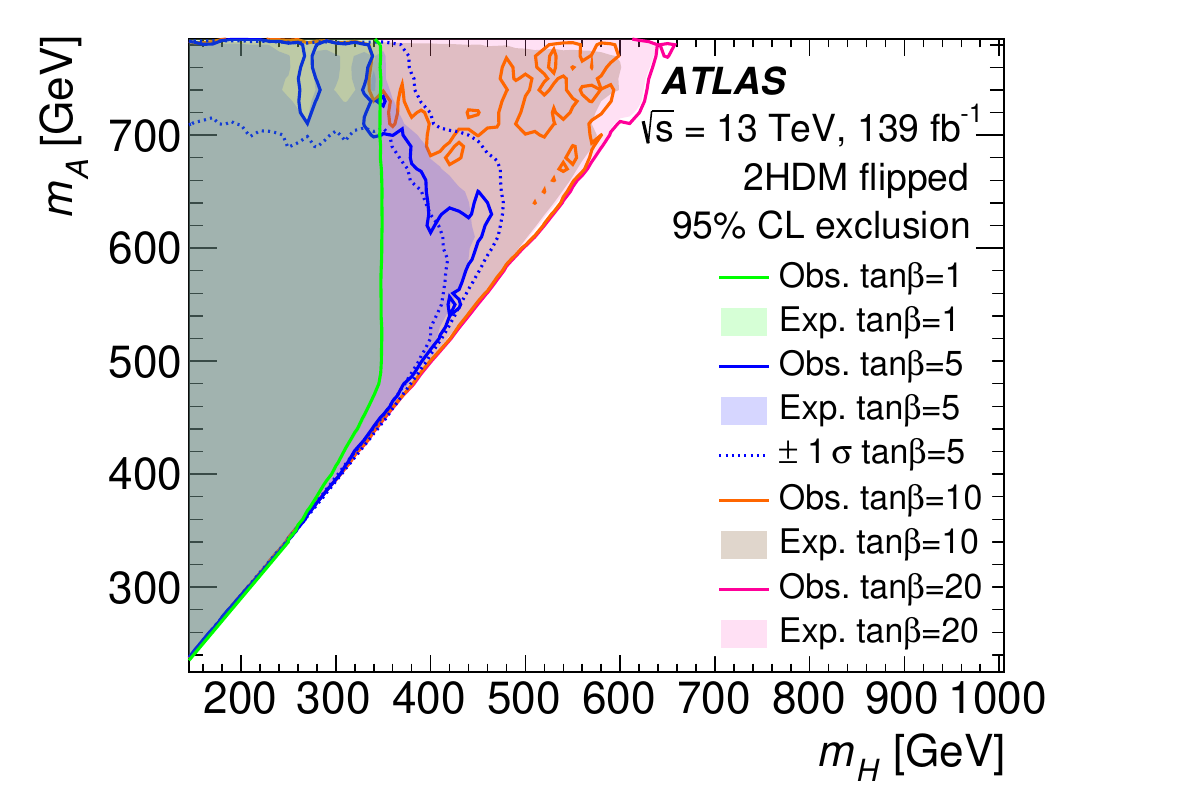}\label{fig:2hdm-type4}}
\caption{Observed and expected 95\% CL exclusion regions for the $\ell\ell bb$ channel in the $(m_H, m_A)$ plane for various $\tan\beta$ values
for the (a) type-I, (b) type-II, (c) lepton-specific and (d) flipped 2HDM, with
$\cos(\beta-\alpha)=0$.}
\label{fig:2hdm}
\end{figure}

The interpretation of the search in the 2HDM is performed in the
$(m_H, m_A)$ plane, as shown in Figure~\ref{fig:2hdm}.
In this plot, colour-shaded areas indicate expected and observed exclusions
for various $\tan\beta$ values. There is one plot for each of the
four 2HDM types. For the type-I and lepton-specific 2HDMs, only gluon--gluon
fusion production is relevant.
The exclusion region reaches $m_H \lesssim$~350~\GeV{} for $\tan\beta=1$
and the sensitivity decreases for larger $\tan\beta$ values.
In type-I 2HDM for instance, for $\tan\beta=10$ the exclusion reaches $m_H \lesssim$~320~\GeV{}
and $m_A \lesssim$~500~\GeV. The limiting value at $m_H \simeq 350$~\GeV\ is
due to the drop of the $H\to bb$ branching ratio, which competes with $H\to t\bar{t}$
at larger $m_H$ values.
The type-II and flipped 2HDMs are dominated by $A$ bosons from $b$-associated production
as $\tan\beta$ increases, although gluon--gluon fusion is still important for
$\tan\beta\approx 1$. Like the type-I and lepton-specific 2HDMs,
the type-II and flipped 2HDMs provide similar constraints because they only
differ in the lepton Yukawa couplings.
The contribution from $b$-associated signal production
increases the sensitivity at large $\tan\beta$ values, excluding
$m_H \lesssim$~650~\GeV{} for $\tan\beta=20$.
The search sensitivity deteriorates at lower $\tan\beta$ values,
excluding $m_H \lesssim$~350~\GeV{}  for $\tan\beta=1$.

\FloatBarrier

\subsection{$A \to ZH \to \ell\ell WW$ results}
\label{sec:res_llWW}
 
The $\mllqqqq$ distributions from different $\mqqqq$ mass
windows are scanned for possible excesses using a procedure similar
to the one in the $\ell \ell bb$ channel. The scan is performed
in steps of 10~\GeV\ for both the $m_A$ range 300--800~\GeV\
and the $m_H$ range 200--700~\GeV, such that
$m_A - m_H \geq 100~\GeV$. This gives in total 51
$\mqqqq$ mass windows and the overall number of
($m_A$, $m_H$) signal hypotheses that are tested is 1326.
 
\begin{figure}
\centering
\subfigure[]{\includegraphics[width=0.49\textwidth]{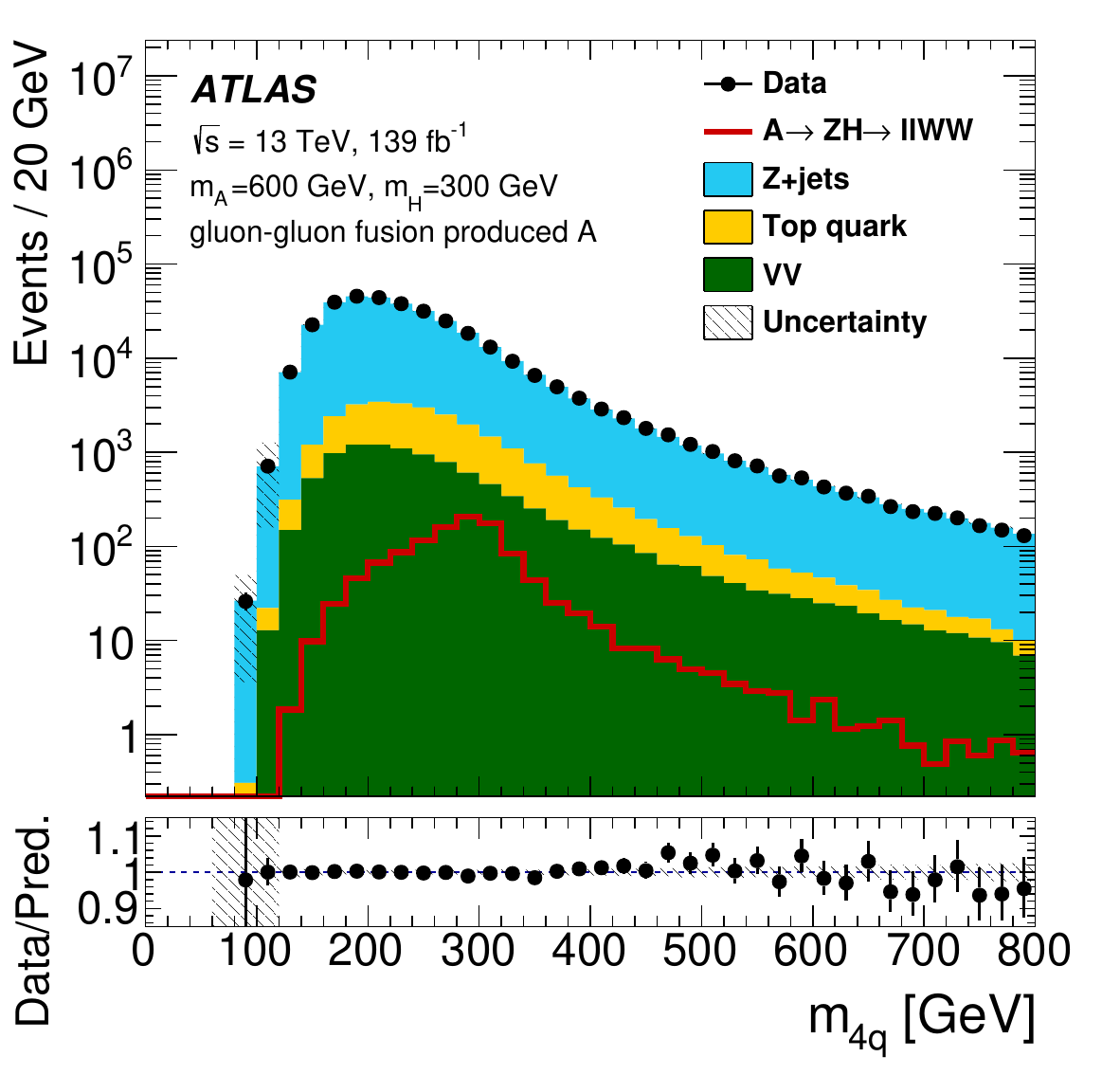}\label{fig:mWW1}}
\caption{The $\mqqqq$ distribution before any $\mqqqq$ window cuts.
The same conventions as in Figure~\ref{fig:Jvar_llWW} are used.
}
\label{fig:mWW}
\end{figure}

\begin{figure}[htb]
\centering
\subfigure[]{\includegraphics[width=0.49\textwidth]{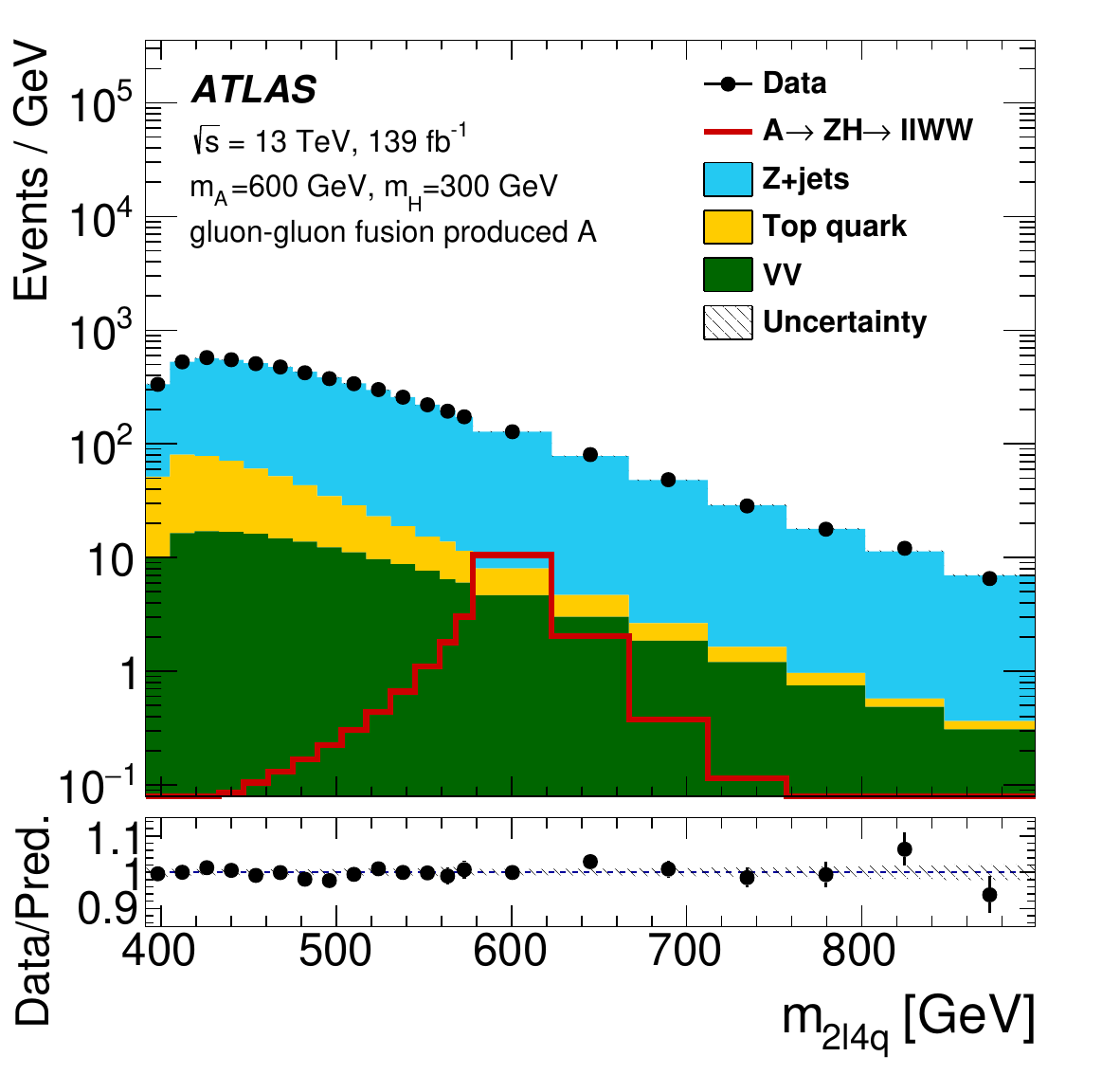}\label{fig:mllWW-1}}
\subfigure[]{\includegraphics[width=0.49\textwidth]{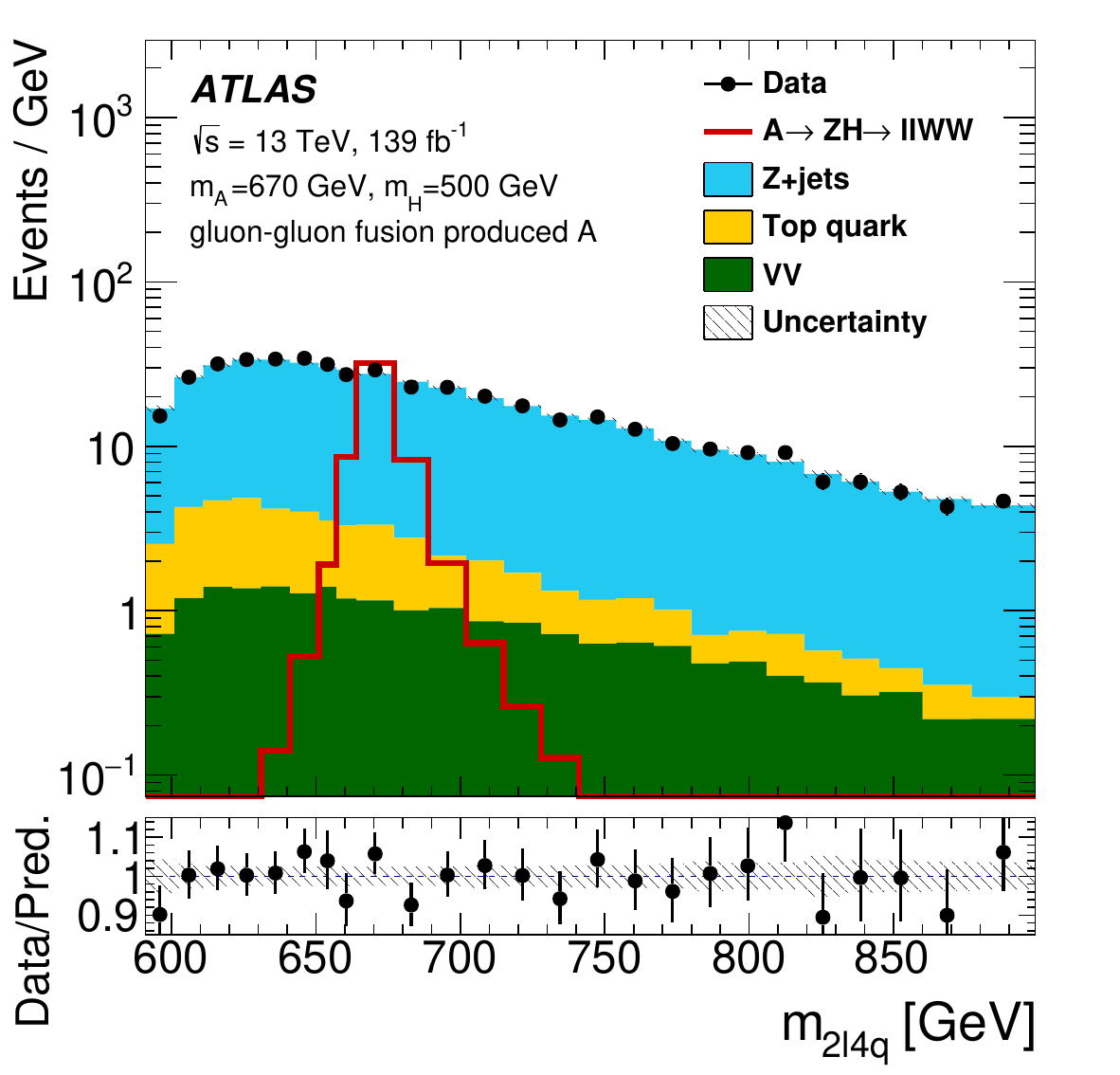}\label{fig:mllWW-2}}
\caption{The $\mllqqqq$ mass distribution for the $\mqqqq$ windows
defined for (a) $m_H=300~\GeV$ and (b) $m_H=500~\GeV$.
Signal distributions are also shown with $(m_A, m_H) = (600, 300)$~\GeV\ and $(m_A, m_H) = (670, 500)$~\GeV.
The number of entries shown in each bin is the number of events in that bin divided by the width of the bin.
The same conventions as in Figure~\ref{fig:Jvar_llWW} are used.
}
\label{fig:mllWW}
\end{figure}

Figure~\ref{fig:mWW} shows the distribution of the $H$ boson
candidate mass $\mqqqq$ before the $\mqqqq$ window requirement.
Typical examples of $\mllqqqq$ distributions after
the application of the $\mqqqq$ window requirement are shown
in Figures~\ref{fig:mllWW-1} and \ref{fig:mllWW-2},
referring to $\mqqqq$ windows defined for $m_H=300~\GeV$
and $m_H=500~\GeV$, respectively.
Signal distributions corresponding to the
$(m_A, m_H) = (600, 300)$~\GeV\ signal point for
Figure~\ref{fig:mllWW-1} and the
$(m_A, m_H) = (670, 500)$~\GeV\ signal point for
Figure~\ref{fig:mllWW-2}
are also shown.

In all cases, the data are found to be well described by the background model.
The most significant excess is at the $(m_A, m_H) = (440, 310)$~\GeV{} signal point, for which the
local (global) significance is $2.9$ ($0.82$)
standard deviations.
 
\FloatBarrier

\begin{figure}
\centering
\subfigure[]{\includegraphics[width=0.49\textwidth]{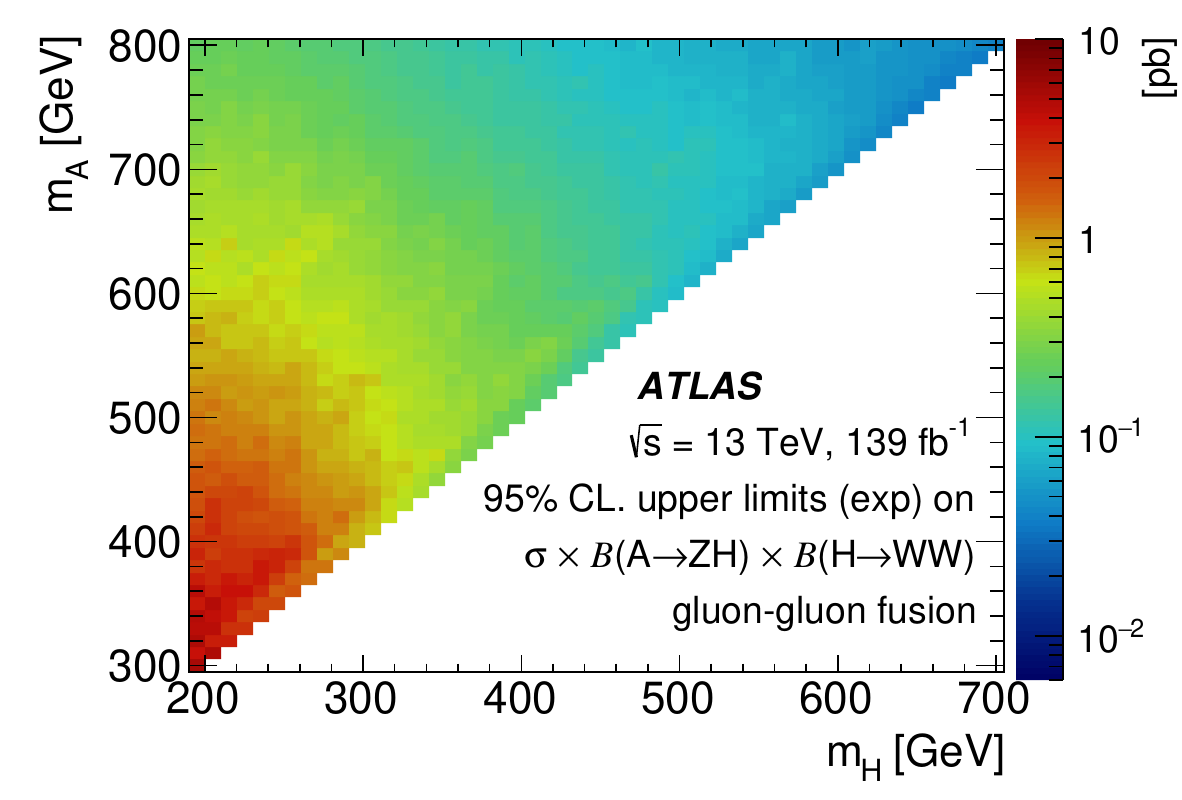}\label{fig:limits:llWW:exp}}
\subfigure[]{\includegraphics[width=0.49\textwidth]{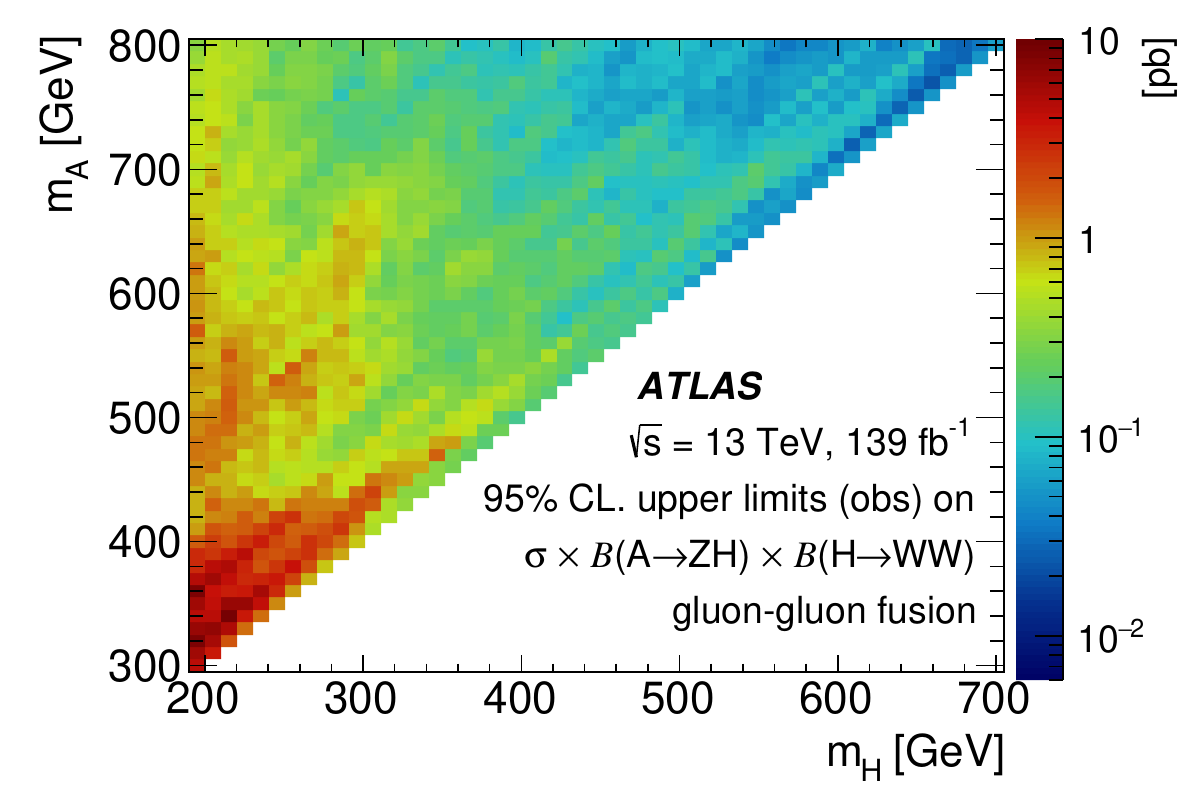}\label{fig:limits:llWW:obs}} \\
 
\caption{Expected (a) and observed (b) upper bounds at 95\% CL on the production cross section times the branching ratio $B(A\to ZH)\times B(H\to WW)$ in pb.}
\label{fig:limitsllWW}
\end{figure}

Using the same method as for the $\ell\ell bb$ channel,
constraints on the production of $A\to ZH$ followed by
$H\to WW$ decay are derived. The 95\% CL upper limits are shown in
Figure~\ref{fig:limitsllWW} for a narrow-width $A$ boson produced
via gluon--gluon fusion.
The upper limit varies from \minGGAlimllWW~pb for the
$(m_A, m_H) = \minGGAlimPointllWW$~\GeV{} signal point
to \maxGGAlimllWW~pb for the
$(m_A, m_H) = \maxGGAlimPointllWW$~\GeV{} signal point.
This is to be compared with the corresponding expected limits
of \minGGAlimExpllWW~pb and \maxGGAlimExpllWW~pb
for these two signal points.
The upper limits deteriorate when the natural width of
the $A$ boson is comparable to, or larger than, the
experimental mass resolution. In particular, for
a natural width that is 10\% of $m_A$
the upper limits decrease on average by a factor of 3.
This factor becomes approximately 5 when the
natural width increases to  20\%.
 
The sensitivity of the $\ell\ell WW$ channel in the
context of the CP-conserving 2HDM was examined.
The same 2HDM calculations as in the $\ell\ell bb$ channel
are used and the only differences are related to the
parameter space of the model that is probed.
In particular, because only $A$ bosons produced by
gluon--gluon fusion are studied in this search,
only type-I and lepton-specific 2HDMs are considered.
In addition, the partial width $\Gamma(H\to WW)$
vanishes when $\cos(\beta-\alpha)=0$ and is maximal
at  $|\cos(\beta-\alpha)|=1$, whereas for the
partial width $\Gamma(A\to ZH)$ the opposite is true, i.e.\
it vanishes when $|\cos(\beta-\alpha)|=1$ and it is maximal
when $\cos(\beta-\alpha)=0$. These observations imply
that this channel should be most sensitive between these
two extreme values of $|\cos(\beta-\alpha)|$.
 
\begin{figure}
\centering
\subfigure[]{\includegraphics[width=0.49\textwidth]{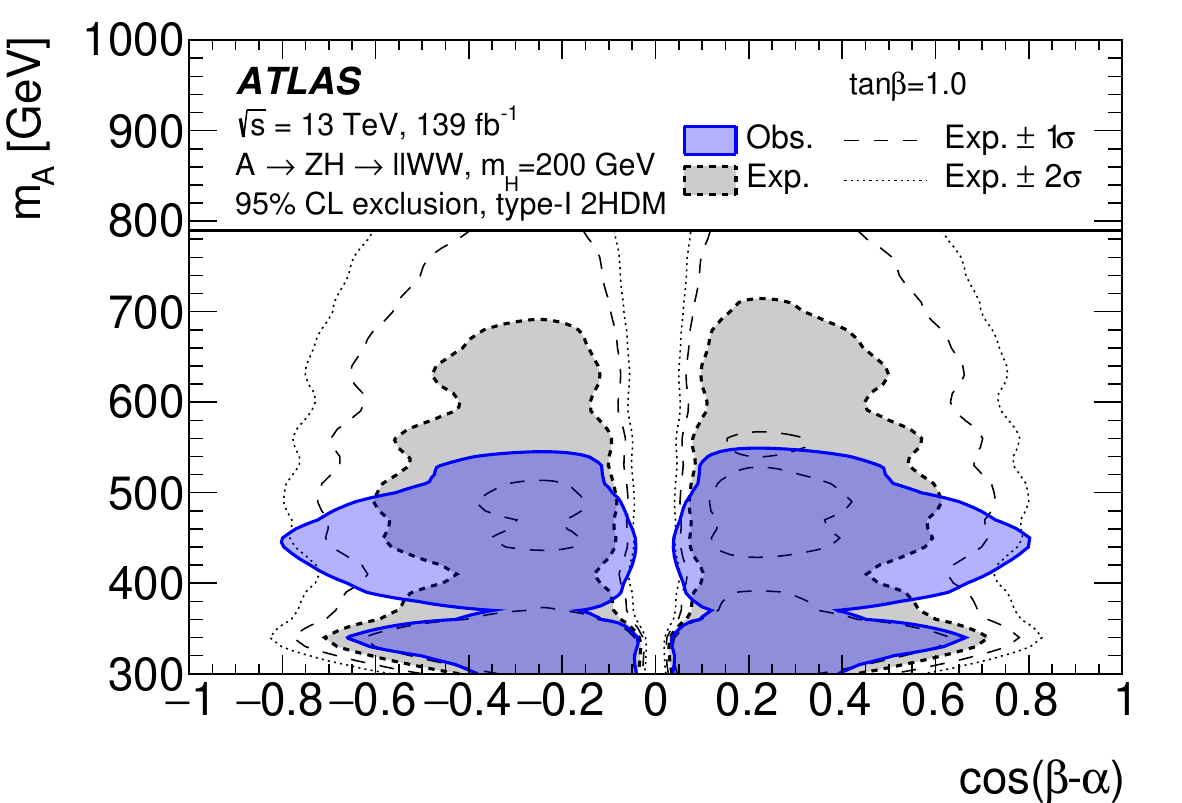}\label{fig:2hdm-1-type1}}
\subfigure[]{\includegraphics[width=0.49\textwidth]{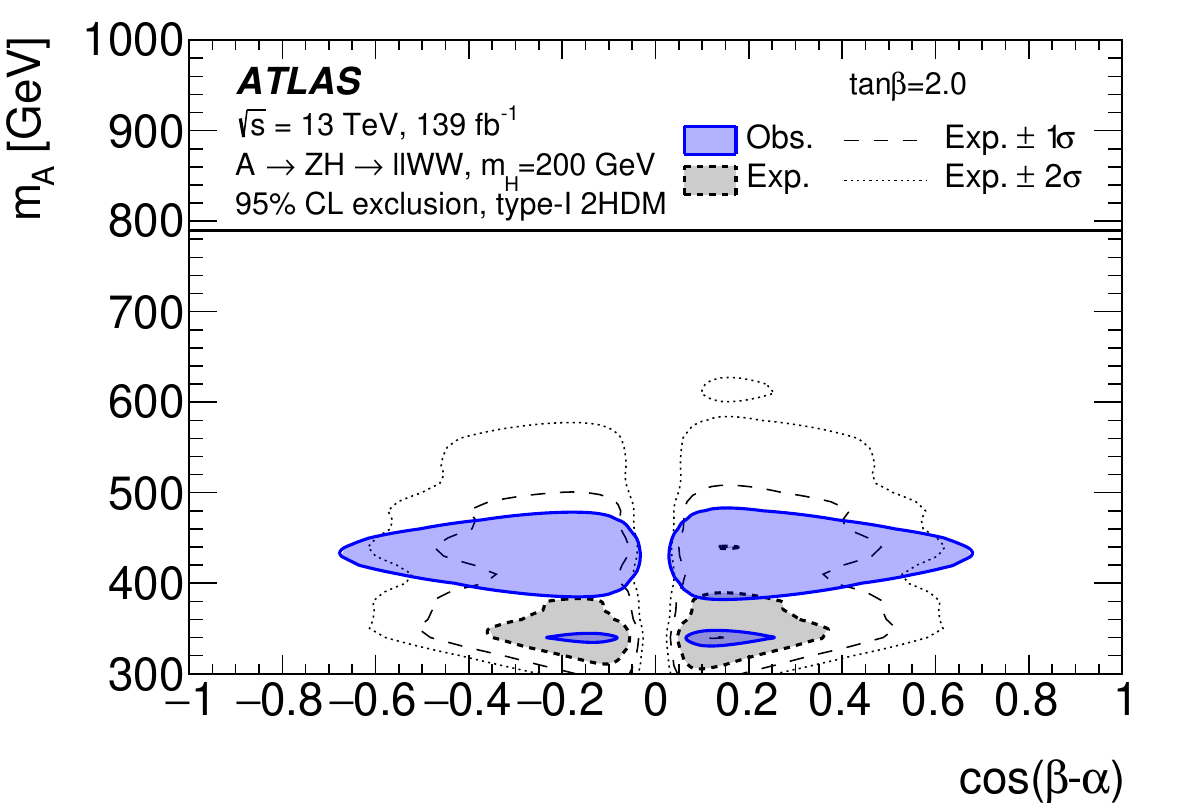}\label{fig:2hdm-2-type1}} \\
\subfigure[]{\includegraphics[width=0.49\textwidth]{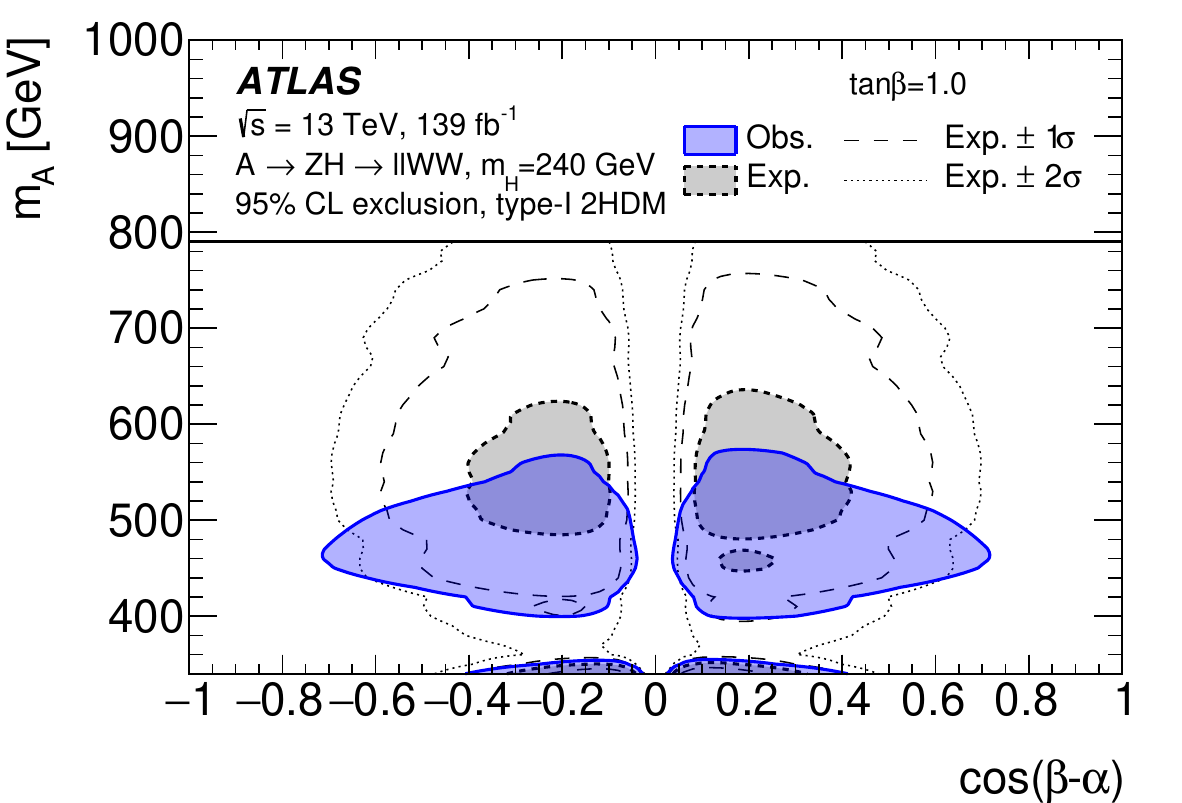}\label{fig:2hdm-3-type1}}
\subfigure[]{\includegraphics[width=0.49\textwidth]{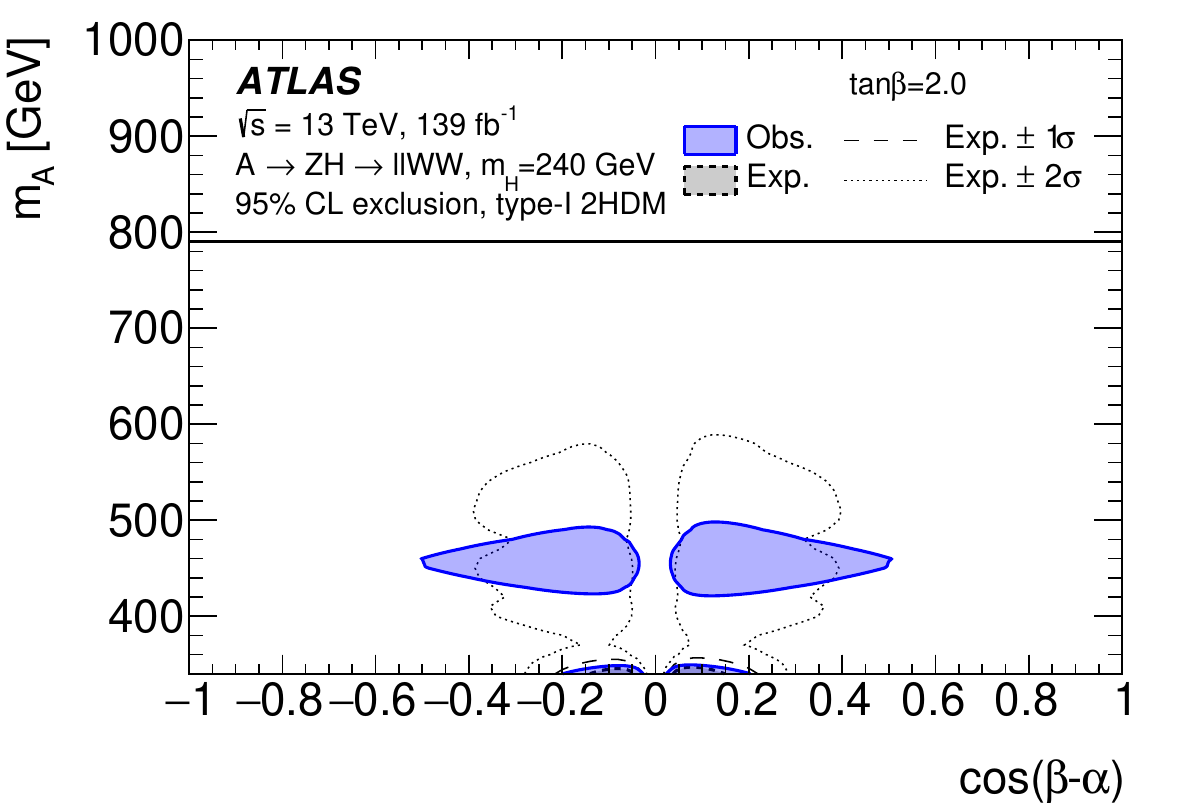}\label{fig:2hdm-4-type1}}
\caption{Observed and expected 95\% CL exclusion regions in the ($\cos(\beta-\alpha)$, $m_A$) plane for various $\tan\beta$ values for (a,b) $m_H=200~\GeV$ and
(c,d) $m_H = 240~\GeV$ in the context of type-I 2HDM for the $\ell\ell WW$ channel.}
\label{fig:2hdm-llWW}
\end{figure}

The interpretation of the observed and expected upper limits
on the cross section times branching ratio
in the context of the type-I and lepton-specific 2HDM
scenarios show that the $\ell\ell WW$ channel
has little sensitivity in regions that are not already
excluded by the 125~\GeV\ Higgs boson
coupling measurements~\cite{HIGG-2018-57}, an analysis that
also provides similar limits in this parameter space.
In particular, for the $m_A$ range considered
in this channel, there is sensitivity up to
$m_H < 250~\GeV$ and for $\tan\beta< 4$.
Some examples of 95\% CL excluded regions
in the plane defined by $m_A$ and $\cos(\beta-\alpha)$
for $m_{H}=200~\GeV$ and $m_{H}=240~\GeV$
are shown in Figure~\ref{fig:2hdm-llWW} for the
type-I 2HDM. The results are very similar for the
lepton-specific 2HDM, since the only difference
between the two 2HDM types is the lepton Yukawa couplings,
which only affect the total width.

\FloatBarrier

% End of text imported from the .//results.tex input file
 
\FloatBarrier
 
\clearpage
\section{Conclusion}
\label{sec:conclusion}
 
% The next lines are included from the .//conclusion.tex input file
Data recorded by the ATLAS experiment at the LHC, corresponding to an
integrated luminosity of 139~$\textrm{fb}^{-1}$ from proton--proton collisions
at a centre-of-mass energy 13~\TeV{}, are used to search for a heavy Higgs
boson, $A$, decaying into $ZH$, where $H$ denotes another heavy Higgs boson with mass
$m_H>125~\GeV$. Two final states were considered, where the $H$ boson
decays into a pair of $b$-quarks or $W$ bosons, and in both cases the $Z$ boson decays into
a pair of electrons or muons.
In the $\ell \ell bb$ channel, the $A$ boson is assumed to be produced via either gluon--gluon
fusion or $b$-associated production. In the $\ell \ell WW$ channel, only gluon--gluon
fusion production is considered.
No significant deviation from the SM background
predictions is observed in the $ZH\to\ell\ell bb$ and $ZH\to\ell\ell WW \to \ell \ell qqqq$ final states that are
considered in this search.
Considering each channel and each production process separately,
upper limits are set at the 95\% confidence level
for $\sigma\times B(A\to ZH)\times B(H\to bb \; \textrm{or} \; H \to WW)$.
For $\ell \ell bb$, upper limits are set in the range \minGGAlim--\maxGGAlim~fb for gluon--gluon fusion and \minBBAlim--\maxBBAlim~fb for
$b$-associated production of a narrow $A$ boson in the mass range 230--800~\GeV, assuming the $H$ boson is in the mass range 130--700~\GeV.
For $\ell \ell WW$, the observed upper limits are in the range \minGGAlimllWW--\maxGGAlimllWW~pb for gluon--gluon fusion production of a narrow $A$ boson
in the mass range 300--800~\GeV, assuming the $H$ boson is in the mass range 200--700~\GeV.
Taking into account both production processes,
the $\ell \ell bb$ search tightens the constraints on the 2HDM scenario in the case of large mass splittings between its heavier neutral Higgs bosons.
The $\ell \ell WW$ channel has not been explored previously at the LHC,
and this search explicitly demonstrates its potential
to constrain 2HDM parameters away from the weak decoupling limit.

\FloatBarrier
 
% End of text imported from the .//conclusion.tex input file

\section*{Acknowledgements}
 
% The next lines are included from the .//acknowledgements/Acknowledgements.tex input file

We thank CERN for the very successful operation of the LHC, as well as the
support staff from our institutions without whom ATLAS could not be
operated efficiently.
 
We acknowledge the support of ANPCyT, Argentina; YerPhI, Armenia; ARC, Australia; BMWFW and FWF, Austria; ANAS, Azerbaijan; SSTC, Belarus; CNPq and FAPESP, Brazil; NSERC, NRC and CFI, Canada; CERN; ANID, Chile; CAS, MOST and NSFC, China; COLCIENCIAS, Colombia; MSMT CR, MPO CR and VSC CR, Czech Republic; DNRF and DNSRC, Denmark; IN2P3-CNRS and CEA-DRF/IRFU, France; SRNSFG, Georgia; BMBF, HGF and MPG, Germany; GSRT, Greece; RGC and Hong Kong SAR, China; ISF and Benoziyo Center, Israel; INFN, Italy; MEXT and JSPS, Japan; CNRST, Morocco; NWO, Netherlands; RCN, Norway; MNiSW and NCN, Poland; FCT, Portugal; MNE/IFA, Romania; JINR; MES of Russia and NRC KI, Russian Federation; MESTD, Serbia; MSSR, Slovakia; ARRS and MIZ\v{S}, Slovenia; DST/NRF, South Africa; MICINN, Spain; SRC and Wallenberg Foundation, Sweden; SERI, SNSF and Cantons of Bern and Geneva, Switzerland; MOST, Taiwan; TAEK, Turkey; STFC, United Kingdom; DOE and NSF, United States of America. In addition, individual groups and members have received support from BCKDF, CANARIE, Compute Canada, CRC and IVADO, Canada; Beijing Municipal Science \& Technology Commission, China; COST, ERC, ERDF, Horizon 2020 and Marie Sk{\l}odowska-Curie Actions, European Union; Investissements d'Avenir Labex, Investissements d'Avenir Idex and ANR, France; DFG and AvH Foundation, Germany; Herakleitos, Thales and Aristeia programmes co-financed by EU-ESF and the Greek NSRF, Greece; BSF-NSF and GIF, Israel; La Caixa Banking Foundation, CERCA Programme Generalitat de Catalunya and PROMETEO and GenT Programmes Generalitat Valenciana, Spain; G\"{o}ran Gustafssons Stiftelse, Sweden; The Royal Society and Leverhulme Trust, United Kingdom.
 
The crucial computing support from all WLCG partners is acknowledged gratefully, in particular from CERN, the ATLAS Tier-1 facilities at TRIUMF (Canada), NDGF (Denmark, Norway, Sweden), CC-IN2P3 (France), KIT/GridKA (Germany), INFN-CNAF (Italy), NL-T1 (Netherlands), PIC (Spain), ASGC (Taiwan), RAL (UK) and BNL (USA), the Tier-2 facilities worldwide and large non-WLCG resource providers. Major contributors of computing resources are listed in Ref.~\cite{ATL-SOFT-PUB-2020-001}.
 
% End of text imported from the .//acknowledgements/Acknowledgements.tex input file

\printbibliography

@article{Oreglia:1980cs,
      author         = "Oreglia, M.",
      title          = "{A Study of the Reactions $\psi^\prime \to \gamma \gamma
                        \psi$}",
      school         = "SLAC",
      journal        = "SLAC-R-0236",
      url            = "http://www.slac.stanford.edu/cgi-wrap/getdoc/slac-r-236.pdf",
      year           = "1980",
      reportNumber   = "SLAC-0236, UMI-81-08973, SLAC-R-0236, SLAC-R-236",
      SLACcitation   = "%%CITATION = SLAC-0236;%%"
}

@article{Ball:2014uwa,
      author         = "Ball, Richard D. and others",
      title          = "{Parton distributions for the LHC Run II}",
      collaboration  = "NNPDF",
      journal        = "JHEP",
      volume         = "04",
      year           = "2015",
      pages          = "040",
      doi            = "10.1007/JHEP04(2015)040",
      eprint         = "1410.8849",
      archivePrefix  = "arXiv",
      primaryClass   = "hep-ph",
      reportNumber   = "EDINBURGH-2014-15, IFUM-1034-FT, CERN-PH-TH-2013-253,
                        OUTP-14-11P, CAVENDISH-HEP-14-11",
      SLACcitation   = "%%CITATION = ARXIV:1410.8849;%%"
}

@article{Ball:2012cx,
      author         = "Ball, Richard D. and others",
      title          = "{Parton distributions with LHC data}",
      journal        = "Nucl. Phys. B",
      volume         = "867",
      year           = "2013",
      pages          = "244-289",
      doi            = "10.1016/j.nuclphysb.2012.10.003",
      eprint         = "1207.1303",
      archivePrefix  = "arXiv",
      primaryClass   = "hep-ph",
      reportNumber   = "EDINBURGH-2012-08, IFUM-FT-997, FR-PHENO-2012-014,
                        RWTH-TTK-12-25, CERN-PH-TH-2012-037, SFB-CPP-12-47",
      SLACcitation   = "%%CITATION = ARXIV:1207.1303;%%"
}

@article{Aaboud:2017itg,
      hideauthor         = "Aaboud, Morad and others",
      author         = "{ATLAS Collaboration}",
      title          = "{Searches for heavy $ZZ$ and $ZW$ resonances in the
                        $\ell\ell qq$ and $\nu\nu qq$ final states in $pp$
                        collisions at $\sqrt{s}=13$ TeV with the ATLAS detector}",
      hidecollaboration  = "ATLAS",
      journal        = "JHEP",
      volume         = "03",
      year           = "2018",
      pages          = "009",
      doi            = "10.1007/JHEP03(2018)009",
      eprint         = "1708.09638",
      archivePrefix  = "arXiv",
      primaryClass   = "hep-ex",
      reportNumber   = "CERN-EP-2017-146",
      SLACcitation   = "%%CITATION = ARXIV:1708.09638;%%"
}

@article{Aaboud:2017xsd,
      hideauthor         = "Aaboud, M. and others",
      author         = "{ATLAS Collaboration}",
      title          = "{Evidence for the $ H\to b\overline{b} $ decay with the
                        ATLAS detector}",
      hidecollaboration  = "ATLAS",
      journal        = "JHEP",
      volume         = "12",
      year           = "2017",
      pages          = "024",
      doi            = "10.1007/JHEP12(2017)024",
      eprint         = "1708.03299",
      archivePrefix  = "arXiv",
      primaryClass   = "hep-ex",
      reportNumber   = "CERN-EP-2017-175",
      SLACcitation   = "%%CITATION = ARXIV:1708.03299;%%"
}

@Article{Coleppa:2014hxa,
      author         = "Coleppa, Baradhwaj and Kling, Felix and Su, Shufang",
      title          = "{Exotic decays of a heavy neutral Higgs through HZ/AZ
                        channel}",
      journal        = "JHEP",
      volume         = "09",
      year           = "2014",
      pages          = "161",
      doi            = "10.1007/JHEP09(2014)161",
      eprint         = "1404.1922",
      archivePrefix  = "arXiv",
      primaryClass   = "hep-ph",
      SLACcitation   = "%%CITATION = ARXIV:1404.1922;%%"
}

@Article{Lee:1973iz,
      author         = "Lee, T. D.",
      title          = "{A Theory of Spontaneous T Violation}",
      journal        = "Phys. Rev. D",
      volume         = "8",
      year           = "1973",
      pages          = "1226-1239",
      doi            = "10.1103/PhysRevD.8.1226",
      hidenote           = "[,516(1973)]",
      SLACcitation   = "%%CITATION = PHRVA,D8,1226;%%"
}

@Article{Branco:2011iw,
      author         = "{G. Branco et al.}",
      title          = "{Theory and phenomenology of two-Higgs-doublet models}",
      journal        = "Phys. Rept.",
      volume         = "516",
      pages          = "1-102",
      doi            = "10.1016/j.physrep.2012.02.002",
      year           = "2012",
      eprint         = "1106.0034",
      archivePrefix  = "arXiv",
      primaryClass   = "hep-ph",
      SLACcitation   = "%%CITATION = ARXIV:1106.0034;%%",
}

@Article{Fromme:2006cm,
      author         = "Fromme, Lars and Huber, Stephan J. and Seniuch, Michael",
      title          = "{Baryogenesis in the two-Higgs doublet model}",
      journal        = "JHEP",
      volume         = "11",
      year           = "2006",
      pages          = "038",
      doi            = "10.1088/1126-6708/2006/11/038",
      eprint         = "hep-ph/0605242",
      archivePrefix  = "arXiv",
%      primaryClass   = "hep-ph",
      reportNumber   = "CERN-PH-TH-2006-094, BI-TP-2006-18",
      SLACcitation   = "%%CITATION = HEP-PH/0605242;%%"
}

@Article{Cohen:1993nk,
      author         = "Cohen, Andrew G. and Kaplan, D. B. and Nelson, A. E.",
      title          = "{Progress in Electroweak Baryogenesis}",
      journal        = "Ann. Rev. Nucl. Part. Sci.",
      volume         = "43",
      year           = "1993",
      pages          = "27-70",
      doi            = "10.1146/annurev.ns.43.120193.000331",
      eprint         = "hep-ph/9302210",
      archivePrefix  = "arXiv",
%      primaryClass   = "hep-ph",
      reportNumber   = "UCSD-PTH-93-02, BUHEP-93-4",
      SLACcitation   = "%%CITATION = HEP-PH/9302210;%%"
}

@article{Dorsch:2014qja,
      author         = "Dorsch, G. C. and Huber, S. J. and Mimasu, K. and No, J.
                        M.",
      title          = "{Echoes of the Electroweak Phase Transition: Discovering
                        a Second Higgs Doublet through $A_0 \rightarrow ZH_0$}",
      journal        = "Phys. Rev. Lett.",
      volume         = "113",
      year           = "2014",
      number         = "21",
      pages          = "211802",
      doi            = "10.1103/PhysRevLett.113.211802",
      eprint         = "1405.5537",
      archivePrefix  = "arXiv",
      primaryClass   = "hep-ph",
      SLACcitation   = "%%CITATION = ARXIV:1405.5537;%%"
}

@article{Basler:2016obg,
      author         = "Basler, P. and Krause, M. and M{\"u}hlleitner, M. and
                        Wittbrodt, J. and Wlotzka, A.",
      title          = "{Strong first order electroweak phase transition in the
                        CP-conserving 2HDM revisited}",
      journal        = "JHEP",
      volume         = "02",
      year           = "2017",
      pages          = "121",
      doi            = "10.1007/JHEP02(2017)121",
      eprint         = "1612.04086",
      archivePrefix  = "arXiv",
      primaryClass   = "hep-ph",
      SLACcitation   = "%%CITATION = ARXIV:1612.04086;%%"
}

@Article{Turok:1990zg,
      author         = "Turok, Neil and Zadrozny, John",
      title          = "{Electroweak baryogenesis in the two-doublet model}",
      journal        = "Nucl. Phys. B",
      volume         = "358",
      year           = "1991",
      pages          = "471-493",
      doi            = "10.1016/0550-3213(91)90356-3",
      reportNumber   = "PUPT-90-1216",
      SLACcitation   = "%%CITATION = NUPHA,B358,471;%%"
}

@Article{PERF-2007-01,
    author         = "{ATLAS Collaboration}",
    title          = "{The ATLAS Experiment at the CERN Large Hadron Collider}",
    journal        = "JINST",
    volume         = "3",
    year           = "2008",
    pages          = "S08003",
    doi            = "10.1088/1748-0221/3/08/S08003",
    primaryClass   = "hep-ex",
}

@Article{Alwall:2014hca,
      author         = "Alwall, J. and Frederix, R. and Frixione, S. and Hirschi,
                        V. and Maltoni, F. and Mattelaer, O. and Shao, H. -S. and
                        Stelzer, T. and Torrielli, P. and Zaro, M.",
      title          = "{The automated computation of tree-level and
                        next-to-leading order differential cross sections, and
                        their matching to parton shower simulations}",
      journal        = "JHEP",
      volume         = "07",
      year           = "2014",
      pages          = "079",
      doi            = "10.1007/JHEP07(2014)079",
      eprint         = "1405.0301",
      archivePrefix  = "arXiv",
      primaryClass   = "hep-ph",
      reportNumber   = "CERN-PH-TH-2014-064, CP3-14-18, LPN14-066, MCNET-14-09,
                        ZU-TH-14-14",
      SLACcitation   = "%%CITATION = ARXIV:1405.0301;%%"
}

@Article{Frixione:2002ik,
      author         = "Frixione, Stefano and Webber, Bryan R.",
      title          = "{Matching NLO QCD computations with parton shower
                        simulations: the POWHEG method}",
      journal        = "JHEP",
      volume         = "06",
      year           = "2002",
      pages          = "029",
      doi            = "10.1088/1126-6708/2002/06/029",
      eprint         = "hep-ph/0204244",
      archivePrefix  = "arXiv",
%      primaryClass   = "hep-ph",
      reportNumber   = "CAVENDISH-HEP-02-01, LAPTH-905-02, GEF-TH-2-2002",
      SLACcitation   = "%%CITATION = HEP-PH/0204244;%%"
}

@Article{Sjostrand:2007gs,
      author         = "Sj{\"o}strand, Torbjorn and Mrenna, Stephen and Skands, Peter
                        Z.",
      title          = "{A brief introduction to PYTHIA 8.1}",
      journal        = "Comput. Phys. Commun.",
      volume         = "178",
      year           = "2008",
      pages          = "852-867",
      doi            = "10.1016/j.cpc.2008.01.036",
      eprint         = "0710.3820",
      archivePrefix  = "arXiv",
      primaryClass   = "hep-ph",
      reportNumber   = "CERN-LCGAPP-2007-04, LU-TP-07-28,
                        FERMILAB-PUB-07-512-CD-T",
      SLACcitation   = "%%CITATION = ARXIV:0710.3820;%%"
}

@Article{Lai:2010vv,
      author         = "Lai, Hung-Liang and Guzzi, Marco and Huston, Joey and Li,
                        Zhao and Nadolsky, Pavel M. and Pumplin, Jon and Yuan, C.
                        -P.",
      title          = "{New parton distributions for collider physics}",
      journal        = "Phys. Rev. D",
      volume         = "82",
      year           = "2010",
      pages          = "074024",
      doi            = "10.1103/PhysRevD.82.074024",
      eprint         = "1007.2241",
      archivePrefix  = "arXiv",
      primaryClass   = "hep-ph",
      reportNumber   = "MSUHEP-100707, SMU-HEP-10-10",
      SLACcitation   = "%%CITATION = ARXIV:1007.2241;%%"
}

@Article{Alwall:2011uj,
      author         = "Alwall, Johan and Herquet, Michel and Maltoni, Fabio and
                        Mattelaer, Olivier and Stelzer, Tim",
      title          = "{MadGraph 5: going beyond}",
      journal        = "JHEP",
      volume         = "06",
      year           = "2011",
      pages          = "128",
      doi            = "10.1007/JHEP06(2011)128",
      eprint         = "1106.0522",
      archivePrefix  = "arXiv",
      primaryClass   = "hep-ph",
      reportNumber   = "FERMILAB-PUB-11-448-T",
      SLACcitation   = "%%CITATION = ARXIV:1106.0522;%%"
}

@Article{Nason:2004rx,
      author         = "Nason, Paolo",
      title          = "{A New method for combining NLO QCD with shower Monte
                        Carlo algorithms}",
      journal        = "JHEP",
      volume         = "11",
      year           = "2004",
      pages          = "040",
      doi            = "10.1088/1126-6708/2004/11/040",
      eprint         = "hep-ph/0409146",
      archivePrefix  = "arXiv",
%      primaryClass   = "hep-ph",
      reportNumber   = "BICOCCA-FT-04-11",
      SLACcitation   = "%%CITATION = HEP-PH/0409146;%%"
}

@Article{Frixione:2007vw,
      author         = "Frixione, Stefano and Nason, Paolo and Oleari, Carlo",
      title          = "{Matching NLO QCD computations with Parton Shower
                        simulations: the POWHEG method}",
      journal        = "JHEP",
      volume         = "11",
      year           = "2007",
      pages          = "070",
      doi            = "10.1088/1126-6708/2007/11/070",
      eprint         = "0709.2092",
      archivePrefix  = "arXiv",
      primaryClass   = "hep-ph",
      reportNumber   = "BICOCCA-FT-07-9, GEF-TH-21-2007",
      SLACcitation   = "%%CITATION = ARXIV:0709.2092;%%"
}

@Article{Alioli:2010xd,
      author         = "Alioli, Simone and Nason, Paolo and Oleari, Carlo and Re,
                        Emanuele",
      title          = "{A general framework for implementing NLO calculations in
                        shower Monte Carlo programs: the POWHEG BOX}",
      journal        = "JHEP",
      volume         = "06",
      year           = "2010",
      pages          = "043",
      doi            = "10.1007/JHEP06(2010)043",
      eprint         = "1002.2581",
      archivePrefix  = "arXiv",
      primaryClass   = "hep-ph",
      reportNumber   = "DESY-10-018, SFB-CPP-10-22, IPPP-10-11, DCPT-10-22",
      SLACcitation   = "%%CITATION = ARXIV:1002.2581;%%"
}

@Article{SOFT-2010-01,
    author         = "{ATLAS Collaboration}",
    title          = "{The ATLAS Simulation Infrastructure}",
    journal        = "Eur. Phys. J. C",
    volume         = "70",
    year           = "2010",
    pages          = "823",
    doi            = "10.1140/epjc/s10052-010-1429-9",
    eprint         = "1005.4568",
    archivePrefix  = "arXiv",
    primaryClass   = "hep-ex",
}

@Article{Agostinelli:2002hh,
      author         = "Agostinelli, S. and others",
      title          = "{GEANT4: A simulation toolkit}",
      collaboration  = "GEANT4",
      journal        = "Nucl. Instrum. Meth. A",
      volume         = "506",
      year           = "2003",
      pages          = "250-303",
      doi            = "10.1016/S0168-9002(03)01368-8",
      reportNumber   = "SLAC-PUB-9350, FERMILAB-PUB-03-339",
      SLACcitation   = "%%CITATION = NUIMA,A506,250;%%"
}

@Article{Cacciari:2008gp,
      author         = "Cacciari, Matteo and Salam, Gavin P. and Soyez, Gregory",
      title          = "{The anti-k$_{\textrm{t}}$ jet clustering algorithm}",
      journal        = "JHEP",
      volume         = "04",
      year           = "2008",
      pages          = "063",
      doi            = "10.1088/1126-6708/2008/04/063",
      eprint         = "0802.1189",
      archivePrefix  = "arXiv",
      primaryClass   = "hep-ph",
      reportNumber   = "LPTHE-07-03",
      SLACcitation   = "%%CITATION = ARXIV:0802.1189;%%"
}

@Article{Cacciari:2011ma,
      author         = "Cacciari, Matteo and Salam, Gavin P. and Soyez, Gregory",
      title          = "{FastJet user manual}",
      journal        = "Eur. Phys. J. C",
      volume         = "72",
      year           = "2012",
      pages          = "1896",
      doi            = "10.1140/epjc/s10052-012-1896-2",
      eprint         = "1111.6097",
      archivePrefix  = "arXiv",
      primaryClass   = "hep-ph",
      reportNumber   = "CERN-PH-TH-2011-297",
      SLACcitation   = "%%CITATION = ARXIV:1111.6097;%%"
}

@Article{Moneta:2010pm,
      author         = "Moneta, Lorenzo and Belasco, Kevin and Cranmer, Kyle S.
                        and Kreiss, S. and Lazzaro, Alfio and Piparo, Danilo and
                        Schott, Gregory and Verkerke, Wouter and Wolf, Matthias",
      title          = "{The RooStats Project}",
      booktitle      = "{Proceedings, 13th International Workshop on Advanced
                        computing and analysis techniques in physics research
                        (ACAT2010): Jaipur, India, February 22-27, 2010}",
      journal        = "PoS",
      volume         = "ACAT2010",
      year           = "2010",
      pages          = "057",
      eprint         = "1009.1003",
      archivePrefix  = "arXiv",
      primaryClass   = "physics.data-an",
      doi = "10.22323/1.093.0057"
}

@Booklet{RooFit,
      author       = "Verkerke, Wouter and Kirkby, David",
      title            = "{The RooFit toolkit for data modeling}",
      year           = "2003",
      eprint         = "physics/0306116",
      archivePrefix  = "arXiv",
      primaryClass   = "physics.data-an",
}

@Article{Cowan:2010js,
      Author = {Cowan, Glen and Cranmer, Kyle and Gross, Eilam and Vitells, Ofer},
      Eprint = "1007.1727",
      Archiveprefix = {arXiv},
      Primaryclass = {physics.data-an},
      Journal = {Eur. Phys. J. C},
      Pages = {1554},
      doi = "10.1140/epjc/s10052-011-1554-0",
      Title = {{Asymptotic formulae for likelihood-based tests of new physics}},
      Volume = {71},
      Year = {2011},
      Addendum = "Erratum: \href{http://dx.doi.org/10.1140/epjc/s10052-013-2501-z}{Eur. Phys. J. C {\textbf{73}} (2013) 2501}"
}

@Article{Read:2002hq,
      author         = "Read, Alexander L.",
      title          = "{Presentation of search results: The CLs technique}",
      booktitle      = "{Advanced Statistical Techniques in Particle Physics.
                        Proceedings, Conference, Durham, UK, March 18-22, 2002}",
      journal        = "J. Phys. G",
      volume         = "28",
      year           = "2002",
      pages          = "2693-2704",
      doi            = "10.1088/0954-3899/28/10/313",
%      note           = "[,11(2002)]",
	  journal="{J. Phys. G}",
	  number="{10}",
  	  year="{2002}",
      SLACcitation   = "%%CITATION = JPAGA,G28,2693;%%"
}

@Article{Das:2016stf,
      author         = "Das, Souvik",
      title          = "{A simple alternative to the Crystal Ball function}",
      year           = "2016",
      eprint         = "1603.08591",
      archivePrefix  = "arXiv",
      primaryClass   = "hep-ex",
      SLACcitation   = "%%CITATION = ARXIV:1603.08591;%%"
}

@Article{Eriksson:2009ws,
      author         = "Eriksson, David and Rathsman, Johan and St{\aa}l, Oscar",
      title          = "{2HDMC - two-Higgs-doublet model calculator physics and
                        manual}",
      journal        = "Comput. Phys. Commun.",
      volume         = "181",
      year           = "2010",
      pages          = "189-205",
      doi            = "10.1016/j.cpc.2009.09.011",
      eprint         = "0902.0851",
      archivePrefix  = "arXiv",
      primaryClass   = "hep-ph",
      SLACcitation   = "%%CITATION = ARXIV:0902.0851;%%"
}

@Article{deFlorian:2016spz,
      author         = "de Florian, D. and others",
      title          = "Handbook of LHC Higgs Cross Sections: 4. Deciphering the
                        Nature of the Higgs Sector",
      collaboration  = "LHC Higgs Cross Section Working Group",
      doi            = "10.23731/CYRM-2017-002",
      year           = "2016",
      eprint         = "1610.07922",
      archivePrefix  = "arXiv",
      primaryClass   = "hep-ph",
      reportNumber   = "FERMILAB-FN-1025-T, CERN-2017-002-M",
      SLACcitation   = "%%CITATION = ARXIV:1610.07922;%%"
}

@Article{Harlander:2012pb,
      author         = "Harlander, Robert V. and Liebler, Stefan and Mantler,
                        Hendrik",
      title          = "{SusHi: A program for the calculation of Higgs production
                        in gluon fusion and bottom-quark annihilation in the
                        Standard Model and the MSSM}",
      journal        = "Comput. Phys. Commun.",
      volume         = "184",
      year           = "2013",
      pages          = "1605-1617",
      doi            = "10.1016/j.cpc.2013.02.006",
      eprint         = "1212.3249",
      archivePrefix  = "arXiv",
      primaryClass   = "hep-ph",
      reportNumber   = "WUB-12-28, LPN12-134",
      SLACcitation   = "%%CITATION = ARXIV:1212.3249;%%"
}

@Article{Harlander:2005rq,
      author         = "Harlander, Robert and Kant, Philipp",
      title          = "{Higgs production and decay: analytic results at
                        next-to-leading order QCD}",
      journal        = "JHEP",
      volume         = "12",
      year           = "2005",
      pages          = "015",
      doi            = "10.1088/1126-6708/2005/12/015",
      eprint         = "hep-ph/0509189",
      archivePrefix  = "arXiv",
%      primaryClass   = "hep-ph",
      reportNumber   = "TTP05-18, SFB-CPP-05-49",
      SLACcitation   = "%%CITATION = HEP-PH/0509189;%%"
}

@Article{Harlander:2003ai,
      author         = "Harlander, Robert V. and Kilgore, William B.",
      title          = "{Higgs boson production in bottom quark fusion at
                        next-to-next-to-leading order}",
      journal        = "Phys. Rev. D",
      volume         = "68",
      year           = "2003",
      pages          = "013001",
      doi            = "10.1103/PhysRevD.68.013001",
      eprint         = "hep-ph/0304035",
      archivePrefix  = "arXiv",
%      primaryClass   = "hep-ph",
      reportNumber   = "BNL-HET-03-4, CERN-TH-2003-067",
      SLACcitation   = "%%CITATION = HEP-PH/0304035;%%"
}

@Article{Harlander:2002wh,
      author         = "Harlander, Robert V. and Kilgore, William B.",
      title          = "{Next-to-Next-to-Leading Order Higgs Production at Hadron
                        Colliders}",
      journal        = "Phys. Rev. Lett.",
      volume         = "88",
      year           = "2002",
      pages          = "201801",
      doi            = "10.1103/PhysRevLett.88.201801",
      eprint         = "hep-ph/0201206",
      archivePrefix  = "arXiv",
%      primaryClass   = "hep-ph",
      reportNumber   = "BNL-HET-02-3, CERN-TH-2002-006",
      SLACcitation   = "%%CITATION = HEP-PH/0201206;%%"
}

@Article{Dawson:2003kb,
      author         = "Dawson, S. and Jackson, C. B. and Reina, L. and
                        Wackeroth, D.",
      title          = "{Exclusive Higgs boson production with bottom quarks at
                        hadron colliders}",
      journal        = "Phys. Rev. D",
      volume         = "69",
      year           = "2004",
      pages          = "074027",
      doi            = "10.1103/PhysRevD.69.074027",
      eprint         = "hep-ph/0311067",
      archivePrefix  = "arXiv",
%      primaryClass   = "hep-ph",
      reportNumber   = "BNL-HET-03-17, FSU-HEP-2003-1015, UB-HET-03-05",
      SLACcitation   = "%%CITATION = HEP-PH/0311067;%%"
}

@Article{Dittmaier:2003ej,
      author         = "Dittmaier, Stefan and Kr{\"a}mer, Michael and Spira,
                        Michael",
      title          = "{Higgs radiation off bottom quarks at the Fermilab Tevatron and
                        the CERN LHC}",
      journal        = "Phys. Rev. D",
      volume         = "70",
      year           = "2004",
      pages          = "074010",
      doi            = "10.1103/PhysRevD.70.074010",
      eprint         = "hep-ph/0309204",
      archivePrefix  = "arXiv",
%      primaryClass   = "hep-ph",
      reportNumber   = "EDINBURGH-2003-13, MPP-2003-49, PSI-PR-03-14",
      SLACcitation   = "%%CITATION = HEP-PH/0309204;%%"
}

@Article{Harlander:2011aa,
      author         = "Harlander, Robert and Kr{\"a}mer, Michael and Schumacher,
                        Markus",
      title          = "{Bottom-quark associated Higgs-boson production:
                        reconciling the four- and five-flavour scheme approach}",
      year           = "2011",
      eprint         = "1112.3478",
      archivePrefix  = "arXiv",
      primaryClass   = "hep-ph",
      reportNumber   = "CERN-PH-TH-2011-134, FR-PHENO-2011-009, TTK-11-17,
                        WUB-11-04",
      SLACcitation   = "%%CITATION = ARXIV:1112.3478;%%"
}

@Article{Duchon1976,
author = {Duchon, Jean},
journal = {ESAIM: Mathematical Modelling and Numerical Analysis - Modélisation Mathématique et Analyse Numérique},
number = {R3},
pages = {5-12},
publisher = {Dunod},
title = {Interpolation des fonctions de deux variables suivant le principe de la flexion des plaques minces},
url = {https://eudml.org/doc/193284},
volume = {10},
year = {1976},
}

@article{PhysRevD.67.075019,
  title = {CP-conserving two-Higgs-doublet model: The approach to the decoupling limit},
  author = {Gunion, John F. and Haber, Howard E.},
  journal = {Phys. Rev. D},
  volume = {67},
  pages = {075019},
  numpages = {26},
  year = {2003},
  month = {Apr},
  publisher = {American Physical Society},
  doi = {10.1103/PhysRevD.67.075019},
%  url = {https://link.aps.org/doi/10.1103/PhysRevD.67.075019}
}

@Article{Gehrmann:2014fva,
      author         = "Gehrmann, T. and Grazzini, M. and Kallweit, S. and
                        Maierhöfer, P. and von Manteuffel, A. and Pozzorini, S.
                        and Rathlev, D. and Tancredi, L.",
      title          = "{$W^+W^-$ Production at Hadron Colliders in Next to Next
                        to Leading Order QCD}",
      journal        = "Phys. Rev. Lett.",
      volume         = "113",
      year           = "2014",
      number         = "21",
      pages          = "212001",
      doi            = "10.1103/PhysRevLett.113.212001",
      eprint         = "1408.5243",
      archivePrefix  = "arXiv",
      primaryClass   = "hep-ph",
      reportNumber   = "ZU-TH-29-14, MITP-14-053",
      SLACcitation   = "%%CITATION = ARXIV:1408.5243;%%"
}

@Article{Grazzini:2016swo,
      author         = "Grazzini, Massimiliano and Kallweit, Stefan and Rathlev,
                        Dirk and Wiesemann, Marius",
      title          = "{$W^{\pm}Z$ production at hadron colliders in NNLO QCD}",
      journal        = "Phys. Lett. B",
      volume         = "761",
      year           = "2016",
      pages          = "179-183",
      doi            = "10.1016/j.physletb.2016.08.017",
      eprint         = "1604.08576",
      archivePrefix  = "arXiv",
      primaryClass   = "hep-ph",
      reportNumber   = "ZU-TH-15-16, MITP-16-037, NSF-KITP-16-046, DESY-16-074",
      SLACcitation   = "%%CITATION = ARXIV:1604.08576;%%"
}

@Article{Grazzini:2015hta,
      author         = "Grazzini, Massimiliano and Kallweit, Stefan and Rathlev,
                        Dirk",
      title          = "{ZZ production at the LHC: Fiducial cross sections and
                        distributions in NNLO QCD}",
      journal        = "Phys. Lett. B",
      volume         = "750",
      year           = "2015",
      pages          = "407-410",
      doi            = "10.1016/j.physletb.2015.09.055",
      eprint         = "1507.06257",
      archivePrefix  = "arXiv",
      primaryClass   = "hep-ph",
      SLACcitation   = "%%CITATION = ARXIV:1507.06257;%%"
}

@Article{Cascioli:2014yka,
      author         = "Cascioli, F. and Gehrmann, T. and Grazzini, M. and
                        Kallweit, S. and Maierhöfer, P. and von Manteuffel, A.
                        and Pozzorini, S. and Rathlev, D. and Tancredi, L. and
                        Weihs, E.",
      title          = "{ZZ production at hadron colliders in NNLO QCD}",
      journal        = "Phys. Lett. B",
      volume         = "735",
      year           = "2014",
      pages          = "311-313",
      doi            = "10.1016/j.physletb.2014.06.056",
      eprint         = "1405.2219",
      archivePrefix  = "arXiv",
      primaryClass   = "hep-ph",
      reportNumber   = "ZU-TH-19-14, MITP-14-033",
      SLACcitation   = "%%CITATION = ARXIV:1405.2219;%%"
}

@Article{Kidonakis:2010tc,
      author         = "Kidonakis, Nikolaos",
      title          = "{Next-to-next-to-leading logarithm resummation for s-channel single top quark production}",
      journal        = "Phys. Rev. D",
      volume         = "{81}",
      year           = "2010",
      pages          = "054028",
      doi            = "10.1103/PhysRevD.81.054028",
      eprint         = "1001.5034",
      archivePrefix  = "arXiv",
      primaryClass   = "hep-ph",
      SLACcitation   = "%%CITATION = ARXIV:1001.5034;%%"
}

@Article{Kidonakis:2011wy,
      author         = "Kidonakis, Nikolaos",
      title          = "{Next-to-next-to-leading-order collinear and soft gluon
                        corrections for t-channel single top quark production}",
      journal        = "Phys. Rev. D",
      volume         = "83",
      year           = "2011",
      pages          = "091503",
      doi            = "10.1103/PhysRevD.83.091503",
      eprint         = "1103.2792",
      archivePrefix  = "arXiv",
      primaryClass   = "hep-ph",
      SLACcitation   = "%%CITATION = ARXIV:1103.2792;%%"
}

@Article{Kidonakis:2010ux,
      author         = "Kidonakis, Nikolaos",
      title          = "{Two-loop soft anomalous dimensions for single top quark
                        associated production with a W- or H-}",
      journal        = "Phys. Rev. D",
      volume         = "82",
      year           = "2010",
      pages          = "054018",
      doi            = "10.1103/PhysRevD.82.054018",
      eprint         = "1005.4451",
      archivePrefix  = "arXiv",
      primaryClass   = "hep-ph",
      SLACcitation   = "%%CITATION = ARXIV:1005.4451;%%"
}

@Article{Vitells:2011da,
      author         = "Vitells, Ofer and Gross, Eilam",
      title          = "{Estimating the significance of a signal in a
                        multi-dimensional search}",
      journal        = "Astropart. Phys.",
      volume         = "35",
      year           = "2011",
      pages          = "230-234",
      doi            = "10.1016/j.astropartphys.2011.08.005",
      eprint         = "1105.4355",
      archivePrefix  = "arXiv",
      primaryClass   = "astro-ph.IM",
      SLACcitation   = "%%CITATION = ARXIV:1105.4355;%%"
}

@article{Wiesemann:2014ioa,
      author         = "Wiesemann, M. and Frederix, R. and Frixione, S. and
                        Hirschi, V. and Maltoni, F. and Torrielli, P.",
      title          = "{Higgs production in association with bottom quarks}",
      journal        = "JHEP",
      volume         = "02",
      year           = "2015",
      pages          = "132",
      doi            = "10.1007/JHEP02(2015)132",
      eprint         = "1409.5301",
      archivePrefix  = "arXiv",
      primaryClass   = "hep-ph",
      reportNumber   = "CERN-PH-TH-2014-182, CP3-14-64, LPN14-114, MCNET-14-20,
                        ZU-TH-33-14",
      SLACcitation   = "%%CITATION = ARXIV:1409.5301;%%"
}

@article{HDBS-2018-10-p,
    author = "{ATLAS Collaboration}",
    title = "{Search for heavy diboson resonances in semileptonic final states in $pp$ collisions at $\sqrt{s}=13$ TeV with the ATLAS detector}",
    eprint = "2004.14636",
    archivePrefix = "arXiv",
    primaryClass = "hep-ex",
    reportNumber = "CERN-EP-2020-049",
    month = "4",
    year = "2020"
}

@article{Aaboud:2017fgj,
     author         = "{ATLAS Collaboration}",
     title = "{Search for $WW/WZ$ resonance production in $\ell \nu qq$ final states in $pp$ collisions at $\sqrt{s} =$ 13 TeV with the ATLAS detector}",
    eprint = "1710.07235",
    archivePrefix = "arXiv",
    primaryClass = "hep-ex",
    reportNumber = "CERN-EP-2017-223",
    doi = "10.1007/JHEP03(2018)042",
    journal = "JHEP",
    volume = "03",
    pages = "042",
    year = "2018"
}

@article{Djouadi:2005gj,
      author         = "Djouadi, Abdelhak",
      title          = "{The anatomy of electroweak symmetry breaking Tome II: The
                        Higgs bosons in the Minimal Supersymmetric Model}",
      journal        = "Phys. Rept.",
      volume         = "459",
      year           = "2008",
      pages          = "1-241",
      doi            = "10.1016/j.physrep.2007.10.005",
      eprint         = "hep-ph/0503173",
      archivePrefix  = "arXiv",
%      primaryClass   = "hep-ph",
      reportNumber   = "LPT-ORSAY-05-18",
      SLACcitation   = "%%CITATION = HEP-PH/0503173;%%"
}

@article{Abdallah:2015ter,
      author         = "Abdallah, Jalal and others",
      title          = "{Simplified models for dark matter searches at the LHC}",
      journal        = "Phys. Dark Univ.",
      volume         = "9-10",
      year           = "2015",
      pages          = "8-23",
      doi            = "10.1016/j.dark.2015.08.001",
      eprint         = "1506.03116",
      archivePrefix  = "arXiv",
      primaryClass   = "hep-ph",
      reportNumber   = "FERMILAB-PUB-15-283-CD, CERN-PH-TH-2015-139",
      SLACcitation   = "%%CITATION = ARXIV:1506.03116;%%"
}

@article{Kim:2008hd,
      author         = "Kim, Jihn E. and Carosi, Gianpaolo",
      title          = "{Axions and the strong CP problem}",
      journal        = "Rev. Mod. Phys.",
      volume         = "82",
      year           = "2010",
      pages          = "557-602",
      doi            = "10.1103/RevModPhys.82.557",
      eprint         = "0807.3125",
      archivePrefix  = "arXiv",
      primaryClass   = "hep-ph",
      SLACcitation   = "%%CITATION = ARXIV:0807.3125;%%"
}

@article{King:2003jb,
      author         = "King, S. F.",
      title          = "{Neutrino mass models}",
      journal        = "Rept. Prog. Phys.",
      volume         = "67",
      year           = "2004",
      pages          = "107-158",
      doi            = "10.1088/0034-4885/67/2/R01",
      eprint         = "hep-ph/0310204",
      archivePrefix  = "arXiv",
%      primaryClass   = "hep-ph",
      reportNumber   = "SHEP-03-15",
      SLACcitation   = "%%CITATION = HEP-PH/0310204;%%"
}

@article{Butterworth:2015oua,
      author         = "Butterworth, Jon and others",
      title          = "{PDF4LHC recommendations for LHC Run II}",
      journal        = "J. Phys. G",
      volume         = "43",
      year           = "2016",
      pages          = "023001",
      doi            = "10.1088/0954-3899/43/2/023001",
      eprint         = "1510.03865",
      archivePrefix  = "arXiv",
      primaryClass   = "hep-ph",
      reportNumber   = "OUTP-15-17P, SMU-HEP-15-12, TIF-UNIMI-2015-14,
                        LCTS-2015-27, CERN-PH-TH-2015-249",
      SLACcitation   = "%%CITATION = ARXIV:1510.03865;%%"
}

@article{Hoeche:2009rj, 
      author         = "H{\"o}che, Stefan and Krauss, Frank and Schumann, Steffen
                        and Siegert, Frank",
      title          = "{QCD matrix elements and truncated showers}",
      journal        = "JHEP",
      volume         = "05",
      pages          = "053",
      doi            = "10.1088/1126-6708/2009/05/053",
      year           = "2009",
      eprint         = "0903.1219",
      archivePrefix  = "arXiv",
      primaryClass   = "hep-ph",
      reportNumber   = "ZU-TH-02-09, IPPP-09-14, DCPT-09-28, HD-THEP-09-2,
                        MCNET-09-04",
      SLACcitation   = "%%CITATION = ARXIV:0903.1219;%%",
}

@article{Schumann:2007mg,
      author         = "Schumann, Steffen and Krauss, Frank",
      title          = "{A parton shower algorithm based on Catani-Seymour dipole
                        factorisation}",
      journal        = "JHEP",
      volume         = "03",
      pages          = "038",
      doi            = "10.1088/1126-6708/2008/03/038",
      year           = "2008",
      eprint         = "0709.1027",
      archivePrefix  = "arXiv",
      primaryClass   = "hep-ph",
}

@article{Hoeche:2012yf,
      author         = "H{\"o}che, Stefan and Krauss, Frank and Sch{\"o}nherr, Marek and
                        Siegert, Frank",
      title          = "{QCD matrix elements + parton showers. The NLO case}",
      journal        = "JHEP",
      volume         = "04",
      year           = "2013",
      pages          = "027",
      doi            = "10.1007/JHEP04(2013)027",
      eprint         = "1207.5030",
      archivePrefix  = "arXiv",
      primaryClass   = "hep-ph",
      reportNumber   = "SLAC-PUB-15191, IPPP-12-52, DCPT-12-104, LPN12-081,
                        FR-PHENO-2012-017, MCNET-12-09, --FR-PHENO-2012-017",
      SLACcitation   = "%%CITATION = ARXIV:1207.5030;%%"
}

@article{Gleisberg:2008fv,
      author         = "Gleisberg, Tanju and H{\"o}che, Stefan",
      title          = "{Comix, a new matrix element generator}",
      journal        = "JHEP",
      volume         = "12",
      pages          = "039",
      doi            = "10.1088/1126-6708/2008/12/039",
      year           = "2008",
      eprint         = "0808.3674",
      archivePrefix  = "arXiv",
      primaryClass   = "hep-ph",
}

@article{Bothmann:2019yzt,
    author = "Bothmann, Enrico and others",
    collaboration = "Sherpa",
    title = "{Event generation with Sherpa 2.2}",
    eprint = "1905.09127",
    archivePrefix = "arXiv",
    primaryClass = "hep-ph",
    reportNumber = "FERMILAB-PUB-19-218-T, SLAC-PUB-17433, IPPP/19/42, MCNET-19-11",
    doi = "10.21468/SciPostPhys.7.3.034",
    journal = "SciPost Phys.",
    volume = "7",
    number = "3",
    pages = "034",
    year = "2019"
}

@article{Hoeche:2011fd,
      author         = "H{\"o}che, Stefan and Krauss, Frank and Sch{\"o}nherr, Marek and
                        Siegert, Frank",
      title          = "{A critical appraisal of NLO+PS matching methods}",
      journal        = "JHEP",
      volume         = "09",
      year           = "2012",
      pages          = "049",
      doi            = "10.1007/JHEP09(2012)049",
      eprint         = "1111.1220",
      archivePrefix  = "arXiv",
      primaryClass   = "hep-ph",
      reportNumber   = "SLAC-PUB-14661, IPPP-11-67, DCPT-11-134, LPN11-58,
                        FR-PHENO-2011-019, MCNET-11-24",
      SLACcitation   = "%%CITATION = ARXIV:1111.1220;%%"
}

@article{Catani:2001cc,
      author         = "Catani, S. and Krauss, F. and Kuhn, R. and Webber, B. R.",
      title          = "{QCD Matrix Elements + Parton Showers}",
      journal        = "JHEP",
      volume         = "11",
      year           = "2001",
      pages          = "063",
      doi            = "10.1088/1126-6708/2001/11/063",
      eprint         = "hep-ph/0109231",
      archivePrefix  = "arXiv",
      reportNumber   = "CERN-TH-2000-367, CAVENDISH-HEP-00-03",
      SLACcitation   = "%%CITATION = HEP-PH/0109231;%%"
}

@article{Sjostrand:2014zea,
      author         = "Sj{\"o}strand, Torbj{\"o}rn and Ask, Stefan and Christiansen,
                        Jesper R. and Corke, Richard and Desai, Nishita and Ilten,
                        Philip and Mrenna, Stephen and Prestel, Stefan and
                        Rasmussen, Christine O. and Skands, Peter Z.",
      title          = "{An introduction to PYTHIA 8.2}",
      journal        = "Comput. Phys. Commun.",
      volume         = "191",
      year           = "2015",
      pages          = "159-177",
      doi            = "10.1016/j.cpc.2015.01.024",
      eprint         = "1410.3012",
      archivePrefix  = "arXiv",
      primaryClass   = "hep-ph",
      reportNumber   = "LU-TP-14-36, MCNET-14-22, CERN-PH-TH-2014-190,
                        FERMILAB-PUB-14-316-CD, DESY-14-178, SLAC-PUB-16122,
                        --FERMILAB-PUB-14-316-CD",
      SLACcitation   = "%%CITATION = ARXIV:1410.3012;%%"
}

@Article{EvtGen,
    author         = "{D. J. Lange}",
    title          = "{The EvtGen particle decay simulation package}",
    journal        = "Nucl. Instrum. Meth. A",
    volume         = "462",
    year           = "2001",
    pages          = "152",
    doi            = "10.1016/S0168-9002(01)00089-4",
    primaryClass   = "hep-ex",
}

@article{Denner:2016kdg,
      author         = "Denner, Ansgar and Dittmaier, Stefan and Hofer, Lars",
      title          = "{Collier: A fortran-based complex one-loop library in
                        extended regularizations}",
      journal        = "Comput. Phys. Commun.",
      volume         = "212",
      year           = "2017",
      pages          = "220-238",
      doi            = "10.1016/j.cpc.2016.10.013",
      eprint         = "1604.06792",
      archivePrefix  = "arXiv",
      primaryClass   = "hep-ph",
      reportNumber   = "FR-PHENO-2016-003, ICCUB-16-016",
      SLACcitation   = "%%CITATION = ARXIV:1604.06792;%%"
}

@article{Cascioli:2011va,
      author         = "Cascioli, Fabio and Maierh{\"o}fer, Philipp and Pozzorini,
                        Stefano",
      title          = "{Scattering Amplitudes with Open Loops}",
      journal        = "Phys. Rev. Lett.",
      volume         = "108",
      year           = "2012",
      pages          = "111601",
      doi            = "10.1103/PhysRevLett.108.111601",
      eprint         = "1111.5206",
      archivePrefix  = "arXiv",
      primaryClass   = "hep-ph",
      reportNumber   = "ZU-TH-23-11, LPN11-66",
      SLACcitation   = "%%CITATION = ARXIV:1111.5206;%%"
}

@article{Alioli:2009je,
    author = "Alioli, Simone and Nason, Paolo and Oleari, Carlo and Re, Emanuele",
    title = "{NLO single-top production matched with shower in POWHEG: s- and t-channel contributions}",
    eprint = "0907.4076",
    archivePrefix = "arXiv",
    primaryClass = "hep-ph",
    doi = "10.1088/1126-6708/2009/09/111",
    journal = "JHEP",
    volume = "09",
    pages = "111",
    year = "2009",
    note = "[Erratum: \href{https://doi.org/10.1007/JHEP02(2010)011}{JHEP 02, 011 (2010)}]"
}

@article{Re:2010bp,
      author         = "Re, Emanuele",
      title          = "{Single-top Wt-channel production matched with parton
                        showers using the POWHEG method}",
      journal        = "Eur. Phys. J. C",
      volume         = "71",
      year           = "2011",
      pages          = "1547",
      doi            = "10.1140/epjc/s10052-011-1547-z",
      eprint         = "1009.2450",
      archivePrefix  = "arXiv",
      primaryClass   = "hep-ph",
      reportNumber   = "IPPP-10-74, DCPT-10-148",
      SLACcitation   = "%%CITATION = ARXIV:1009.2450;%%"
}

@article{Frixione:2008yi,
      author         = "Frixione, Stefano and Laenen, Eric and Motylinski,
                        Patrick and Webber, Bryan R. and White, Chris D.",
      title          = "{Single-top hadroproduction in association with a W
                        boson}",
      journal        = "JHEP",
      volume         = "07",
      year           = "2008",
      pages          = "029",
      doi            = "10.1088/1126-6708/2008/07/029",
      eprint         = "0805.3067",
      archivePrefix  = "arXiv",
      primaryClass   = "hep-ph",
      reportNumber   = "CERN-TH-2008-104, ITP-UU-08-27, NIKHEF-2008-005,
                        ITFA-2008-16, CAVENDISH-HEP-08-06",
      SLACcitation   = "%%CITATION = ARXIV:0805.3067;%%"
}

@article{Anastasiou:2003ds,
      author         = "Anastasiou, Charalampos and Dixon, Lance J. and Melnikov,
                        Kirill and Petriello, Frank",
      title          = "{High-precision QCD at hadron colliders: Electroweak
                        gauge boson rapidity distributions at next-to-next-to leading order}",
      journal        = "Phys. Rev. D",
      volume         = "69",
      year           = "2004",
      pages          = "094008",
      doi            = "10.1103/PhysRevD.69.094008",
      eprint         = "hep-ph/0312266",
      archivePrefix  = "arXiv",
      reportNumber   = "SLAC-PUB-10288, UH-511-1042-03",
      SLACcitation   = "%%CITATION = HEP-PH/0312266;%%"
}

@article{Frixione:2007nw,
      author         = "Frixione, Stefano and Nason, Paolo and Ridolfi, Giovanni",
      title          = "{A positive-weight next-to-leading-order Monte Carlo for
                        heavy flavour hadroproduction}",
      journal        = "JHEP",
      volume         = "09",
      pages          = "126",
      doi            = "10.1088/1126-6708/2007/09/126",
      year           = "2007",
      eprint         = "0707.3088",
      archivePrefix  = "arXiv",
      primaryClass   = "hep-ph",
      SLACcitation   = "%%CITATION = ARXIV:0707.3088;%%",
}

@article{Nason:2009ai,
      author         = "Nason, Paolo and Oleari, Carlo",
      title          = "{NLO Higgs boson production via vector-boson fusion
                        matched with shower in POWHEG}",
      journal        = "JHEP",
      volume         = "02",
      year           = "2010",
      pages          = "037",
      doi            = "10.1007/JHEP02(2010)037",
      eprint         = "0911.5299",
      archivePrefix  = "arXiv",
      primaryClass   = "hep-ph",
      SLACcitation   = "%%CITATION = ARXIV:0911.5299;%%"
}

@article{Ciccolini:2003jy,
      author         = "Ciccolini, M. L. and Dittmaier, S. and Kr{\"a}mer, M.",
      title          = "{Electroweak radiative corrections to associated $WH$ and
                        $ZH$ production at hadron colliders}",
      journal        = "Phys. Rev. D",
      volume         = "68",
      year           = "2003",
      pages          = "073003",
      doi            = "10.1103/PhysRevD.68.073003",
      eprint         = "hep-ph/0306234",
      archivePrefix  = "arXiv",
      primaryClass   = "hep-ph",
      reportNumber   = "EDINBURGH-2003-05, MPI-PHT-2003-24",
      SLACcitation   = "%%CITATION = HEP-PH/0306234;%%"
}

@article{Brein:2003wg,
      author         = "Brein, Oliver and Djouadi, Abdelhak and Harlander,
                        Robert",
      title          = "{NNLO QCD corrections to the Higgs-strahlung processes at
                        hadron colliders}",
      journal        = "Phys. Lett. B",
      volume         = "579",
      year           = "2004",
      pages          = "149-156",
      doi            = "10.1016/j.physletb.2003.10.112",
      eprint         = "hep-ph/0307206",
      archivePrefix  = "arXiv",
      XprimaryClass   = "hep-ph",
      reportNumber   = "MPP-2003-35, CERN-TH-2003-161, PM-03-16",
      SLACcitation   = "%%CITATION = HEP-PH/0307206;%%"
}

@article{Brein:2011vx,
      author         = "Brein, Oliver and Harlander, Robert and Wiesemann, Marius
                        and Zirke, Tom",
      title          = "{Top-quark mediated effects in hadronic Higgs-Strahlung}",
      journal        = "Eur. Phys. J. C",
      volume         = "72",
      year           = "2012",
      pages          = "1868",
      doi            = "10.1140/epjc/s10052-012-1868-6",
      eprint         = "1111.0761",
      archivePrefix  = "arXiv",
      primaryClass   = "hep-ph",
      reportNumber   = "CERN-PH-TH-2011-268, FR-PHENO-2011-016, WUB-11-15",
      SLACcitation   = "%%CITATION = ARXIV:1111.0761;%%"
}

@article{Altenkamp:2012sx,
      author         = "Altenkamp, Lukas and Dittmaier, Stefan and Harlander,
                        Robert V. and Rzehak, Heidi and Zirke, Tom J. E.",
      title          = "{Gluon-induced Higgs-strahlung at next-to-leading order
                        QCD}",
      journal        = "JHEP",
      volume         = "02",
      year           = "2013",
      pages          = "078",
      doi            = "10.1007/JHEP02(2013)078",
      eprint         = "1211.5015",
      archivePrefix  = "arXiv",
      primaryClass   = "hep-ph",
      reportNumber   = "CERN-PH-TH-2012-312, FR-PHENO-2012-023, WUB-12-21",
      SLACcitation   = "%%CITATION = ARXIV:1211.5015;%%"
}

@article{Denner:2014cla,
      author         = "Denner, Ansgar and Dittmaier, Stefan and Kallweit, Stefan
                        and M{\"u}ck, Alexander",
      title          = "{HAWK 2.0: A Monte Carlo program for Higgs production
                        in vector-boson fusion and Higgs strahlung at hadron
                        colliders}",
      journal        = "Comput. Phys. Commun.",
      volume         = "195",
      year           = "2015",
      pages          = "161-171",
      doi            = "10.1016/j.cpc.2015.04.021",
      eprint         = "1412.5390",
      archivePrefix  = "arXiv",
      primaryClass   = "hep-ph",
      reportNumber   = "FR-PHENO-2014-013, MITP-14-101, TTK-14-36",
      SLACcitation   = "%%CITATION = ARXIV:1412.5390;%%"
}

@article{Brein:2012ne,
      author         = "Brein, Oliver and Harlander, Robert V. and Zirke, Tom J.
                        E.",
      title          = "{vh@nnlo -- Higgs Strahlung at hadron colliders}",
      journal        = "Comput. Phys. Commun.",
      volume         = "184",
      year           = "2013",
      pages          = "998-1003",
      doi            = "10.1016/j.cpc.2012.11.002",
      eprint         = "1210.5347",
      archivePrefix  = "arXiv",
      primaryClass   = "hep-ph",
      SLACcitation   = "%%CITATION = ARXIV:1210.5347;%%"
}

@article{Harlander:2014wda,
      author         = "Harlander, Robert V. and Kulesza, Anna and Theeuwes,
                        Vincent and Zirke, Tom",
      title          = "{Soft gluon resummation for gluon-induced Higgs
                        Strahlung}",
      journal        = "JHEP",
      volume         = "11",
      year           = "2014",
      pages          = "082",
      doi            = "10.1007/JHEP11(2014)082",
      eprint         = "1410.0217",
      archivePrefix  = "arXiv",
      primaryClass   = "hep-ph",
      reportNumber   = "MS-TP-14-17, WUB-14-10, LPN14-112",
      SLACcitation   = "%%CITATION = ARXIV:1410.0217;%%"
}

@article{PERF-2018-01,
      author         = "{ATLAS Collaboration}",
      title          = "{Electron and photon performance measurements with the
                        ATLAS detector using the 2015-2017 LHC proton-proton
                        collision data}",
      collaboration  = "ATLAS",
      doi            = "10.1088/1748-0221/14/12/P12006",
      year           = "2019",
      eprint         = "1908.00005",
      archivePrefix  = "arXiv",
      primaryClass   = "hep-ex",
      reportNumber   = "CERN-EP-2019-145",
      journal        = "JINST",
      volume         = "14",
      pages          = "P12006",
      SLACcitation   = "%%CITATION = ARXIV:1908.00005;%%"
}

@article{Aad:2020uyd,
    author = "{ATLAS Collaboration}",
    collaboration = "ATLAS",
    title = "{Performance of the ATLAS muon triggers in Run 2}",
    eprint = "2004.13447",
    archivePrefix = "arXiv",
    primaryClass = "hep-ex",
    reportNumber = "CERN-EP-2020-031",
    doi = "10.1088/1748-0221/15/09/p09015",
    journal = "JINST",
    volume = "15",
    number = "09",
    pages = "P09015",
    year = "2020"
}

@article{PhysRevLett.13.321,
  title = {Broken Symmetry and the Mass of Gauge Vector Mesons},
  author = {Englert, F. and Brout, R.},
  journal = {Phys. Rev. Lett.},
  volume = {13},
  pages = {321--323},
  numpages = {0},
  year = {1964},
  month = {Aug},
  publisher = {American Physical Society},
  doi = {10.1103/PhysRevLett.13.321},
%  url = {https://link.aps.org/doi/10.1103/PhysRevLett.13.321}
}

@article{HIGGS1964132,
title = "Broken symmetries, massless particles and gauge fields",
journal = "Phys. Lett.",
volume = "12",
number = "2",
pages = "132 - 133",
year = "1964",
%issn = "0031-9163",
doi = "https://doi.org/10.1016/0031-9163(64)91136-9",
%url = "http://www.sciencedirect.com/science/article/pii/0031916364911369",
author = {Higgs, Peter W.}
}

@article{PhysRevLett.13.508,
  title = {Broken Symmetries and the Masses of Gauge Bosons},
  author = {Higgs, Peter W.},
  journal = {Phys. Rev. Lett.},
  volume = {13},
  pages = {508--509},
  numpages = {0},
  year = {1964},
  month = {Oct},
  publisher = {American Physical Society},
  doi = {10.1103/PhysRevLett.13.508},
%  url = {https://link.aps.org/doi/10.1103/PhysRevLett.13.508}
}

@article{PhysRevLett.13.585,
  title = {Global Conservation Laws and Massless Particles},
  author = {Guralnik, G. S. and Hagen, C. R. and Kibble, T. W. B.},
  journal = {Phys. Rev. Lett.},
  volume = {13},
  pages = {585--587},
  numpages = {0},
  year = {1964},
  month = {Nov},
  publisher = {American Physical Society},
  doi = {10.1103/PhysRevLett.13.585},
%  url = {https://link.aps.org/doi/10.1103/PhysRevLett.13.585}
}

@article{PhysRev.145.1156,
  title = {Spontaneous Symmetry Breakdown without Massless Bosons},
  author = {Higgs, Peter W.},
  journal = {Phys. Rev.},
  volume = {145},
  pages = {1156--1163},
  numpages = {0},
  year = {1966},
  month = {May},
  publisher = {American Physical Society},
  doi = {10.1103/PhysRev.145.1156},
%  url = {https://link.aps.org/doi/10.1103/PhysRev.145.1156}
}

@article{PhysRev.155.1554,
  title = {Symmetry Breaking in Non-Abelian Gauge Theories},
  author = {Kibble, T. W. B.},
  journal = {Phys. Rev.},
  volume = {155},
  pages = {1554--1561},
  numpages = {0},
  year = {1967},
  month = {Mar},
  publisher = {American Physical Society},
  doi = {10.1103/PhysRev.155.1554},
%  url = {https://link.aps.org/doi/10.1103/PhysRev.155.1554}
}

@article{Aad:2020ivc,
    author = "{ATLAS Collaboration}",
    title = "{$CP$ Properties of Higgs Boson Interactions with Top Quarks in the $t\bar{t}H$ and $tH$ Processes Using $H \rightarrow \gamma\gamma$ with the ATLAS Detector}",
    eprint = "2004.04545",
    archivePrefix = "arXiv",
    primaryClass = "hep-ex",
    reportNumber = "CERN-EP-2020-046",
    doi = "10.1103/PhysRevLett.125.061802",
    journal = "Phys. Rev. Lett.",
    volume = "125",
    pages = "061802",
    year = "2020"
}

@article{Sirunyan:2018sgc,
    author = "{CMS Collaboration}",
    collaboration = "CMS",
    title = "{Measurement and interpretation of differential cross sections for Higgs boson production at $\sqrt{s} =$ 13 TeV}",
    eprint = "1812.06504",
    archivePrefix = "arXiv",
    primaryClass = "hep-ex",
    reportNumber = "CMS-HIG-17-028, CERN-EP-2018-304",
    doi = "10.1016/j.physletb.2019.03.059",
    journal = "Phys. Lett. B",
    volume = "792",
    pages = "369--396",
    year = "2019"
}

@article{Sirunyan:2018koj,
    author = "{CMS Collaboration}",
    collaboration = "CMS",
    title = "{Combined measurements of Higgs boson couplings in proton--proton collisions at $\sqrt{s}=13\,\text {Te}\text {V} $}",
    eprint = "1809.10733",
    archivePrefix = "arXiv",
    primaryClass = "hep-ex",
    reportNumber = "CMS-HIG-17-031, CERN-EP-2018-263",
    doi = "10.1140/epjc/s10052-019-6909-y",
    journal = "Eur. Phys. J. C",
    volume = "79",
    number = "5",
    pages = "421",
    year = "2019"
}

@article{Sirunyan:2020sum,
    author = "{CMS Collaboration}",
    collaboration = "CMS",
    title = "{Measurements of $\mathrm{t\bar{t}}$H production and the CP structure of the Yukawa interaction between the Higgs boson and top quark in the diphoton decay channel}",
    eprint = "2003.10866",
    archivePrefix = "arXiv",
    primaryClass = "hep-ex",
    reportNumber = "CMS-HIG-19-013, CERN-EP-2020-028",
    doi = "10.1103/PhysRevLett.125.061801",
    journal = "Phys. Rev. Lett.",
    volume = "125",
    number = "6",
    pages = "061801",
    year = "2020"
}

@article{Khachatryan:2016vau,
    author = "{ATLAS and CMS Collaborations}",
    collaboration = "ATLAS, CMS",
    title = "{Measurements of the Higgs boson production and decay rates and constraints on its couplings from a combined ATLAS and CMS analysis of the LHC pp collision data at $ \sqrt{s}=7 $ and 8 TeV}",
    eprint = "1606.02266",
    archivePrefix = "arXiv",
    primaryClass = "hep-ex",
    reportNumber = "CERN-EP-2016-100, ATLAS-HIGG-2015-07, CMS-HIG-15-002",
    doi = "10.1007/JHEP08(2016)045",
    journal = "JHEP",
    volume = "08",
    pages = "045",
    year = "2016"
}

@Report{ATLAS-TDR-2010-19-with-add,
    author         = "{ATLAS Collaboration}",
      title         = "{ATLAS Insertable B-Layer Technical Design Report}",
      collaboration = "ATLAS Collaboration",
      number        = "CERN-LHCC-2010-013. ATLAS-TDR-19",
      month         = "Sep",
      year          = "2010",
      reportNumber  = "CERN-LHCC-2010-013",
      url           = "https://cds.cern.ch/record/1291633",
      addendum     = "{\textit{ATLAS Insertable B-Layer Technical Design Report Addendum}}, ATLAS-TDR-19-ADD-1, 2012, {\scriptsize{URL}}: \url{https://cds.cern.ch/record/1451888}",
}

@article{Abbott:2018ikt,
%    author = "{ATLAS IBL Collaboration}",
    author = "{B. Abbott et al}",
%    collaboration = "ATLAS IBL",
    title = "{Production and Integration of the ATLAS Insertable B-Layer}",
    eprint = "1803.00844",
    archivePrefix = "arXiv",
    primaryClass = "physics.ins-det",
    doi = "10.1088/1748-0221/13/05/T05008",
    journal = "JINST",
    volume = "13",
    number = "05",
    pages = "T05008",
    year = "2018"
}

@article{Aad:2020flx,
    author = "{ATLAS Collaboration}",
    collaboration = "ATLAS",
    title = "{Jet energy scale and resolution measured in proton-proton collisions at $\sqrt{s}=13$ TeV with the ATLAS detector}",
    eprint = "2007.02645",
    archivePrefix = "arXiv",
    primaryClass = "hep-ex",
    reportNumber = "CERN-EP-2020-083",
    doi = "10.3204/PUBDB-2020-02665",
    month = "7",
    year = "2020"
}

@article{LUCID2,
  author={G. Avoni and others},
  title={The new LUCID-2 detector for luminosity measurement and monitoring in ATLAS},
  journal={JINST},
  volume={13},
  number={07},
  pages={P07017},
  doi="10.1088/1748-0221/13/07/P07017",
  year={2018},
}

@Article{DAPR-2010-01,
    author         = "{ATLAS Collaboration}",
    title          = "{Luminosity determination in \(pp\) collisions at \(\sqrt{s} = 7\,\text{TeV}\) using the ATLAS detector at the LHC}",
    journal        = "Eur. Phys. J. C",
    volume         = "71",
    year           = "2011",
    pages          = "1630",
    doi            = "10.1140/epjc/s10052-011-1630-5",
    reportNumber   = "CERN-PH-EP-2010-069",
    eprint         = "1101.2185",
    archivePrefix  = "arXiv",
    primaryClass   = "hep-ex",
}

@Article{DAPR-2011-01,
    author         = "{ATLAS Collaboration}",
    title          = "{Improved luminosity determination in \(pp\) collisions at \(\sqrt{s} = 7\,\text{TeV}\) using the ATLAS detector at the LHC}",
    journal        = "Eur. Phys. J. C",
    volume         = "73",
    year           = "2013",
    pages          = "2518",
    doi            = "10.1140/epjc/s10052-013-2518-3",
    reportNumber   = "CERN-PH-EP-2013-026",
    eprint         = "1302.4393",
    archivePrefix  = "arXiv",
    primaryClass   = "hep-ex",
}

@Article{HIGG-2012-27,
    author         = "{ATLAS Collaboration}",
    title          = "{Observation of a new particle in the search for the Standard Model Higgs boson with the ATLAS detector at the LHC}",
    journal        = "Phys. Lett. B",
    volume         = "716",
    year           = "2012",
    pages          = "1",
    doi            = "10.1016/j.physletb.2012.08.020",
    reportNumber   = "CERN-PH-EP-2012-218",
    eprint         = "1207.7214",
    archivePrefix  = "arXiv",
    primaryClass   = "hep-ex",
}

@Article{STDM-2012-23,
    author         = "{ATLAS Collaboration}",
    title          = "{Measurement of the \(Z/\gamma^*\) boson transverse momentum distribution in \(pp\) collisions at \(\sqrt{s} = 7\,\text{TeV}\) with the ATLAS detector}",
    journal        = "JHEP",
    volume         = "09",
    year           = "2014",
    pages          = "145",
    doi            = "10.1007/JHEP09(2014)145",
    reportNumber   = "CERN-PH-EP-2014-075",
    eprint         = "1406.3660",
    archivePrefix  = "arXiv",
    primaryClass   = "hep-ex",
}

@Article{DAPR-2013-01,
    author         = "{ATLAS Collaboration}",
    title          = "{Luminosity determination in \(pp\) collisions at \(\sqrt{s} = 8\,\text{TeV}\) using the ATLAS detector at the LHC}",
    journal        = "Eur. Phys. J. C",
    volume         = "76",
    year           = "2016",
    pages          = "653",
    doi            = "10.1140/epjc/s10052-016-4466-1",
    eprint         = "1608.03953",
    archivePrefix  = "arXiv",
    primaryClass   = "hep-ex",
}

@Article{HIGG-2013-17,
    author         = "{ATLAS Collaboration}",
    title          = "{Study of the spin and parity of the Higgs boson in diboson decays with the ATLAS detector}",
    journal        = "Eur. Phys. J. C",
    volume         = "75",
    year           = "2015",
    pages          = "476",
    doi            = "10.1140/epjc/s10052-015-3685-1",
    reportNumber   = "CERN-PH-EP-2015-114",
    eprint         = "1506.05669",
    archivePrefix  = "arXiv",
    primaryClass   = "hep-ex",
    related        = "HIGG-2013-17-err",
    relatedstring  = "Erratum:",
}

@Article{PERF-2014-03,
    author         = "{ATLAS Collaboration}",
    title          = "{Performance of pile-up mitigation techniques for jets in \(pp\) collisions at \(\sqrt{s} = 8\,\text{TeV}\) using the ATLAS detector}",
    journal        = "Eur. Phys. J. C",
    volume         = "76",
    year           = "2016",
    pages          = "581",
    doi            = "10.1140/epjc/s10052-016-4395-z",
    reportNumber   = "CERN-PH-EP-2015-206",
    eprint         = "1510.03823",
    archivePrefix  = "arXiv",
    primaryClass   = "hep-ex",
}

@Article{PERF-2014-07,
    author         = "{ATLAS Collaboration}",
    title          = "{Topological cell clustering in the ATLAS calorimeters and its performance in LHC Run~1}",
    journal        = "Eur. Phys. J. C",
    volume         = "77",
    year           = "2017",
    pages          = "490",
    doi            = "10.1140/epjc/s10052-017-5004-5",
    reportNumber   = "CERN-PH-EP-2015-304",
    eprint         = "1603.02934",
    archivePrefix  = "arXiv",
    primaryClass   = "hep-ex",
}

@Article{PERF-2015-10,
    author         = "{ATLAS Collaboration}",
    title          = "{Muon reconstruction performance of the ATLAS detector in proton--proton collision data at \(\sqrt{s} = 13\,\text{TeV}\)}",
    journal        = "Eur. Phys. J. C",
    volume         = "76",
    year           = "2016",
    pages          = "292",
    doi            = "10.1140/epjc/s10052-016-4120-y",
    reportNumber   = "CERN-EP-2016-033",
    eprint         = "1603.05598",
    archivePrefix  = "arXiv",
    primaryClass   = "hep-ex",
}

@Article{STDM-2015-05,
    author         = "{ATLAS Collaboration}",
    title          = "{Measurement of the Inelastic Proton--Proton Cross Section at \(\sqrt{s} = 13\,\text{TeV}\) with the ATLAS Detector at the LHC}",
    journal        = "Phys. Rev. Lett.",
    volume         = "117",
    year           = "2016",
    pages          = "182002",
    doi            = "10.1103/PhysRevLett.117.182002",
    reportNumber   = "CERN-EP-2016-140",
    eprint         = "1606.02625",
    archivePrefix  = "arXiv",
    primaryClass   = "hep-ex",
}

@Article{EXOT-2016-10,
    author         = "{ATLAS Collaboration}",
    title          = "{Search for heavy resonances decaying into a \(W\) or \(Z\) boson and a Higgs boson in final states with leptons and \(b\)-jets in \(36\,\text{fb}^{-1}\) of \(\sqrt{s} = 13\,\text{TeV}\) \(pp\) collisions with the ATLAS detector}",
    journal        = "JHEP",
    volume         = "03",
    year           = "2018",
    pages          = "174",
    doi            = "10.1007/JHEP03(2018)174",
    reportNumber   = "CERN-EP-2017-250",
    eprint         = "1712.06518",
    archivePrefix  = "arXiv",
    primaryClass   = "hep-ex",
    related        = "EXOT-2016-10-err",
    relatedstring  = "Erratum:",
}

@Article{EXOT-2016-34,
    author         = "{ATLAS Collaboration}",
    title          = "{Search for a heavy Higgs boson decaying into a \(Z\) boson and another heavy Higgs boson in the \(\ell\ell bb\) final state in \(pp\) collisions at \(\sqrt{s} = 13\,\text{TeV}\) with the ATLAS detector}",
    journal        = "Phys. Lett. B",
    volume         = "783",
    year           = "2018",
    pages          = "392",
    doi            = "10.1016/j.physletb.2018.07.006",
    reportNumber   = "CERN-EP-2018-030",
    eprint         = "1804.01126",
    archivePrefix  = "arXiv",
    primaryClass   = "hep-ex",
}

@Article{PERF-2016-07,
    author         = "{ATLAS Collaboration}",
    title          = "{Performance of missing transverse momentum reconstruction with the ATLAS detector using proton--proton collisions at \(\sqrt{s} = 13\,\text{TeV}\)}",
    journal        = "Eur. Phys. J. C",
    volume         = "78",
    year           = "2018",
    pages          = "903",
    doi            = "10.1140/epjc/s10052-018-6288-9",
    reportNumber   = "CERN-EP-2017-274",
    eprint         = "1802.08168",
    archivePrefix  = "arXiv",
    primaryClass   = "hep-ex",
}

@Article{TRIG-2016-01,
    author         = "{ATLAS Collaboration}",
    title          = "{Performance of the ATLAS trigger system in 2015}",
    journal        = "Eur. Phys. J. C",
    volume         = "77",
    year           = "2017",
    pages          = "317",
    doi            = "10.1140/epjc/s10052-017-4852-3",
    reportNumber   = "CERN-EP-2016-241",
    eprint         = "1611.09661",
    archivePrefix  = "arXiv",
    primaryClass   = "hep-ex",
}

@Article{DAPR-2018-01,
    author         = "{ATLAS Collaboration}",
    title          = "{ATLAS data quality operations and performance for 2015--2018 data-taking}",
    journal        = "JINST",
    volume         = "15",
    year           = "2020",
    pages          = "P04003",
    doi            = "10.1088/1748-0221/15/04/P04003",
    reportNumber   = "CERN-EP-2019-207",
    eprint         = "1911.04632",
    archivePrefix  = "arXiv",
    primaryClass   = "physics.ins-det",
}

@Article{FTAG-2018-01,
    author         = "{ATLAS Collaboration}",
    title          = "{ATLAS \(b\)-jet identification performance and efficiency measurement with \(t\bar{t}\) events in \(pp\) collisions at \(\sqrt{s} = 13\,\text{TeV}\)}",
    journal        = "Eur. Phys. J. C",
    volume         = "79",
    year           = "2019",
    pages          = "970",
    doi            = "10.1140/epjc/s10052-019-7450-8",
    reportNumber   = "CERN-EP-2019-132",
    eprint         = "1907.05120",
    archivePrefix  = "arXiv",
    primaryClass   = "hep-ex",
}

@Article{HDBS-2018-31,
    author         = "{ATLAS Collaboration}",
    title          = "{Search for diboson resonances in hadronic final states in \(139\,\text{fb}^{-1}\) of \(pp\) collisions at \(\sqrt{s} = 13\,\text{TeV}\) with the ATLAS detector}",
    journal        = "JHEP",
    volume         = "09",
    year           = "2019",
    pages          = "091",
    doi            = "10.1007/JHEP09(2019)091",
    reportNumber   = "CERN-EP-2019-044",
    eprint         = "1906.08589",
    archivePrefix  = "arXiv",
    primaryClass   = "hep-ex",
}

@Article{HDBS-2018-46,
    author         = "{ATLAS Collaboration}",
    title          = "{Search for Heavy Higgs Bosons Decaying into Two Tau Leptons with the ATLAS Detector Using \(pp\) Collisions at \(\sqrt{s} = 13\,\text{TeV}\)}",
    journal        = "Phys. Rev. Lett.",
    volume         = "125",
    year           = "2020",
    pages          = "051801",
    doi            = "10.1103/PhysRevLett.125.051801",
    reportNumber   = "CERN-EP-2020-014",
    eprint         = "2002.12223",
    archivePrefix  = "arXiv",
    primaryClass   = "hep-ex",
}

@Article{HDBS-2018-58,
    author         = "{ATLAS Collaboration}",
    title          = "{Combination of searches for Higgs boson pairs in \(pp\) collisions at \(\sqrt{s} = 13\,\text{TeV}\) with the ATLAS detector}",
    journal        = "Phys. Lett. B",
    volume         = "800",
    year           = "2020",
    pages          = "135103",
    doi            = "10.1016/j.physletb.2019.135103",
    reportNumber   = "CERN-EP-2019-099",
    eprint         = "1906.02025",
    archivePrefix  = "arXiv",
    primaryClass   = "hep-ex",
}

@Article{HIGG-2018-14,
    author         = "{ATLAS Collaboration}",
    title          = "{Test of CP invariance in vector-boson fusion production of the Higgs boson in the \(H\to\tau\tau\) channel in proton--proton collisions at \(\sqrt{s} = 13\,\text{TeV}\) with the ATLAS detector}",
    journal        = "Phys. Lett. B",
    volume         = "805",
    year           = "2020",
    pages          = "135426",
    doi            = "10.1016/j.physletb.2020.135426",
    reportNumber   = "CERN-EP-2020-009",
    eprint         = "2002.05315",
    archivePrefix  = "arXiv",
    primaryClass   = "hep-ex",
}

@Article{HIGG-2018-57,
    author         = "{ATLAS Collaboration}",
    title          = "{Combined measurements of Higgs boson production and decay using up to \(80\,\text{fb}^{-1}\) of proton--proton collision data at \(\sqrt{s} = 13\,\text{TeV}\) collected with the ATLAS experiment}",
    journal        = "Phys. Rev. D",
    volume         = "101",
    year           = "2020",
    pages          = "012002",
    doi            = "10.1103/PhysRevD.101.012002",
    reportNumber   = "CERN-EP-2019-097",
    eprint         = "1909.02845",
    archivePrefix  = "arXiv",
    primaryClass   = "hep-ex",
}

@Article{TRIG-2018-05,
    author         = "{ATLAS Collaboration}",
    title          = "{Performance of electron and photon triggers in ATLAS during LHC Run~2}",
    journal        = "Eur. Phys. J. C",
    volume         = "80",
    year           = "2020",
    pages          = "47",
    doi            = "10.1140/epjc/s10052-019-7500-2",
    reportNumber   = "CERN-EP-2019-169",
    eprint         = "1909.00761",
    archivePrefix  = "arXiv",
    primaryClass   = "hep-ex",
}

@Booklet{ATL-SOFT-PUB-2020-001,
    author         = "{ATLAS Collaboration}",
    title          = "{ATLAS Computing Acknowledgements}",
    howpublished   = "{ATL-SOFT-PUB-2020-001}",
    url            = "https://cds.cern.ch/record/2717821",
}

@Article{CMS-HIG-12-028,
    author         = "{CMS Collaboration}",
    title          = "{Observation of a new boson at a mass of 125 GeV with the CMS experiment at the LHC}",
    journal        = "Phys. Lett. B",
    volume         = "716",
    year           = "2012",
    pages          = "30",
    doi            = "10.1016/j.physletb.2012.08.021",
    reportNumber   = "CERN-PH-EP-2012-220",
    eprint         = "1207.7235",
    archivePrefix  = "arXiv",
    primaryClass   = "hep-ex",
}

@Article{CMS-HIG-15-001,
    author         = "{CMS Collaboration}",
    title          = "{Search for neutral resonances decaying into a \(Z\) boson and a pair of \(b\) jets or \(\tau\) leptons}",
    journal        = "Phys. Lett. B",
    volume         = "759",
    year           = "2016",
    pages          = "369",
    doi            = "10.1016/j.physletb.2016.05.087",
    reportNumber   = "CERN-EP-2016-039",
    eprint         = "1603.02991",
    archivePrefix  = "arXiv",
    primaryClass   = "hep-ex",
}

@Article{CMS-HIG-16-018,
    author         = "{CMS Collaboration}",
    title          = "{Search for beyond the standard model Higgs bosons decaying into a \(b\bar{b}\) pair in \(pp\) collisions at \(\sqrt{s} = 13\,\text{TeV}\)}",
    journal        = "JHEP",
    volume         = "08",
    year           = "2018",
    pages          = "113",
    doi            = "10.1007/JHEP08(2018)113",
    reportNumber   = "CERN-EP-2018-124",
    eprint         = "1805.12191",
    archivePrefix  = "arXiv",
    primaryClass   = "hep-ex",
}

@Article{CMS-HIG-17-012,
    author         = "{CMS Collaboration}",
    title          = "{Search for a new scalar resonance decaying to a pair of \(Z\) bosons in proton--proton collisions at \(\sqrt{s} = 13\,\text{TeV}\)}",
    journal        = "JHEP",
    volume         = "06",
    year           = "2018",
    pages          = "127",
    doi            = "10.1007/JHEP06(2018)127",
    reportNumber   = "CERN-EP-2018-009",
    eprint         = "1804.01939",
    archivePrefix  = "arXiv",
    primaryClass   = "hep-ex",
}

@Article{CMS-HIG-17-020,
    author         = "{CMS Collaboration}",
    title          = "{Search for additional neutral MSSM Higgs bosons in the \(\tau\tau\) final state in proton--proton collisions at \(\sqrt{s} = 13\,\text{TeV}\)}",
    journal        = "JHEP",
    volume         = "09",
    year           = "2018",
    pages          = "007",
    doi            = "10.1007/JHEP09(2018)007",
    reportNumber   = "CERN-EP-2018-026",
    eprint         = "1803.06553",
    archivePrefix  = "arXiv",
    primaryClass   = "hep-ex",
}

@Article{CMS-HIG-17-030,
    author         = "{CMS Collaboration}",
    title          = "{Combination of Searches for Higgs Boson Pair Production in Proton--Proton Collisions at \(\sqrt{s} =  13\,\text{TeV}\)}",
    journal        = "Phys. Rev. Lett.",
    volume         = "122",
    year           = "2019",
    pages          = "121803",
    doi            = "10.1103/PhysRevLett.122.121803",
    reportNumber   = "CERN-EP-2018-292",
    eprint         = "1811.09689",
    archivePrefix  = "arXiv",
    primaryClass   = "hep-ex",
}

@Article{CMS-HIG-17-033,
    author         = "{CMS Collaboration}",
    title          = "{Search for a heavy Higgs boson decaying to a pair of \(W\) bosons in proton--proton collisions at \(\sqrt{s} = 13\,\text{TeV}\)}",
    journal        = "JHEP",
    volume         = "03",
    year           = "2020",
    pages          = "034",
    doi            = "10.1007/JHEP03(2020)034",
    reportNumber   = "CERN-EP-2019-230",
    eprint         = "1912.01594",
    archivePrefix  = "arXiv",
    primaryClass   = "hep-ex",
}

@Article{CMS-HIG-18-012,
    author         = "{CMS Collaboration}",
    title          = "{Search for new neutral Higgs bosons through the \(H \to ZA \to \ell^{+}\ell^{-} b\bar{b}\) process in \(pp\) collisions at \(\sqrt{s} = 13\,\text{TeV}\)}",
    journal        = "JHEP",
    volume         = "03",
    year           = "2020",
    pages          = "055",
    doi            = "10.1007/JHEP03(2020)055",
    reportNumber   = "CERN-EP-2019-254",
    eprint         = "1911.03781",
    archivePrefix  = "arXiv",
    primaryClass   = "hep-ex",
}

@Article{CMS-HIG-18-023,
    author         = "{CMS Collaboration}",
    title          = "{Search for a heavy pseudoscalar Higgs boson decaying into a \(125\,\text{GeV}\) Higgs boson and a \(Z\) boson in final states with two tau and two light leptons at \(\sqrt{s} = 13\,\text{TeV}\)}",
    journal        = "JHEP",
    volume         = "03",
    year           = "2020",
    pages          = "065",
    doi            = "10.1007/JHEP03(2020)065",
    reportNumber   = "CERN-EP-2019-231",
    eprint         = "1910.11634",
    archivePrefix  = "arXiv",
    primaryClass   = "hep-ex",
}

@Booklet{ATLAS-CONF-2018-023,
    author         = "{ATLAS Collaboration}",
    title          = "{\(E_{\text{T}}^{\text{miss}}\) performance in the ATLAS detector using 2015--2016 LHC \(pp\) collisions}",
    howpublished   = "{ATLAS-CONF-2018-023}",
    url            = "https://cds.cern.ch/record/2625233",
    year           = "2018",
}

@Booklet{ATLAS-CONF-2019-021,
    author         = "{ATLAS Collaboration}",
    title          = "{Luminosity determination in \(pp\) collisions at \(\sqrt{s} = 13\,\text{TeV}\) using the ATLAS detector at the LHC}",
    howpublished   = "{ATLAS-CONF-2019-021}",
    url            = "https://cds.cern.ch/record/2677054",
    year           = "2019",
}

@Booklet{ATLAS-CONF-2020-030,
    author         = "{ATLAS Collaboration}",
    title          = "{Muon reconstruction and identification efficiency in ATLAS using the full Run~2 \(pp\) collision data set at \(\sqrt{s} = 13\,\text{TeV}\)}",
    howpublished   = "{ATLAS-CONF-2020-030}",
    url            = "https://cds.cern.ch/record/2725736",
    year           = "2020",
}

@Booklet{ATL-PHYS-PUB-2014-021,
    author         = "{ATLAS Collaboration}",
    title          = "{ATLAS Pythia~8 tunes to \(7~\text{TeV}\) data}",
    howpublished   = "{ATL-PHYS-PUB-2014-021}",
    url            = "https://cds.cern.ch/record/1966419",
    year           = "2014",
}

@Booklet{ATL-PHYS-PUB-2016-017,
    author         = "{ATLAS Collaboration}",
    title          = "{The Pythia~8 A3 tune description of ATLAS minimum bias and inelastic measurements incorporating the Donnachie--Landshoff diffractive model}",
    howpublished   = "{ATL-PHYS-PUB-2016-017}",
    url            = "https://cds.cern.ch/record/2206965",
    year           = "2016",
}

@Booklet{ATL-PHYS-PUB-2016-020,
    author         = "{ATLAS Collaboration}",
    title          = "{Studies on top-quark Monte Carlo modelling for Top2016}",
    howpublished   = "{ATL-PHYS-PUB-2016-020}",
    url            = "https://cds.cern.ch/record/2216168",
    year           = "2016",
}

@Booklet{ATL-PHYS-PUB-2017-013,
    author         = "{ATLAS Collaboration}",
    title          = "{Optimisation and performance studies of the ATLAS \(b\)-tagging algorithms for the 2017-18 LHC run}",
    howpublished   = "{ATL-PHYS-PUB-2017-013}",
    url            = "https://cds.cern.ch/record/2273281",
    year           = "2017",
}

\clearpage % ATLAS Collaboration author list
% Reference date of HDBS-2018-13 is 2020-08-11
% Author list last updated on date 01-MAR-21
% Data extracted on 01-Mar-2021 for paper reference HDBS-2018-13
% at 2:13pm
 
\begin{flushleft}
\hypersetup{urlcolor=black}
{\Large The ATLAS Collaboration}

\bigskip

\AtlasOrcid[0000-0002-6665-4934]{G.~Aad}$^\textrm{\scriptsize 102}$,    
\AtlasOrcid[0000-0002-5888-2734]{B.~Abbott}$^\textrm{\scriptsize 128}$,    
\AtlasOrcid[0000-0002-7248-3203]{D.C.~Abbott}$^\textrm{\scriptsize 103}$,    
\AtlasOrcid[0000-0002-2788-3822]{A.~Abed~Abud}$^\textrm{\scriptsize 36}$,    
\AtlasOrcid[0000-0002-1002-1652]{K.~Abeling}$^\textrm{\scriptsize 53}$,    
\AtlasOrcid[0000-0002-2987-4006]{D.K.~Abhayasinghe}$^\textrm{\scriptsize 94}$,    
\AtlasOrcid[0000-0002-8496-9294]{S.H.~Abidi}$^\textrm{\scriptsize 29}$,    
\AtlasOrcid[0000-0002-8279-9324]{O.S.~AbouZeid}$^\textrm{\scriptsize 40}$,    
\AtlasOrcid{N.L.~Abraham}$^\textrm{\scriptsize 156}$,    
\AtlasOrcid[0000-0001-5329-6640]{H.~Abramowicz}$^\textrm{\scriptsize 161}$
% ...
\bigskip
\\

% ...
$^{29}$Physics Department, Brookhaven National Laboratory, Upton NY; United States of America.\\
% ...
$^{36}$CERN, Geneva; Switzerland.\\
% ...
$^{40}$Niels Bohr Institute, University of Copenhagen, Copenhagen; Denmark.\\
% ...
$^{53}$II. Physikalisches Institut, Georg-August-Universit\"{a}t G\"ottingen, G\"ottingen; Germany.\\
% ...
$^{94}$Department of Physics, Royal Holloway University of London, Egham; United Kingdom.\\
% ...
$^{102}$CPPM, Aix-Marseille Universit\'e, CNRS/IN2P3, Marseille; France.\\
$^{103}$Department of Physics, University of Massachusetts, Amherst MA; United States of America.\\
% ...
$^{128}$Homer L. Dodge Department of Physics and Astronomy, University of Oklahoma, Norman OK; United States of America.\\
% ...
$^{156}$Department of Physics and Astronomy, University of Sussex, Brighton; United Kingdom.\\
% ...
$^{161}$Raymond and Beverly Sackler School of Physics and Astronomy, Tel Aviv University, Tel Aviv; Israel.\\
% ...

\end{flushleft}

% Created with Glance <Atlas.Glance@cern.ch>

\end{document}